\documentclass[aos]{imsart}
\usepackage{amsthm,amsmath}
\newtheorem{theorem}{Theorem}
\newtheorem{lemma}{Lemma}
\newtheorem{corollary}{Corollary}
\RequirePackage[colorlinks,citecolor=blue,urlcolor=blue]{hyperref}
\usepackage{amsmath,rotating,subfigure}
\usepackage{amsfonts}
\usepackage{amsmath}
\usepackage{amssymb}
\usepackage{bbm}
\usepackage{array}
\usepackage{multirow,makecell,booktabs}
\usepackage{pdflscape}

\usepackage{times}
\usepackage{bm, color}
\usepackage{algorithm}
\usepackage[graphicx]{realboxes}
\usepackage{rotating}

\newcommand{\blue}{\textcolor{blue}}

  \newcommand{\3}{{(3)}}

\RequirePackage[colorlinks,citecolor=blue,urlcolor=blue]{hyperref}

\startlocaldefs
\endlocaldefs

\begin{document}

\begin{frontmatter}
\title{Distribution and correlation free two-sample test of high-dimensional means}
\runtitle{Distribution and correlation free two-sample mean test}

\begin{aug}
  \author{\fnms{Kaijie}  \snm{Xue}\corref{}\ead[label=e1]{kaijie@utstat.toronto.edu}}
  \and
  \author{\fnms{Fang} \snm{Yao}\ead[label=e2]{fyao@utstat.toronto.edu}}

%

  \affiliation{University of Texas MD Anderson Cancer Center, and University of Toronto}

 \address{K. Xue \\ Department of Biostatistics \\ University of Texas MD Anderson Cancer Center \\ 1400 Pressler Street \\ Houston, Texas 77030, U.S.A.\\
\printead{e1}}

\address{F. Yao \\  Corresponding author \\ Department of Statistical Sciences\\ University of Toronto \\ Toronto, Ontario M5S 3G3, Canada\\
\printead{e2}}



\end{aug}

\begin{abstract}
We propose a two-sample test for high-dimensional means that requires neither distributional nor correlational assumptions, besides some weak conditions on the moments and tail properties of the elements in the random vectors.
This two-sample test based on a nontrivial extension of the one-sample central limit theorem \cite{chern:16:1} provides a practically useful procedure with rigorous theoretical guarantees on its size and power assessment. In particular, the proposed test is easy to compute and does not require the independently and identically distributed assumption, which is allowed to have different distributions and arbitrary correlation structures.
Further desired features include weaker moments and tail conditions than existing methods, allowance for highly unequal sample sizes, consistent power behavior under fairly general alternative, data dimension allowed to be exponentially high under the umbrella of such general conditions.
Simulated and real data examples have demonstrated favorable numerical performance over existing methods.
\end{abstract}


\end{frontmatter}

\section{Introduction}\label{sec:intro}
Two-sample test of high dimensional means as one of the key issues has attracted a great deal of attention due to its importance in  various applications, including \cite{bai:96},
\cite{chen:10:1}, \cite{sri:13:1}, \cite{cai:14:1}, \cite{yama:15:1}, \cite{feng:15:1}, \cite{greg:15:1}, \cite{zhan:15:1}, \cite{zhang:16:1},  \cite{zhu:16:1}, \cite{hu:17:1}, \cite{chanj:17:1} and \cite{xu:17:1}, among others. In this article, we tackle this problem with the theoretical advance brought by a high-dimensional two-sample central limit theorem.  Based on this, we propose a {new type of} testing procedure, called distribution and correlation free (DCF) two-sample mean test, which requires neither distributional nor correlational assumptions and greatly enhances its generality in  practice.

We denote two samples by $X^n=\{X_1,\dots,X_n\}$ and $Y^m=\{Y_1,\dots,Y_m\}$ respectively, where $X^n$ is  a collection of mutually independent ({\em not necessarily identically distributed}) random vectors in $\mathbb{R}^p$ with $X_i=(X_{i1},\dots,X_{ip})'$  and $E(X_i)=\mu^X=(\mu_1^X,\dots,\mu_p^X)'$, $i=1,\dots,n$, and $Y^m$ is defined in a similar fashion with $E(Y_i)=\mu^Y=(\mu_1^Y,\dots,\mu_p^Y)'$ for all $i=1,\dots,m$. The normalized sums $S_n^X$ and $S_m^Y$ are denoted by
$S_n^X=n^{-1/2}\sum_{i=1}^n X_i=(S_{n1}^X,\dots,S_{np}^X)'$ and  $S_m^Y=m^{-1/2}\sum_{i=1}^m Y_i=(S_{m1}^Y,\dots,S_{mp}^Y)'$, respectively. Note that we only assume independent observations, and each sample with a common mean. The hypothesis of interest is
\begin{align*}
H_0: \mu^X=\mu^Y\quad \text{v.s.} \quad H_a: \mu^X\neq\mu^Y,
\end{align*}
and the proposed  two-sample DCF mean test is such that we reject $H_0: \mu^X=\mu^Y$ at significance level $\alpha\in(0, 1)$, provided that
\begin{align}
T_n=\|S_n^X-n^{1/2}m^{-1/2}S_m^Y\|_{\infty}\geq c_B(\alpha), \nonumber
\end{align}
where $T_n=\|S_n^X-n^{1/2}m^{-1/2}S_m^Y\|_{\infty}$ is the test statistic that only depends on the infinity norm of the sample mean difference,  and $c_B(\alpha)$ that plays a central role in this test is a data-driven critical value defined in (\ref{eq:cb}) of Theorem~\ref{theorem:term5}. It is worth mentioning that $c_B(\alpha)$ is easy to compute via a multiplier bootstrap based on a set of independently and identically distributed (i.i.d.) standard normal random variables that are independent of the data, where the explicit calculation is described after (\ref{inequality:term131}). Note that the computation of the proposed test is of an order $O\{ n(p+N)\}$, more efficient than $O(Nnp)$ that is usually demanded by a general resampling method. In spite of the simple structure of  $T_n$, we shall illustrate its desirable theoretical properties and superior numerical performance in the rest of the article.

We emphasize that our {\em main contributions} reside on developing a practically useful test that is computationally efficient with rigorous theoretical guarantees given in Theorem~\ref{theorem:term5}--\ref{theorem:term7}. We begin with deriving nontrivial two-sample extensions of the one-sample central limit theorems and its corresponding bootstrap approximation theorems in high dimensions \cite{chern:16:1}, where we do not require the ratio between sample sizes $n/(n+m)$ to converge but merely reside within any open interval $(c_1, c_2)$, $0<c_1\le c_2<1$, as $n, m\rightarrow \infty$. Further, Theorem~\ref{theorem:term5} lays down a foundation for conducting the two-sample DCF mean test uniformly over all $\alpha\in(0, 1)$. The power of the proposed test is assessed in Theorem~\ref{theorem:term6} that establishes the asymptotic equivalence between the estimated and true versions. Moreover, the asymptotic power is shown consistent in Theorem~\ref{theorem:term7} under some general alternatives with no sparsity or correlation constraints.

The proposed test sets itself apart from existing methods by allowing for non i.i.d.  random vectors in both samples. The distribution-free feature is in the sense that, under the umbrella of some mild assumptions on the moments and tail properties of the coordinates, there is no other restriction on the distributions of those random vectors. {In contrast, existing literature require the random vectors within sample to be i.i.d.\cite{chen:10:1, xiao:18, cai:14:1, chanj:17:1},} and some methods further restrict the coordinates to follow a certain type of distribution, such as Gaussian or sub-Gaussian \cite{zhan:15:1, zhu:16:1}.  This feature sets the proposed test free of making assumptions such as i.i.d. or sub-Gaussianity, which is desirable as distributions of real data are often confounded by numerous factors unknown to researchers.
Another key feature is correlation-free in the sense  that individual random vectors may have different and arbitrary correlation structures. By contrast, most previous works assume not only a common within-sample correlation matrix, but also some structural conditions, such as those on trace  \cite{chen:10:1}, mixing conditions \cite{xu:17:1}, or bounded eigenvalues from below \cite{cai:14:1}. It is worth noting that our  assumptions on the moments and tail properties of the coordinates in random vectors are also weaker than those adopted in  literature, e.g., \cite{cai:14:1}, \cite{greg:15:1} and \cite{xu:17:1} assumed a common fixed upper bound to those moments, \cite{chen:10:1} and \cite{sri:13:1} allowed a portion of those moments to grow but paid a price on correlation assumptions.

We also stress that the proposed test possesses consistent power behavior under fairly general alternative (a mild separation lower bound on $\mu^X-\mu^Y$ in Theorem~\ref{theorem:term7}) with neither sparsity nor correlation conditions, while previous work requiring either sparsity  \cite{zhan:15:1} or structural assumption on signal strength \cite{chen:10:1, greg:15:1} or correlation \cite{xu:17:1},
or both \cite{cai:14:1}. Lastly, we point out that the data dimension $p$ can be exponentially high relative to the sample size under the umbrella of such mild assumptions. This is also favorable compared to previous work, as
\cite{chen:10:1}, \cite{cai:14:1} and \cite{xu:17:1} allowed such  ultrahigh dimensions under nontrivial conditions on either the distribution type (e.g., sub-Gaussian) or the  correlation structure (or both) as a tradeoff.


We conclude the introduction by noting relevant work on one-sample high-dimensional mean test, such as \cite{sriv:07:1}, \cite{sriv:08:1}, \cite{sriv:09:1}, \cite{zhon:13:1}, \cite{park:13:1}, \cite{yagi:14:1}, \cite{shen:15:1}, \cite{wang:15:1}, \cite{zhao:17:1}, and \cite{ayya:17:1}, among others. It is relatively easier to develop a one-sample DCF mean test with similar advantages based on results in \cite{chern:16:1}, thus is not pursued here.
The rest of the article is organized as follows.
In Section~\ref{sec:theo},  we present the two-sample high-dimensional central limit theorem, and the result on multiplier bootstrap for evaluating the Gaussian approximation.
In section~\ref{sec:theo2}, we establish the main result Theorem~\ref{theorem:term5} for conducting the proposed test,
and Theorem~\ref{theorem:term6} to approximate its power function, {followed by Theorem~\ref{theorem:term7} to analyze its asymptotic power under alternatives.}
Simulation study is carried out in Section \ref{sec:simulation} to compare with existing methods, and an application to
a real data example is presented in  Section~\ref{sec:realdata}.
{We collect the auxiliary lemmas and the proofs of the main results, Theorems~\ref{theorem:term5}--\ref{theorem:term7} in the Appendix, and delegate the proofs of Theorems~\ref{theorem:term2}--\ref{theorem:term4}, Corollary~\ref{lemma:term9},
 and the auxiliary lemmas to an online Supplementary Material for space economy.}


\section{Two-sample central limit theorem and multiplier bootstrap in high dimensions}\label{sec:theo}

In this section, we first present  an intelligible two-sample central limit theorem in high dimensions, which is derived from its more abstract version  in Lemma~\ref{theorem:term1} in the Appendix.
Then the result on the asymptotic equivalence between the Gaussian approximation  appeared in the two-sample central limit theorem and its multiplier bootstrap term  is also elaborated, whose abstract version can be referred to Lemma~\ref{theorem:term3}.

We first list some notations used throughout the paper. For two vectors $x=(x_1,\dots,x_p)\in \mathbb{R}^p$ and $y=(y_1,\dots,y_p)'\in \mathbb{R}^p$, write $x\leq y$ if $x_j\leq y_j$ for all $j=1,\dots,p$. For any $x=(x_1,\dots,x_p)'\in \mathbb{R}^p$ and $a\in \mathbb{R}$,  denote $x+a=(x_1+a,\dots,x_p+a)'$. For any $a, b\in \mathbb{R}$, use the notations $a\vee b=\max\{a, b\}$ and $a\wedge b=\min\{a, b\}$. For any two sequences of constants $a_n$ and $b_n$, write $a_n\lesssim b_n$ if $a_n \leq C b_n$ up to a universal constant $C>0$, and $a_n\sim b_n$ if $a_n \lesssim b_n$ and $b_n \lesssim a_n$.  For any matrix $A=(a_{ij})$, define $\|A\|_{\infty}=\max_{i, j}|a_{ij}|$. For any function $f:\mathbb{R}\to\mathbb{R}$, write $\|f\|_{\infty}=\sup_{z\in \mathbb{R}}|f(z)|$. For a smooth function $g:\mathbb{R}^p\to\mathbb{R}$, we adopt indices to represent the partial derivatives for brevity, for example, $\partial_j\partial_k\partial_l g=g_{jkl}$. For any $\alpha>0$, define the function $\psi_{\alpha}(x)=\exp(x^\alpha)-1$ for $x\in [0, \infty)$, then for any random variable $X$, define
\begin{align}\label{inequality:term1}
\|X\|_{\psi_{\alpha}}=\inf\{\lambda>0: E\{\psi_{\alpha}(|X|/\lambda)\}\leq 1\},
\end{align}
which is an Orlicz norm for $\alpha\in [1, \infty)$ and a quasi-norm for $\alpha\in (0, 1)$.

Denote $F^n=\{F_1,\dots,F_n\}$ as a set of mutually independent random vectors in $\mathbb{R}^p$ such that $F_i=(F_{i1},\dots,F_{ip})'$  and $F_i\sim N_p(\mu^X, E\{(X_i-\mu^X)(X_i-\mu^X)'\})$ for all $i=1,\dots,n$, which denotes a Gaussian approximation to $X^n$.
 Likewise, define a set of mutually independent random vectors $G^m=\{G_1,\dots,G_m\}$ in $\mathbb{R}^p$ such that $G_i=(G_{i1},\dots,G_{ip})'$  and $G_i\sim N_p(\mu^Y, E\{(Y_i-\mu^Y)(Y_i-\mu^Y)'\})$ for all $i=1,\dots,m$ to approximate $Y^m$.
The sets $X^n$, $Y^m$, $F^n$ and $G^m$ are assumed to be independent of each other.  To this end, denote the normalized sums $S_n^X$, $S_n^F$, $S_m^Y$ and $S_m^G$ by
$S_n^X=n^{-1/2}\sum_{i=1}^n X_i=(S_{n1}^X,\dots,S_{np}^X)'$, $S_n^F=n^{-1/2}\sum_{i=1}^n F_i=(S_{n1}^F,\dots,S_{np}^F)'$, $S_m^Y=m^{-1/2}\sum_{i=1}^m Y_i=(S_{m1}^Y,\dots,S_{mp}^Y)'$ and $S_m^G=m^{-1/2}\sum_{i=1}^m G_i=(S_{m1}^G,\dots,S_{mp}^G)'$, where $S_n^F$ and $S_m^G$ serve as the Gaussian approximations for $S_n^X$ and $S_m^Y$, respectively.
Lastly, denote a set of independent standard normal random variables $e^{n+m}=\{e_1,\dots,e_{n+m}\}$ that is independent of any of $X^n$, $F^n$, $Y^m$ and $G^m$.
\subsection{Two-sample central limit theorem in high dimensions}\label{subsec:1}

To introduce Theorem~\ref{theorem:term2}, a list of useful notations are given as follows. Denote
\begin{align*}
&L_n^X=\max_{1\leq j\leq p}\sum_{i=1}^nE(|X_{ij}-\mu_j^X|^3)/n, \qquad L_m^Y=\max_{1\leq j\leq p}\sum_{i=1}^mE(|Y_{ij}-\mu_j^Y|^3)/m.
\end{align*}
We denote the key quantity $\rho_{n, m}^{**}$ by
\begin{eqnarray} \label{eq:rho}
\rho_{n, m}^{**}&=&\sup_{A\in \mathcal{A}^{Re}}\big|P(S_n^X-n^{1/2}\mu^X+\delta_{n, m}S_m^Y-\delta_{n, m}m^{1/2}\mu^Y\in A)\\
&&-P(S_n^F-n^{1/2}\mu^X+\delta_{n, m}S_m^G-\delta_{n, m}m^{1/2}\mu^Y\in A)\big|, \nonumber
\end{eqnarray}
where $P(S_n^X-n^{1/2}\mu^X+\delta_{n, m}S_m^Y-\delta_{n, m}m^{1/2}\mu^Y\in A)$ represents the unknown probability of interest, and $P(S_n^F-n^{1/2}\mu^X+\delta_{n, m}S_m^G-\delta_{n, m}m^{1/2}\mu^Y\in A)$ serves as a Gaussian approximation to this probability of interest, and $\rho_{n, m}^{**}$ measures the error of approximation over all  hyperrectangles $A\in \mathcal{A}^{Re}$. Note that $\mathcal{A}^{Re}$ is the class of all hyperrectangles in $\mathbb{R}^p$ of the form $\{w\in \mathbb{R}^p: a_j\leq w_j \leq b_j \quad \text{for all} \quad j=1,\dots,p\}$ with $-\infty \leq a_j\leq b_j\leq \infty$ for all $j=1,\dots,p$.
 {By assuming more specific conditions, Theorem~\ref{theorem:term2}  gives a more explicit bound on $\rho_{n, m}^{**}$ compared to Lemma~\ref{theorem:term1}.}
\begin{theorem}\label{theorem:term2} 
For any sequence of constants $\delta_{n, m}$, assume we have the following conditions (a)--(e),
\begin{enumerate}
\item[(a)]There exist universal constants $\delta_1>\delta_2>0$ such that $\delta_2< |\delta_{n, m}|<\delta_1$.
\item[(b)]There exists a universal constant $b>0$ such that $$\min_{1\leq j \leq p}E\{(S_{nj}^X-n^{1/2}\mu_j^X+\delta_{n, m}S_{mj}^Y-\delta_{n, m}m^{1/2}\mu_j^Y)^2\}\geq b.$$
\item[(c)]There exists a sequence of constants $B_{n, m}\geq 1$ such that $L_n^X\leq B_{n, m}$ and $L_m^Y\leq B_{n, m}$.
\item[(d)]The sequence of constants $B_{n, m}$ defined in (c) also satisfies
\begin{align*}
\max_{1\leq i \leq n}\max_{1\leq j \leq p}E\{\exp(|X_{ij}-\mu_j^X|/B_{n, m})\}\leq& 2, \\
\max_{1\leq i \leq m}\max_{1\leq j \leq p}E\{\exp(|Y_{ij}-\mu_j^Y|/B_{n, m})\}\leq& 2.
\end{align*}
\item[(e)]There exists a universal constant $c_1>0$ such that
\begin{align*}
(B_{n, m})^2\{\log(pn)\}^7/n\leq c_1, \quad (B_{n, m})^2\{\log(pm)\}^7/m\leq c_1.
\end{align*}
\end{enumerate}	
Then we have the following property, where $\rho_{m, n}^{**}$ is defined in (\ref{eq:rho}),
\begin{align*}
\rho_{n,m}^{**}\leq& K_3\Big(\big[(B_{n, m})^2\{\log(pn)\}^7/n\big]^{1/6}+\big[(B_{n, m})^2\{\log(pm)\}^7/m\big]^{1/6}\Big),
\end{align*}	
for a universal constant $K_3>0$.
\end{theorem}

Conditions (a)--(c) correspond to the moment properties of the coordinates, and (d) concerns the tail properties. It follows from (a) and (b) that the moments {\em  on average} are bounded below away from zero, hence allowing certain proportion of these moments to converge to zero. This is weaker than previous work that usually require a uniform lower bound on all moments \cite{cai:14:1, greg:15:1, xu:17:1}. Condition (c) implies that the moments {\em on average} has an upper bound $B_{n, m}$ that can diverge to infinity without restriction on correlation, thus offers more flexibility than those in literature that demands either a fixed 
upper bound or a certain correlation structure or both.
To appreciate this, letting $B_{n, m}\sim n^{1/3}$, one notes that all the variances of the coordinates are allowed to be uniformly as large as $B_{n, m}^{2/3}\sim n^{2/9}\to \infty$ under condition (c), while no restriction on correlation is needed. As a comparison, if we assign a common covariance to two samples, say $\Sigma=(\Sigma_{jk})_{1\leq j, k\leq p}$ with each $\Sigma_{jk}=n^{2/9}\rho^{1\{j\neq k\}}$ for some constant $\rho\in (0, 1)$, then the trace condition in \cite{chen:10:1} implies that $p=o(1)$.
Compared with a fixed upper bound on the tails of the coordinates \cite{cai:14:1, xu:17:1}, condition (d) allows for uniformly diverging tails as long as $B_{n, m}\to \infty$. Condition (e) indicates that the data dimension $p$ can grow exponentially in $n$, provided that $B_{n, m}$ is of some appropriate order. These conditions as a whole set the basis for the so-called ``distribution and correlation free'' features. 

\subsection{Two-sample multiplier bootstrap in high dimensions}\label{subsec:2}
Due to the unknown probability in $\rho_{n,m}^{**}$ (\ref{eq:rho}) denoting the Gaussian approximation, it limits the applicability of the central limit theorem for inference. The idea is to adopt a multiplier bootstrap to approximate its Gaussian approximation, and quantify its approximation error bound.
Denote
\begin{align*}
&\Sigma^X=n^{-1}\sum_{i=1}^nE\{(X_i-\mu^X)(X_i-\mu^X)'\}, \quad \hat{\Sigma}^X=n^{-1}\sum_{i=1}^n(X_{i}-\bar{X})(X_{i}-\bar{X})',
\end{align*}
where $\bar{X}=n^{-1}\sum_{i=1}^nX_{i}=(\bar{X}_1,\dots,\bar{X}_p)'$. Analogously, denote $\Sigma^Y$, $\hat{\Sigma}^Y$ and $\bar{Y}$.
Now we introduce the multiplier bootstrap approximation in this context.
Let $e^{n+m}=\{e_1,\dots,e_{n+m}\}$ be a set of i.i.d. standard normal random variables independent of the data,  we further denote
\begin{equation} \label{eq:Se}
S_n^{eX}=n^{-1/2}\sum_{i=1}^n e_i(X_i-\bar{X}), \qquad S_m^{eY}=m^{-1/2}\sum_{i=1}^m e_{i+n}(Y_i-\bar{Y}),
\end{equation}
and it is obvious that $E_e(S_n^{eX}{S_n^{eX}}')=\hat{\Sigma}^X$ and  $E_e(S_n^{eY}{S_n^{eY}}')=\hat{\Sigma}^Y$,  where $E_e(\cdot)$ means the expectation with respect to $e^{n+m}$ only.
Then,   for any sequence of constants $\delta_{n, m}$ that depends on both $n$ and $m$, we denote the quantity of interest $\rho_{n, m}^{MB}$ by
\begin{eqnarray}
\label{eq:rhoMB} \rho_{n, m}^{MB}&=&\sup_{A\in \mathcal{A}^{Re}}\big|P_e(S_n^{eX}+\delta_{n, m}S_m^{eY}\in A)-  \\ &&P(S_n^F-n^{1/2}\mu^X+\delta_{n, m}S_m^G-\delta_{n, m}m^{1/2}\mu^Y\in A)\big|, \nonumber
\end{eqnarray}
where $P_e(\cdot)$ means the probability with respect to $e^{n+m}$ only, and $P_e(S_n^{eX}+\delta_{n, m}S_m^{eY}\in A)$ acts as the multiplier bootstrap approximation for the Gaussian approximation $P(S_n^F-n^{1/2}\mu^X+\delta_{n, m}S_m^G-\delta_{n, m}m^{1/2}\mu^Y\in A)$. In particular, $\rho_{n, m}^{MB}$ can be understood as a measure of  error between the two approximations over all  hyperrectangles $A\in \mathcal{A}^{Re}$.
%
The following theorem provides a more explicit bound on $\rho_{n,m}^{MB}$ in contrast to its abstract version stated in Lemma~\ref{theorem:term3} in the Appendix.
\begin{theorem}\label{theorem:term4}
For any sequence of constants $\delta_{n, m}$, assume we have the following conditions (a)--(e),
\begin{enumerate}
\item[(a)]There exists a universal constant $\delta_1>0$ such that $|\delta_{n, m}|<\delta_1$.
\item[(b)]There exists a universal constant $b>0$ such that $$\min_{1\leq j \leq p}E\{(S_{nj}^X-n^{1/2}\mu_j^X+\delta_{n, m}S_{mj}^Y-\delta_{n, m}m^{1/2}\mu_j^Y)^2\}\geq b.$$
\item[(c)]There exists a sequence of constants $B_{n, m}\geq 1$ such that
\begin{align*}
\max_{1\leq j \leq p}\sum_{i=1}^n E\{(X_{ij}-\mu_j^X)^4\}/n&\leq B_{n, m}^2, \\
\max_{1\leq j \leq p}\sum_{i=1}^m E\{(Y_{ij}-\mu_j^Y)^4\}/m&\leq B_{n, m}^2.
\end{align*}
\item[(d)]The sequence of constants $B_{n, m}$ defined in (c) also satisfies
\begin{align*}
\max_{1\leq i \leq n}\max_{1\leq j \leq p}E\{\exp(|X_{ij}-\mu_j^X|/B_{n, m})\}&\leq 2, \\
\max_{1\leq i \leq m}\max_{1\leq j \leq p}E\{\exp(|Y_{ij}-\mu_j^Y|/B_{n, m})\}&\leq 2.
\end{align*}
\item[(e)]There exists a sequence of constants $\alpha_{n, m}\in (0, e^{-1})$ such that
\begin{align*}
B_{n, m}^2\log^5(pn)\log^2(1/\alpha_{n, m})/n&\leq 1, \\
B_{n, m}^2\log^5(pm)\log^2(1/\alpha_{n, m})/m&\leq 1.
\end{align*}
\end{enumerate}	
Then there exists a universal constant $c^*>0$ such that with probability at least $1-\gamma_{n, m}$ where
\begin{align*}
\gamma_{n, m}=&(\alpha_{n, m})^{\log(pn)/3}+3(\alpha_{n, m})^{\log^{1/2}(pn)/c_*}+(\alpha_{n, m})^{\log(pm)/3}+\\
&3(\alpha_{n, m})^{\log^{1/2}(pm)/c_*}+(\alpha_{n, m})^{\log^3(pn)/6}+3(\alpha_{n, m})^{\log^3(pn)/c_*}+\\
&(\alpha_{n, m})^{\log^3(pm)/6}+3(\alpha_{n, m})^{\log^3(pm)/c_*},
\end{align*}
we have the following property, where $\rho_{n,m}^{MB}$ is defined in (\ref{eq:rhoMB}),
\begin{align*}
\rho_{n,m}^{MB}\lesssim&\{ B_{n, m}^2\log^5(pn)\log^2(1/\alpha_{n, m})/n\}^{1/6}+\\
&\{ B_{n, m}^2\log^5(pm)\log^2(1/\alpha_{n, m})/m\}^{1/6}.
\end{align*}	
\end{theorem}

Conditions (a)--(c) pertain to the moment properties of the coordinates, condition (d) concerns the tail properties, and condition (e) characterizes the order of $p$.
These conditions have the desirable features as those in Theorem~\ref{theorem:term2}, such as allowing for uniformly diverging moments and tails and so on. Moreover, by combining Theorem~\ref{theorem:term4} with a two-sample Borel-Cantelli Lemma (i.e., Lemma~\ref{lemma:term8}), where condition (f) is  needed for Lemma~\ref{lemma:term8}, one can deduce Corollary~\ref{lemma:term9} below,
which facilitates the derivation of  our main result in Theorem~\ref{theorem:term5}.
\begin{corollary}\label{lemma:term9}
For any sequence of constants $\delta_{n, m}$, assume the conditions (a)--(e) in Theorem~\ref{theorem:term4} hold. Also suppose that the condition (f) holds as follows,
\begin{enumerate}
\item[(f)]The sequence of constants $\gamma_{n, m}$ defined in Theorem~\ref{theorem:term4} also satisfies $$\sum_{n}\sum_{m}\gamma_{n, m}<\infty.$$
\end{enumerate}	
Then with probability one, we have the following property, where $\rho_{n, m}^{MB}$ is defined in (\ref{eq:rhoMB}),
\begin{align*}
\rho_{n,m}^{MB}\lesssim&\{ B_{n, m}^2\log^5(pn)\log^2(1/\alpha_{n, m})/n\}^{1/6}+\\
&\{ B_{n, m}^2\log^5(pm)\log^2(1/\alpha_{n, m})/m\}^{1/6}.
\end{align*}	
\end{corollary}


\section{Two-sample mean test  in high dimensions}\label{sec:theo2}
In this section, based on the  theoretical results from the preceding section, we first establish the main result, Theorem~\ref{theorem:term5}, which gives a confidence region for the mean difference $(\mu^X-\mu^Y)$ and, equivalently, the DCF test procedure. We note that the theoretical guarantee is uniform for all $\alpha \in (0, 1)$ with probability one.
\begin{theorem}\label{theorem:term5}
Assume we have the following conditions (a)--(e),
\begin{enumerate}
\item[(a)]$n/(n+m)\in (c_1, c_2)$, for some universal constants $0<c_1<c_2<1$.
\item[(b)]There exists a universal constant $b>0$ such that $$\min_{1\leq j \leq p}\big[E\{(S_{nj}^X-n^{1/2}\mu_j^X)^2\}+E\{(S_{mj}^Y-m^{1/2}\mu_j^Y)^2\}\big]\geq b.$$
\item[(c)]There exists a sequence of constants $B_{n, m}\geq 1$ such that
\begin{align*}
\max_{1\leq j \leq p}\sum_{i=1}^n E(|X_{ij}-\mu_j^X|^{k+2})/n\leq& B_{n, m}^k, \\
\max_{1\leq j \leq p}\sum_{i=1}^m E(|Y_{ij}-\mu_j^Y|^{k+2})/m\leq& B_{n, m}^k,
\end{align*}
for all $k=1,2.$
\item[(d)]The sequence of constants $B_{n, m}$ defined in (c) also satisfies
\begin{align*}
\max_{1\leq i \leq n}\max_{1\leq j \leq p}E\{\exp(|X_{ij}-\mu_j^X|/B_{n, m})\}\leq& 2, \\
 \max_{1\leq i \leq m}\max_{1\leq j \leq p}E\{\exp(|Y_{ij}-\mu_j^Y|/B_{n, m})\}\leq& 2.
\end{align*}
\item[(e)]$B_{n, m}^2\log^7(pn)/n\to 0$ as $n\to\infty$.
\end{enumerate}	
Then with probability one, the Kolmogorov distance between the distributions of the quantity $\|S_n^X-n^{1/2}m^{-1/2}S_m^Y-n^{1/2}(\mu^X-\mu^Y)\|_{\infty}$ and the quantity $\|S_n^{eX}-n^{1/2}m^{-1/2}S_m^{eY}\|_{\infty}$ satisfies
\begin{align*}
&\sup_{t\geq 0}\big|P(\|S_n^X-n^{1/2}m^{-1/2}S_m^Y-n^{1/2}(\mu^X-\mu^Y)\|_{\infty}\leq t)-\\
&P_e(\|S_n^{eX}-n^{1/2}m^{-1/2}S_m^{eY}\|_{\infty}\leq t)\big|
\lesssim\{B_{n, m}^2\log^7(pn)/n\}^{1/6},
\end{align*}	
where $S_n^{eX}$ and $S_m^{eY}$ are as in (\ref{eq:Se}), and $P_e(\cdot)$ means the probability with respect to $e^{n+m}$ only. Consequently,
\begin{align*}
&\sup_{\alpha\in (0, 1)}\big|P\{\|S_n^X-n^{1/2}m^{-1/2}S_m^Y-n^{1/2}(\mu^X-\mu^Y)\|_{\infty}\leq c_B(\alpha)\}-(1-\alpha)\big|\\
&\lesssim \{B_{n, m}^2\log^7(pn)/n\}^{1/6},
\end{align*}	
where
\begin{equation} \label{eq:cb}
c_B(\alpha)= \inf \{t \in \mathbb{R}: P_e(||S_n^{eX}-n^{1/2}m^{-1/2}S_m^{eY}||_\infty \leq t)\geq 1-\alpha \},
\end{equation}
for $\alpha\in (0, 1)$, where $S_n^{eX}$ and $S_m^{eY}$ are as in (\ref{eq:Se}), and $P_e(\cdot)$ denotes the probability with respect to $e^{n+m}$ only.
\end{theorem}
Note that condition (a) is on the relative sample sizes that allows the ratio $n/(n+m)$  to diverge within any  open interval $(c_1, c_2)$ for $0<c_1<c_2<1$, rather than demanding  convergence as in existing work.
Conditions (b) and (c) concern the moment properties of the coordinates,
while condition (d) is associated with the tail properties,
 and condition (e) quantifies the order of $p$. By inspection, these conditions are slightly stronger than those in Theorems~\ref{theorem:term2} and \ref{theorem:term4}, but still maintain all desired advantages. To appreciate such benefits, consider the following example.
\begin{align*}
&n/(n+m)\in (.1, .9), \quad B_{n, m}\sim n^{1/9},\quad \log p\sim n^{\alpha}, \quad \alpha\in (0, 1/9),\notag\\
&X_1,\dots, X_{\lfloor n/2\rfloor} \stackrel{i.i.d.}{\sim} N(0_p, \Sigma), \quad
X_{\lfloor n/2\rfloor+1},\dots, X_n \stackrel{i.i.d.}{\sim} N(0_p, 2\Sigma), \notag\\
&Y_1,\dots, Y_{\lfloor m/3\rfloor}\stackrel{i.i.d.}{\sim}  N(1_p, 3\Sigma), \quad
Y_{\lfloor m/3\rfloor+1},\dots, Y_m \stackrel{i.i.d.}{\sim} N(1_p, 4\Sigma),
\end{align*}
where $1_p$ is the vector of ones, and the covariance matrix $\Sigma=(\Sigma_{jk})\in \mathbb{R}^{p\times p}$ with each $\Sigma_{jk}=n^{2/27}\rho^{1\{j\ne k\}}$ for some constant $\rho\in (0, 1)$. Then, one can verify that this example fulfills all conditions in Theorem~\ref{theorem:term5}, but violates the assumptions in most existing articles which requires i.i.d samples or  trace conditions \cite{chen:10:1}.

From Theorem~\ref{theorem:term5}, the $100(1-\alpha)\%$ confidence region for $(\mu^X-\mu^Y)$ can be expressed as
\begin{align*}
\mbox{CR}_{1-\alpha}=\{\mu^X-\mu^Y: \|S_n^X-n^{1/2}m^{-1/2}S_m^Y-n^{1/2}(\mu^X-\mu^Y)\|_{\infty}\leq c_B(\alpha)\}.
\end{align*}
Equivalently, the proposed test procedure in (\ref{inequality:term131}) is such that, we reject $H_0: \mu^X=\mu^Y$ at significance level $\alpha\in(0, 1)$, if
\begin{align} \label{inequality:term131}
T_n=\|S_n^X-n^{1/2}m^{-1/2}S_m^Y\|_{\infty}\geq c_B(\alpha),
\end{align}
where the data-driven critical value $c_B(\alpha)$ in (\ref{eq:cb}) admits fast computation via the multiplier bootstrap using independent set of i.i.d. standard normal random variables, which is implemented as follows.
\begin{itemize}
\item Generate $N$ sets of standard normal random variables, each of size $(n+m)$, denoted by $e_1^{n+m}$, $\dots$, $e_N^{n+m}$  as random copies of  $e^{n+m}=\{e_1, \ldots, e_{n+m}\}$. Then calculate $N$ times of $T_n^e=||S_n^{eX}-n^{1/2}m^{-1/2}S_m^{eY}||_\infty$ while keeping $X^n$ and $Y^m$ fixed, where $S_n^{eX}$ and $S_m^{eY}$ are in (\ref{eq:Se}). These values are denoted as $\{T_{n}^{e1},\dots,T_{n}^{eN}\}$ whose $100(1-\alpha)$th quantile is used to approximate $c_B(\alpha)$.
\end{itemize}
It is easy to see that the computation of the DCF test is of the order $O\{ n(p+N)\}$, compared with $O(Nnp)$ that is usually demanded by a general resampling method.

According to (\ref{inequality:term131}), the true power function for the test can be formulated as
\begin{align}\label{inequality:term132}
Power(\mu^X-\mu^Y)=P\{\|S_n^X-n^{1/2}m^{-1/2}S_m^Y\|_{\infty}\geq c_B(\alpha) \, \big| \, \mu^X-\mu^Y\}.
\end{align}
To quantify the power of the DCF test, the expression (\ref{inequality:term132}) is not directly applicable since the distribution of $(S_n^X-n^{1/2}m^{-1/2}S_m^Y)$ is unknown. Motivated by Theorem~\ref{theorem:term5}, we propose another multiplier bootstrap approximation for $Power(\mu^X-\mu^Y)$, based on a different set of standard normal random variables $e^{*\,n+m}=\{e_1^*, \ldots, e_{n+m}^*\}$ independent of $e^{n+m}$ that are used to calculate $c_B(\alpha)$,
\begin{align}\label{inequality:term133}
&Power^*(\mu^X-\mu^Y)\\ \notag
=&P_{e^*}\{\|S_n^{e^*X}-n^{1/2}m^{-1/2}S_m^{e^*Y}+n^{1/2}(\mu^X-\mu^Y)\|_{\infty}\geq c_B(\alpha)\},
\end{align}
where $S_n^{e^*X}$ and $S_m^{e^*Y}$ are as defined in (\ref{eq:Se}) with $e^{* \, n+m}$ instead of $e^{n+m}$, and $P_{e^*}(\cdot)$ means the probability with respect to $e^{* \, n+m}$ only.
The following theorem  is devoted to establishing the asymptotic equivalence between $Power(\mu^X-\mu^Y)$ and $Power^*(\mu^X-\mu^Y)$ under the same conditions as those in Theorem~\ref{theorem:term5}. 

\begin{theorem}\label{theorem:term6}
Assume the conditions (a)--(e) in Theorem \ref{theorem:term5} hold,
then for any $\mu^X-\mu^Y\in \mathbb{R}^p$,  we have with probability one,
\begin{align*}
\big|Power^*(\mu^X-\mu^Y)-Power(\mu^X-\mu^Y)\big|\lesssim\{B_{n, m}^2\log^7(pn)/n\}^{1/6}.
\end{align*}
\end{theorem}

By inspection of the conditions in Theorem \ref{theorem:term6}, it is worth mentioning that neither sparsity nor correlation restriction is required, as opposed to previous work requiring  sparsity \cite{cai:14:1} for instance.
To appreciate this point, the asymptotic power under fairly general alternatives specified by  condition ($f$)  is analyzed in the theorem below.
\begin{theorem}\label{theorem:term7}
Assume the conditions (a)--(e) in Theorem \ref{theorem:term5} and that
\begin{enumerate}
\item[(f)]$\mathcal{F}_{n,m,p}=\{\mu^X\in \mathbb{R}^p, \mu^Y \in \mathbb{R}^p: \|\mu^X-\mu^Y\|_\infty\geq K_s \{B_{n, m}\log (pn)/n\}^{1/2}\}$, for a sufficiently large universal constant $K_s>0$.
\end{enumerate}
Then for any $\mu^X-\mu^Y\in \mathcal{F}_{n,m,p}$,  we have with probability tending to one,
$$ Power^*(\mu^X-\mu^Y)\to 1, \quad \text{as}\quad n\to \infty.$$
\end{theorem}
The set $\mathcal{F}_{n,m,p}$ in ($f$) imposes a lower bound on the separation between $\mu^X$ and $\mu^Y$, which is comparable to the assumption $\max_i |\delta_i/\sigma_{i, i}^{1/2}|\geq \{2\beta\log(p)/n\}^{1/2}$ in Theorem~2 in \cite{cai:14:1}. The latter is in fact a special case of condition ($f$) when the sequence $B_{n, m}$ is constant. It is worth mentioning that the asymptotic power converges to 1 under neither sparsity nor correlation assumptions in the context of our theorem.  In contrast, Theorem~2 in  \cite{cai:14:1} requires not only sparse alternatives, but also restrictions on the correlation structure, e.g., condition~1 in that theorem such that the eigenvalues of the correlation matrix $\text{diag}{(\Sigma)}^{-1/2}\Sigma \, \text{diag}{(\Sigma)}^{-1/2}$ is lower bounded by a positive universal constant. These comparisons
  reveal  that the proposed DCF is powerful for a broader range of alternatives. {We conclude this section by noting that the theory for the DCF-type test based on $L_2$-norm can also be of interest but is not yet established, which needs further investigation.}

\section{Simulation Studies}\label{sec:simulation}
In the two-sample test for high-dimensional means, methods that are frequently used and/or recently proposed  include those proposed by \cite{chen:10:1} (abbreviated as CQ, an $L_2$ norm test), \cite{cai:14:1} (abbreviated as CL, an $L_\infty$ norm test) and \cite{xu:17:1} (abbreviated as XL, a test combining $L_2$ and $L_\infty$ norms) tests. We conduct comprehensive simulation studies to compare our DCF test with these existing methods in terms of size and power under various settings. The two samples $X^n=\{X_i\}_{i=1}^n$ and $Y^m=\{Y_i\}_{i=1}^m$  have  sizes $(n, m)$, while the data dimension is chosen to be $p=1000$. Without loss of generality, we  let $\mu^X=0\in \mathbb{R}^p$. The structure of  $\mu^Y\in \mathbb{R}^p$ is controlled by  a signal strength parameter $\delta>0$ and a  sparsity level parameter $\beta\in [0, 1]$. To construct $\mu^Y$, in each scenario, we first generate a sequence of $i.i.d$ random variables $\theta_k\sim U(-\delta, \delta)$ for $k=1,\dots,p$ and keep them fixed in the  simulation under that scenario. {We set $\delta(r)=\{2r\log (p)/ (n\vee m)\}^{1/2}$ that gives appropriate scale of signal strength \cite{zhon:13:1,cai:14:1,chen:10:1}.} We take $\mu^Y=(\theta_1,\dots,\theta_{\lfloor \beta p \rfloor}, 0_{p-\lfloor \beta p \rfloor}')'\in \mathbb{R}^p$, where $\lfloor a \rfloor$ denotes the nearest integer no more than $a$, and $0_q$ is the $q$-dimensional vector of $0$'s.  Thus the signal becomes sparser for a smaller value of $\beta$, with  $\beta=0$ corresponding to the null hypothesis and {$\beta=1$} representing the fully dense alternative.
The covariance matrices of the random vectors are denoted by $\text{cov}(X_i)=\Sigma^{X_i}$,  $\text{cov}(Y_{i'})=\Sigma^{Y_{i'}}$ for all $i=1,\dots,n, i'=1,\dots,m$. The nominal significance level is $\alpha=.05$, and the DCF test is conducted based on the multiplier bootstrap of size $N=10^4$.

To have comprehensive comparison, we first consider the following {six} different settings. The first setting is standard with $(n, m, p)=(200, 300, 1000)$, where the elements in each sample are i.i.d Gaussian, and the two samples  share a common covariance matrix $\Sigma=(\Sigma_{jk})_{1\leq j, k\leq p}$. The matrix $\Sigma$ is specified by a dependence structure such that   $\Sigma_{jk}=(1+|j-k|)^{-1/4}$. {Beginning with $\delta=.1$, where the implicit chosen value $r=.217$ corresponds to quite weak signal according to  \cite{zhon:13:1,cai:14:1},  we calculate the rejection proportions of the four tests  based on $1000$ Monte Carlo runs over a full range of sparsity levels from $\beta=0$ (corresponding to null hypothesis) to $\beta=1$ (corresponding to fully dense alternative).}
Then the the signals  are {gradually strengthened  to $\delta=.15, .2, .25, .3$}. The second setting is similar to the first, except for
$\Sigma^{Y_i}=2\Sigma^{X_{i'}}=2\Sigma$ for all $i=1,\dots,n, i'=1,\dots,m$, where $\Sigma$ is defined in the first setting.
These two settings are denoted by ``$i.i.d$ equal (respectively, unequal) covariance setting''.

In the third setting, the random vectors in each sample have completely different distributions and covariance matrices from one another. The procedure to generate the two samples is as follows.
First, a set of parameters $\{\phi_{ij}: i=1,\dots,m, j=1,\dots,p\}$ are generated from the uniform distribution $U(1, 2)$ independently, and are kept fixed for all Monte Carlo runs.  In a similar fashion, $\{\phi_{ij}^*: i=1,\dots,m, j=1,\dots,p\}$ are generated from $U(1, 3)$ independently.
Then, for every $i=1,\dots,n$, we define a $p\times p$ matrix $\Omega_i=(\omega_{ijk})_{1\leq j, k\leq p}$ with each $\omega_{ijk}=(\phi_{ij}\phi_{ik})^{1/2}(1+|j-k|)^{-1/4}$.
 Likewise, for every $i=1,\dots,m$, define a $p\times p$ matrix $\Omega_i^*=(\omega_{ijk}^*)_{1\leq j, k\leq p}$ with each $\omega_{ijk}^*=(\phi_{ij}^*\phi_{ik}^*)^{1/2}(1+|j-k|)^{-1/4}$.
Subsequently, we generate a set of $i.i.d$ random vectors $\breve{X}^n=\{\breve{X}_i\}_{i=1}^n$  with each  $\breve{X}_i=(\breve{X}_{i1},\dots,\breve{X}_{ip})'\in \mathbb{R}^p$, such that $\{\breve{X}_{i1},\dots,\breve{X}_{i, 2p/5}\}$ are $i.i.d$ standard normal random variables, $\{\breve{X}_{i, 2p/5+1},\dots,\breve{X}_{i, p}\}$ are $i.i.d$ centered Gamma$(16, 1/4)$ random variables , and they are independent of each other.
Accordingly, we construct each $X_i$ by letting $X_i=\mu^X+\Omega_i^{1/2}\breve{X}_{i}$ for all $i=1,\dots,n$. It is worth noting that $\Sigma^{X_i}=\Omega_i$ for all $i=1,\dots,n$, i.e., $X_i$'s have different covariance matrices and distributions. The other sample $Y^m=\{Y_i\}_{i=1}^m$ is constructed in the same way with $\Sigma^{Y_i}=\Omega_i^*$ for all $i=1,\dots,m$.
Then we obtained the results for various signal  strength levels of $\delta$ over a full range of sparsity levels of $\beta$, and we denote this setting as ``completely relaxed''. The fourth setting is analogous  to the third, except that we set $(n, m, p)=(100, 400, 1000)$, where two sample sizes deviates substantially from each other. Since this setting is concerned with highly unequal sample sizes, and  is therefore denoted as ``completely relaxed and highly unequal setting''.
{The fifth setting is similar to the third, except that we replace the standard normal innovations in $\breve{X}_i$ and $\breve{Y}_{i'}$ by   independent and heavy-tailed  innovations $(5/3)^{-1/2}t(5)$ with mean zero and unit variances,} referred to as ``completely relaxed and heavy-tailed setting''.
{The sixth setting is also analogous to the third, while independent and skewed  innovations $8^{-1/2}\{\chi^2(4)-4\}$ with mean zero and unit variances are used, denoted by ``completely relaxed and skewed setting''.}

\begin{figure}[htbp]
\begin{minipage}[t]{0.19\textwidth}
\centering
\includegraphics[width=0.95in]{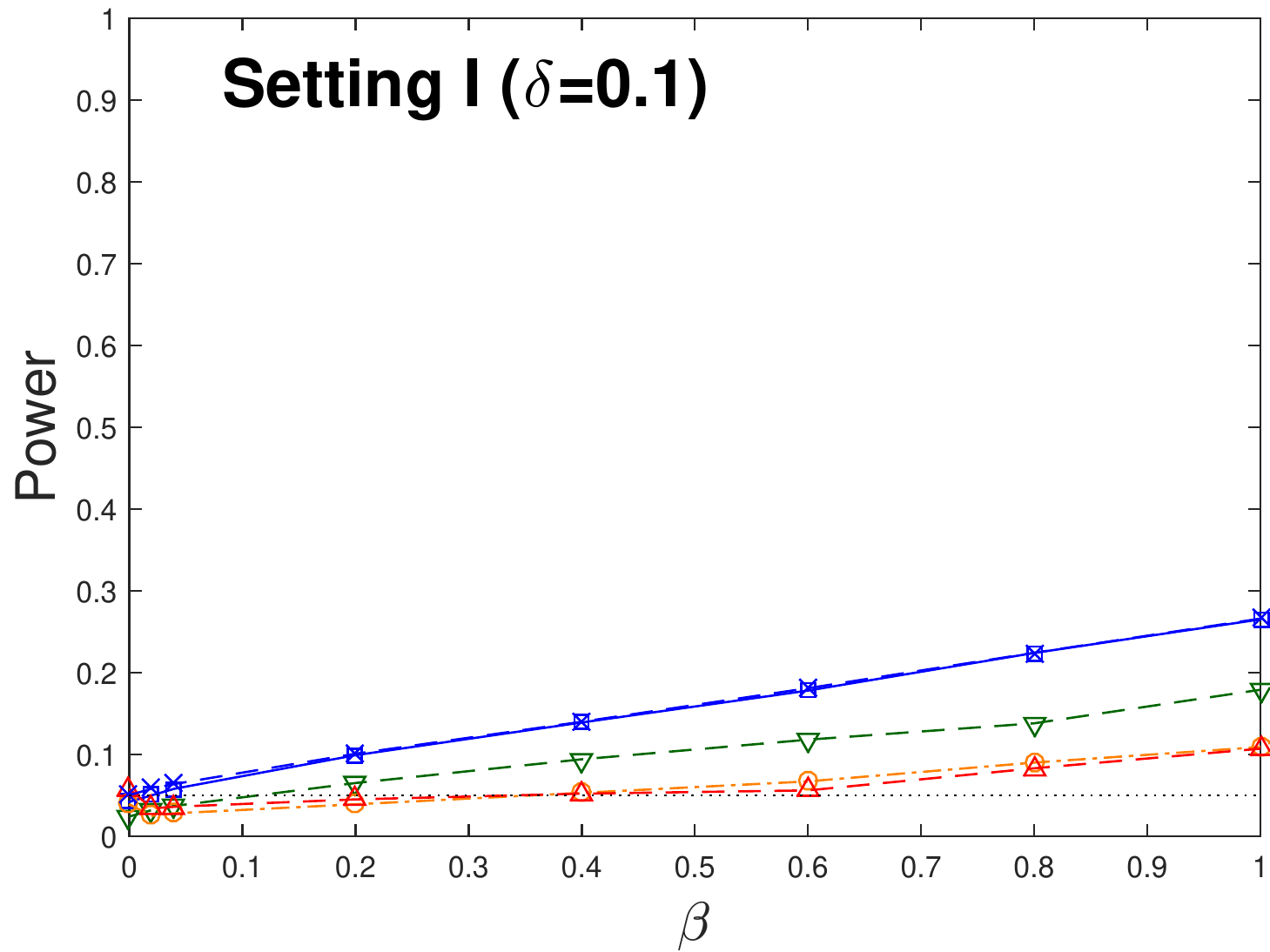}
\end{minipage}%
\begin{minipage}[t]{0.19\textwidth}
\centering
\includegraphics[width=0.95in]{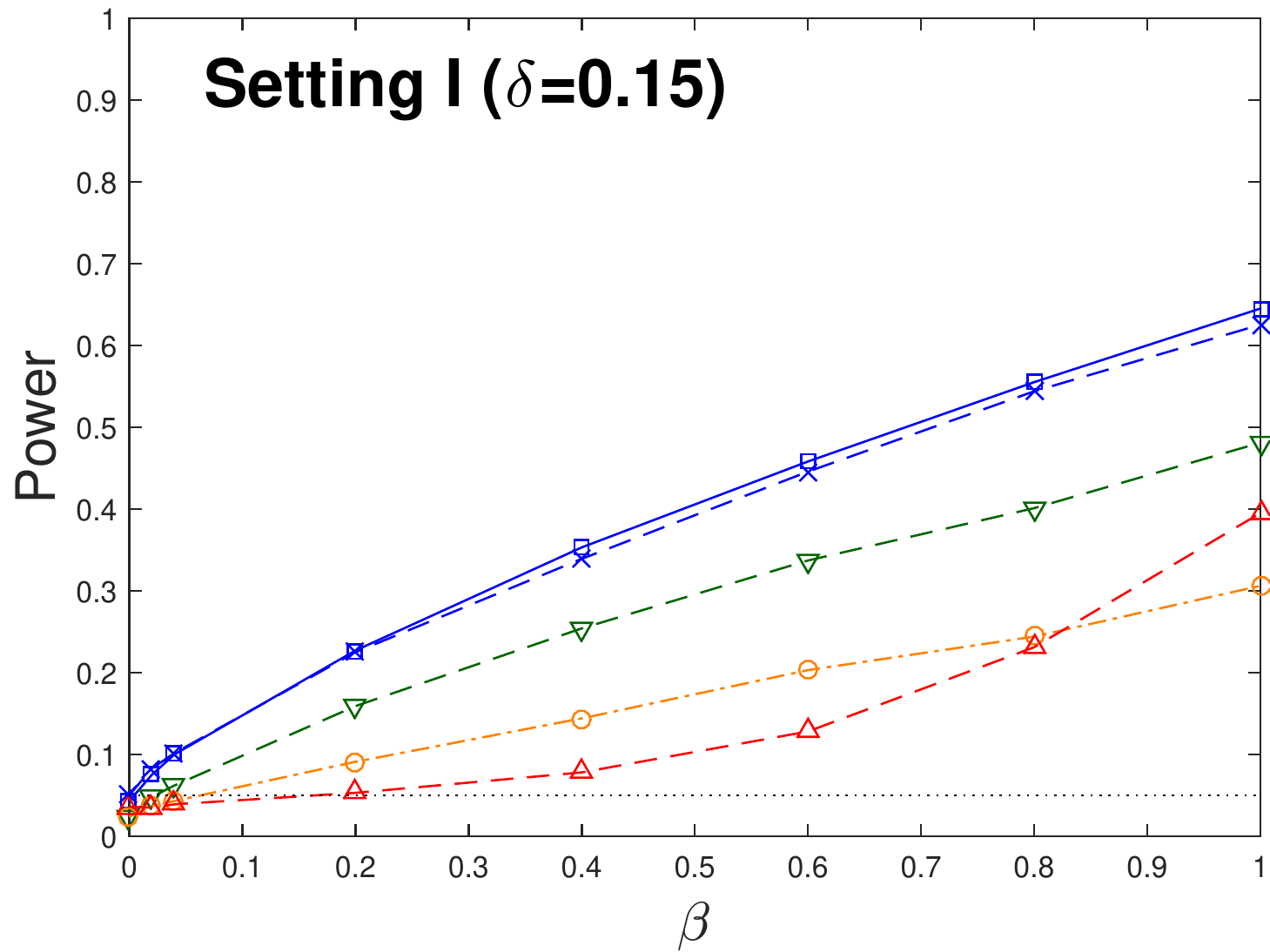}
\end{minipage}%
\begin{minipage}[t]{0.19\textwidth}
\centering
\includegraphics[width=0.95in]{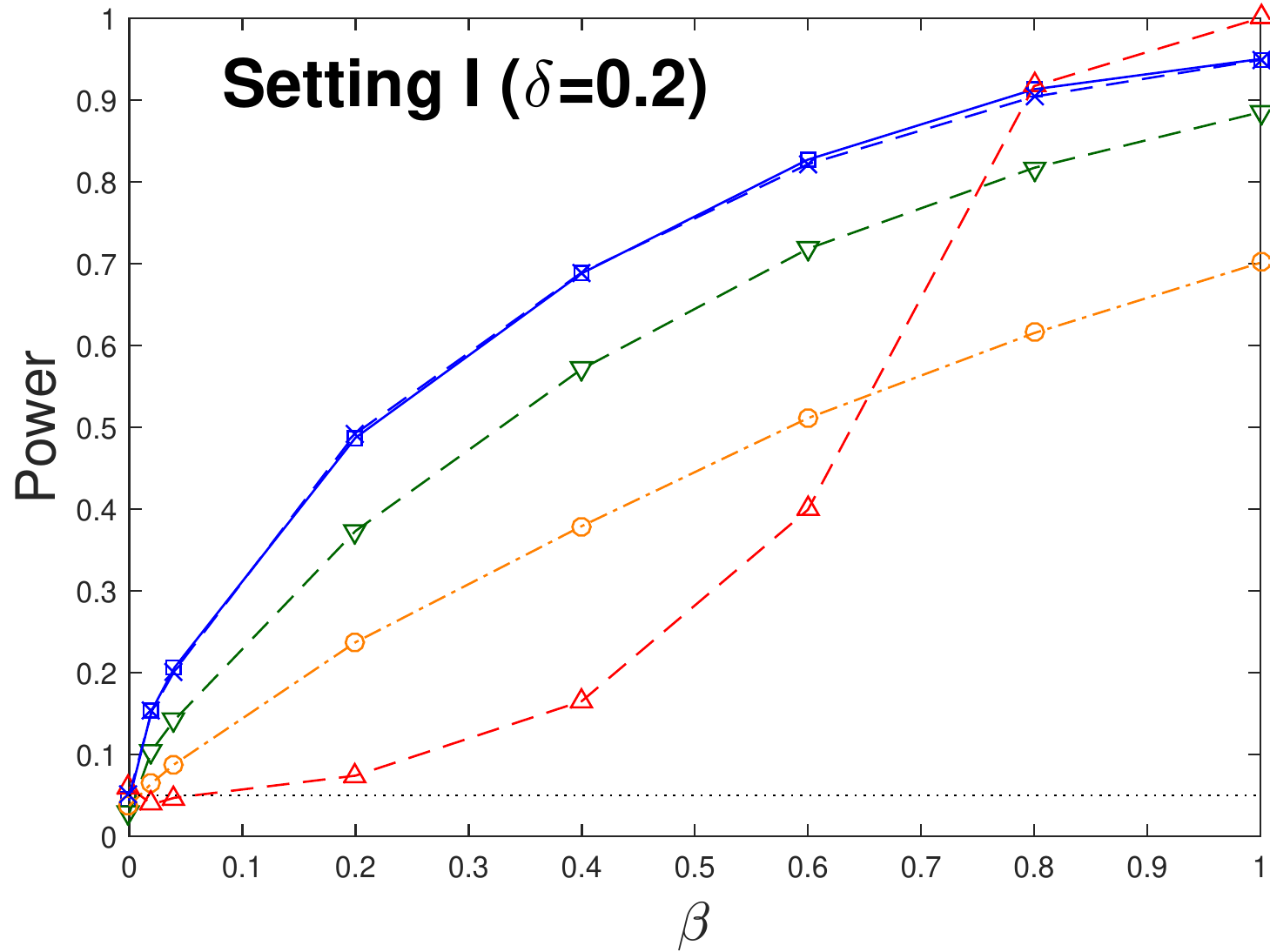}
\end{minipage}%
\begin{minipage}[t]{0.19\textwidth}
\centering
\includegraphics[width=0.95in]{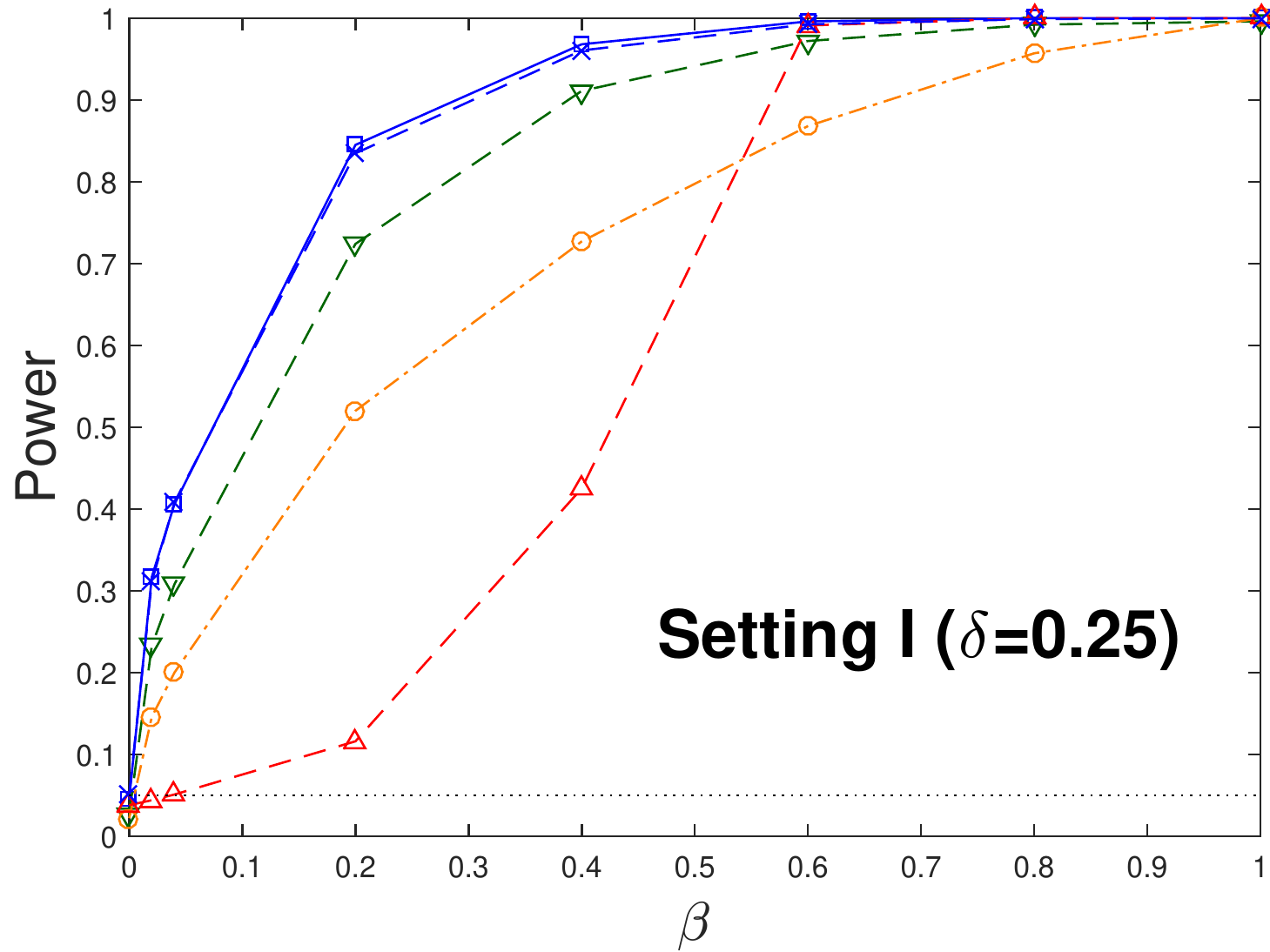}
\end{minipage}%
\begin{minipage}[t]{0.19\textwidth}
\centering
\includegraphics[width=0.95in]{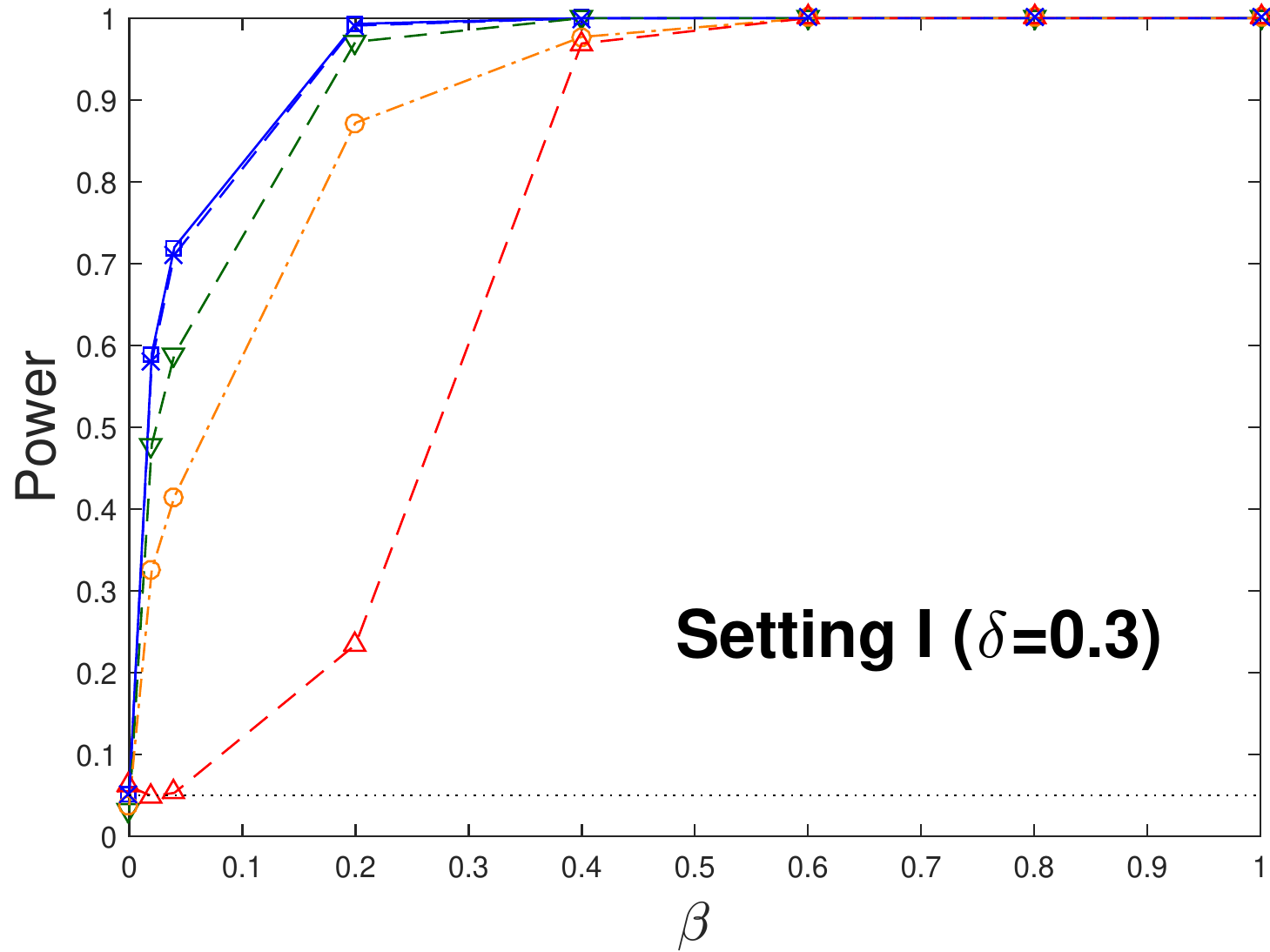}
\end{minipage}%

\begin{minipage}[t]{0.19\textwidth}
\centering
\includegraphics[width=0.95in]{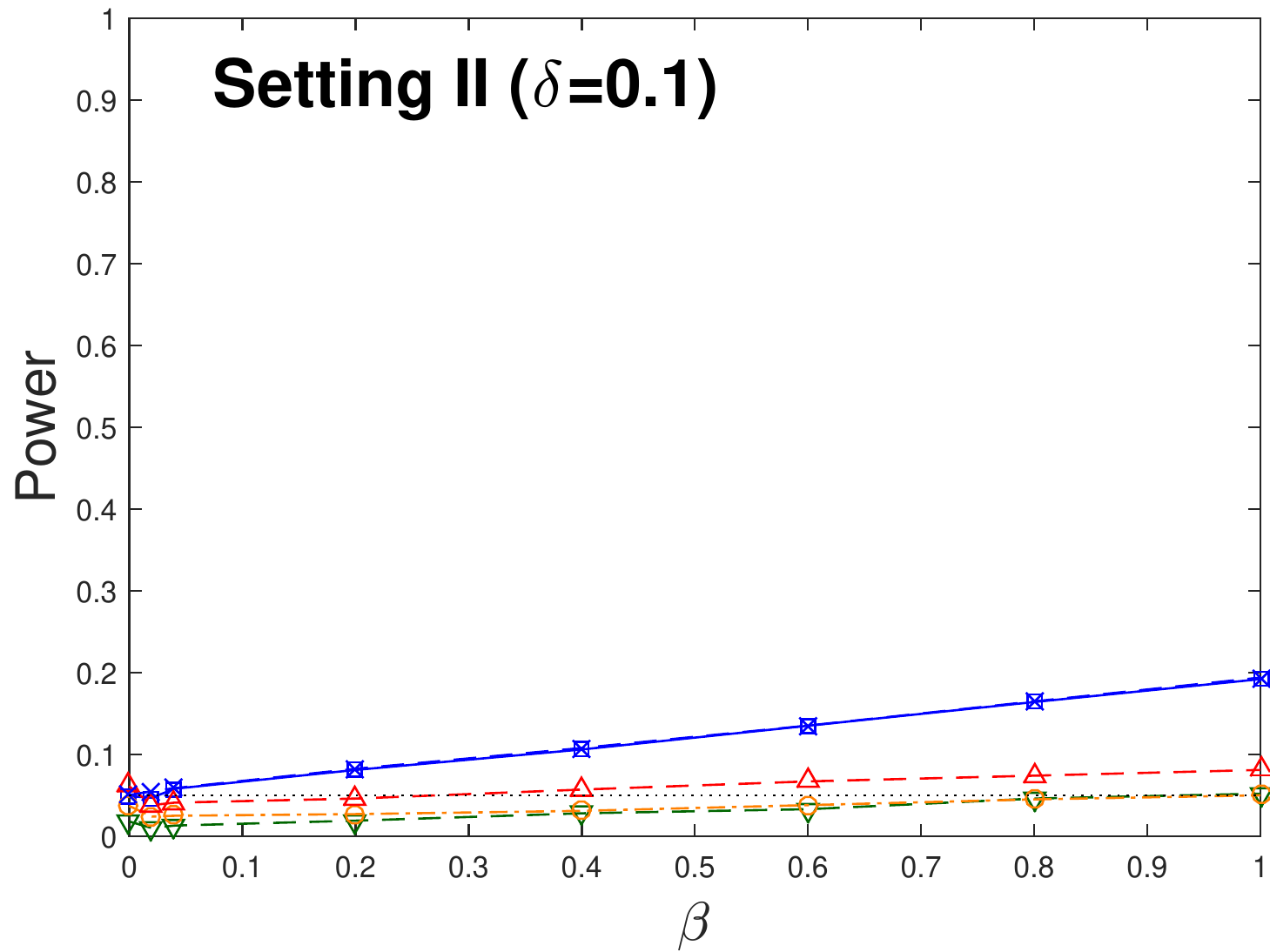}
\end{minipage}%
\begin{minipage}[t]{0.19\textwidth}
\centering
\includegraphics[width=0.95in]{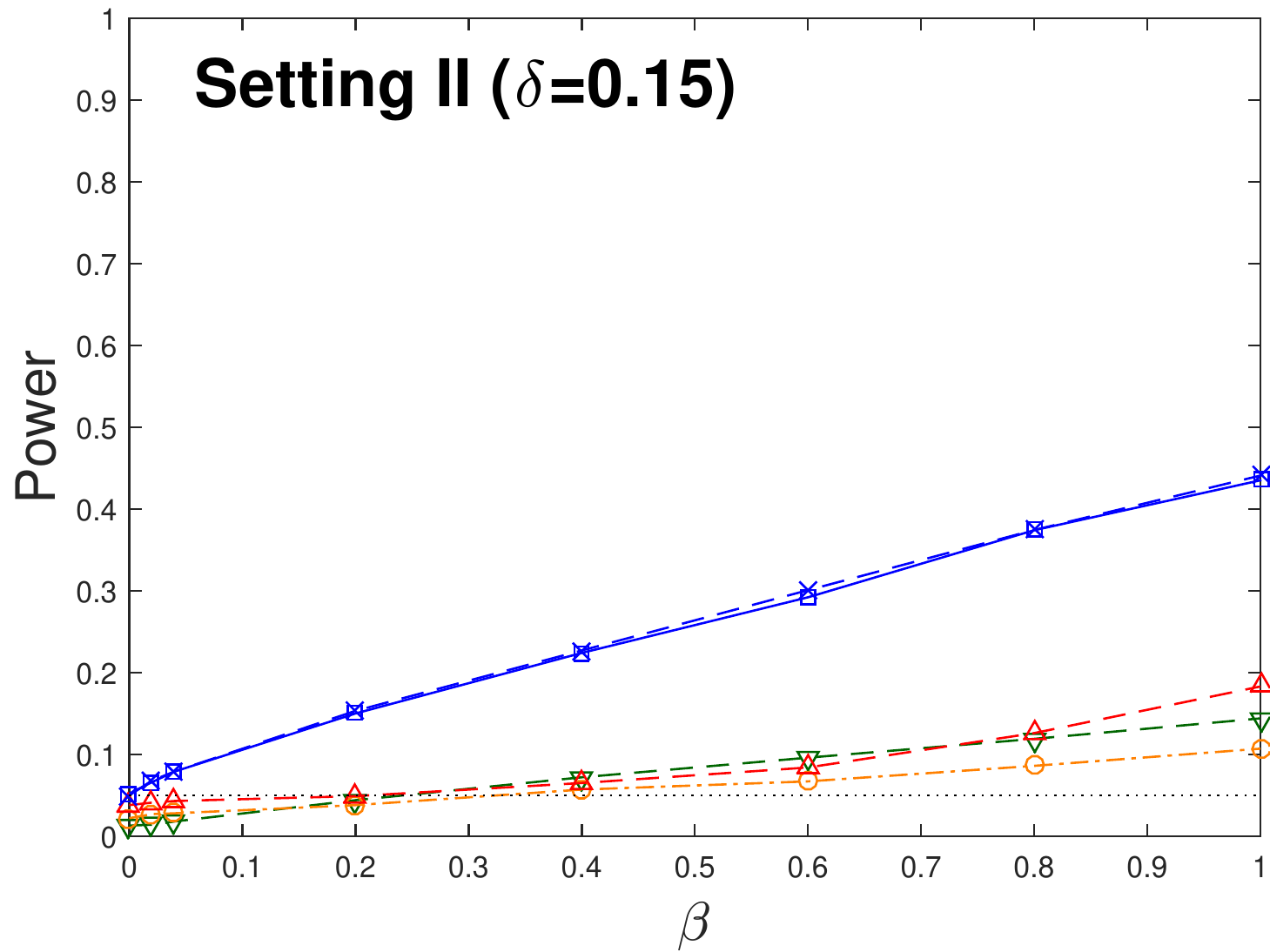}
\end{minipage}%
\begin{minipage}[t]{0.19\textwidth}
\centering
\includegraphics[width=0.95in]{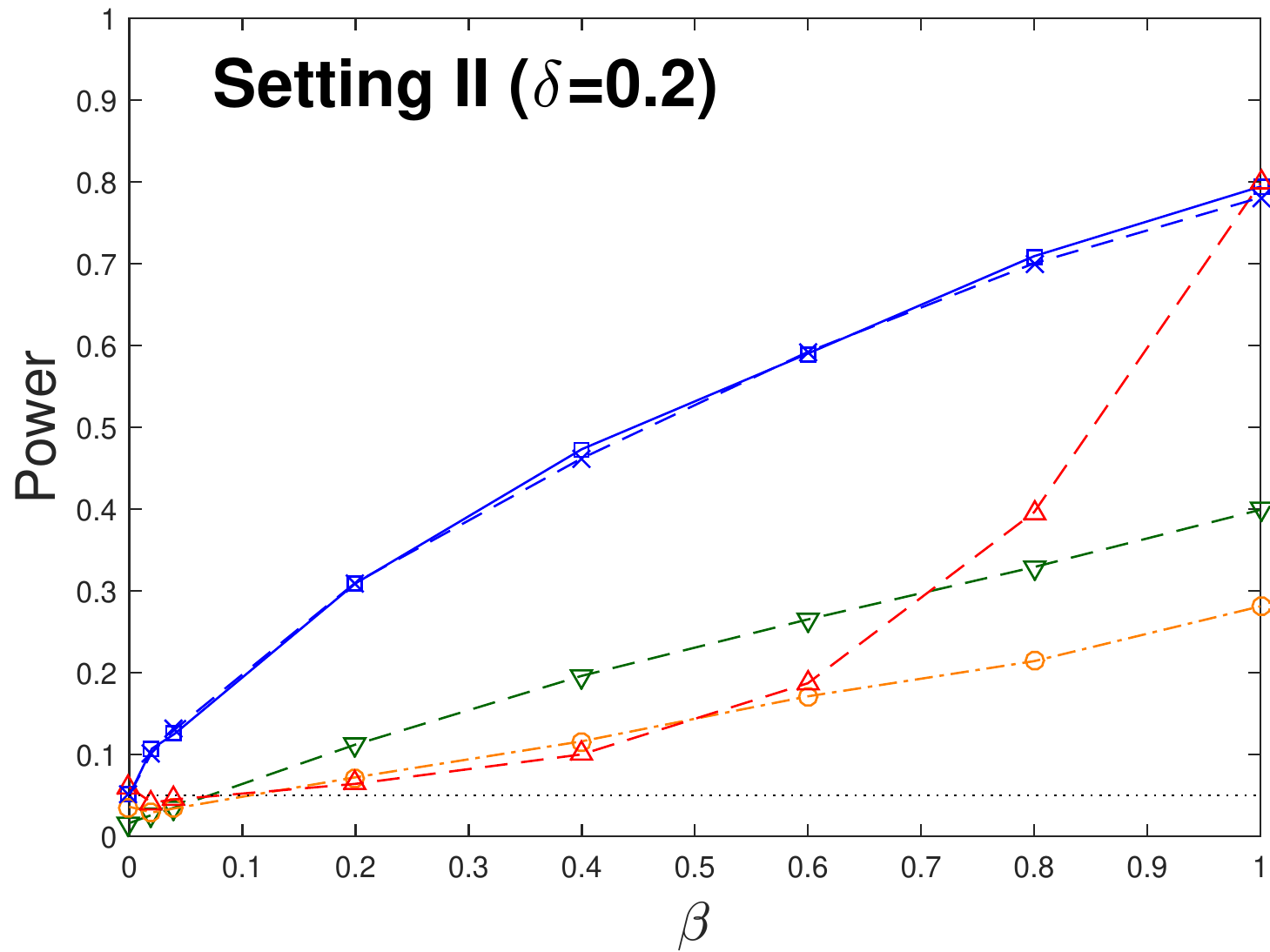}
\end{minipage}%
\begin{minipage}[t]{0.19\textwidth}
\centering
\includegraphics[width=0.95in]{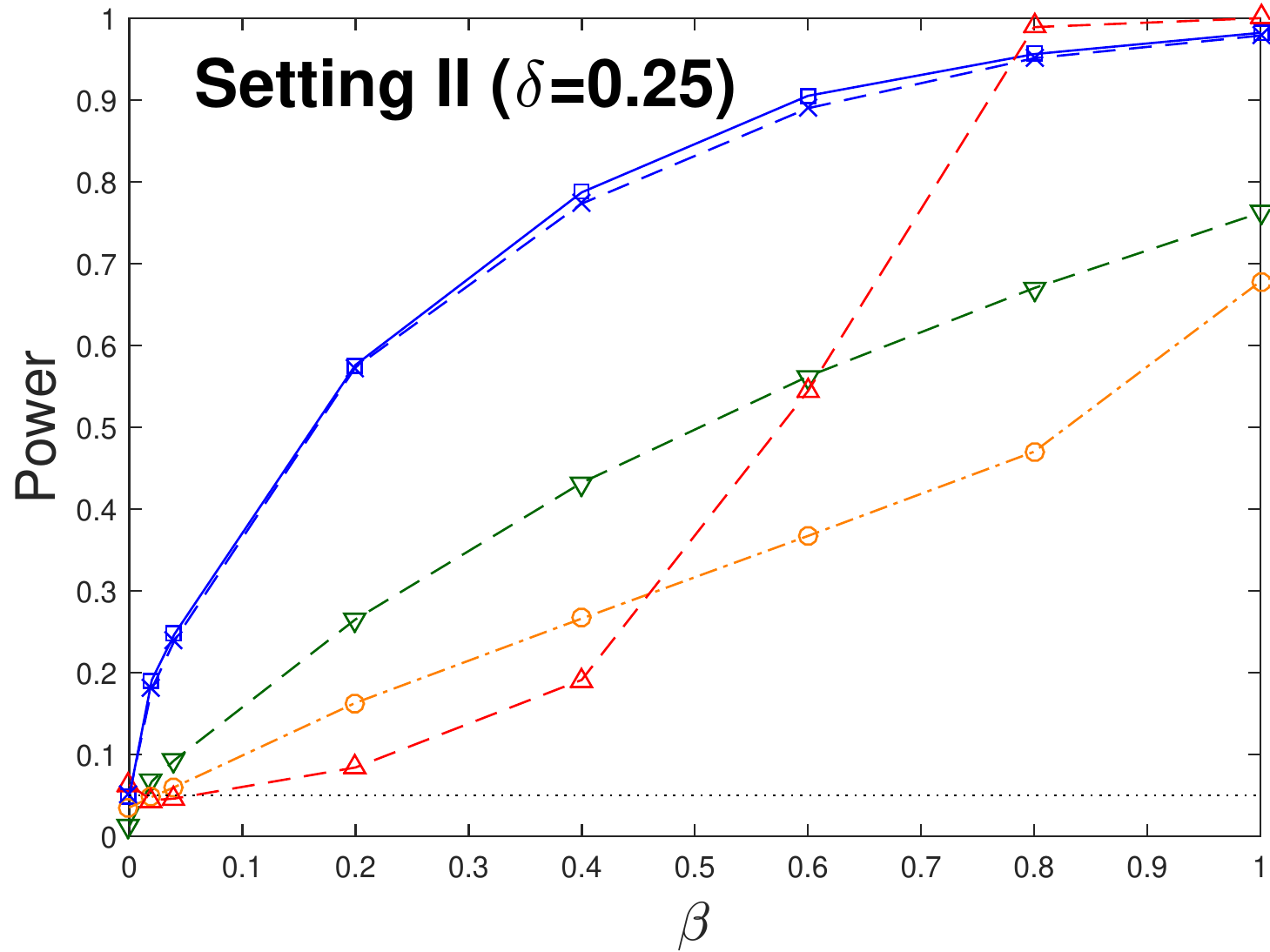}
\end{minipage}%
\begin{minipage}[t]{0.19\textwidth}
\centering
\includegraphics[width=0.95in]{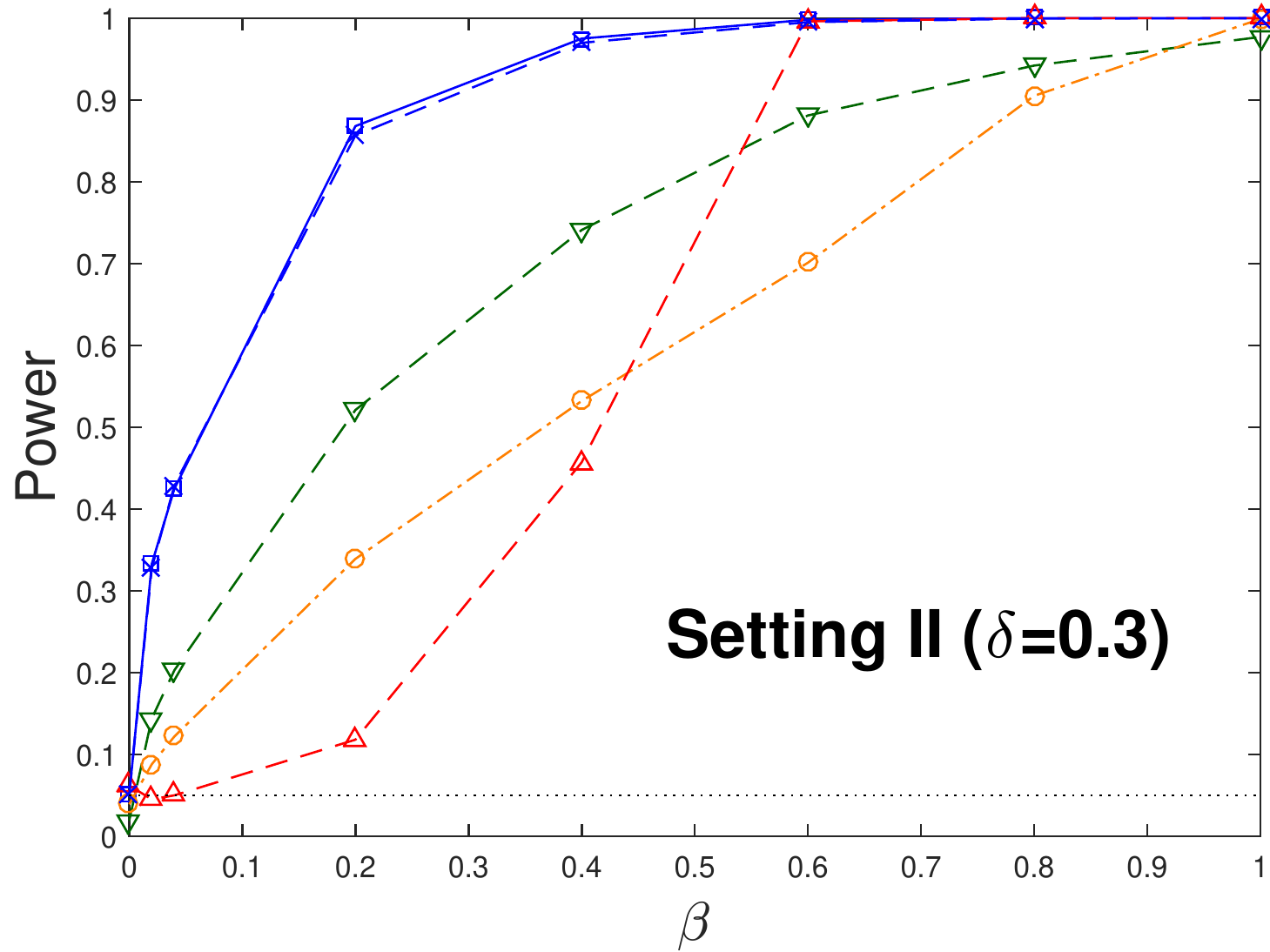}
\end{minipage}%

\begin{minipage}[t]{0.19\textwidth}
\centering
\includegraphics[width=0.95in]{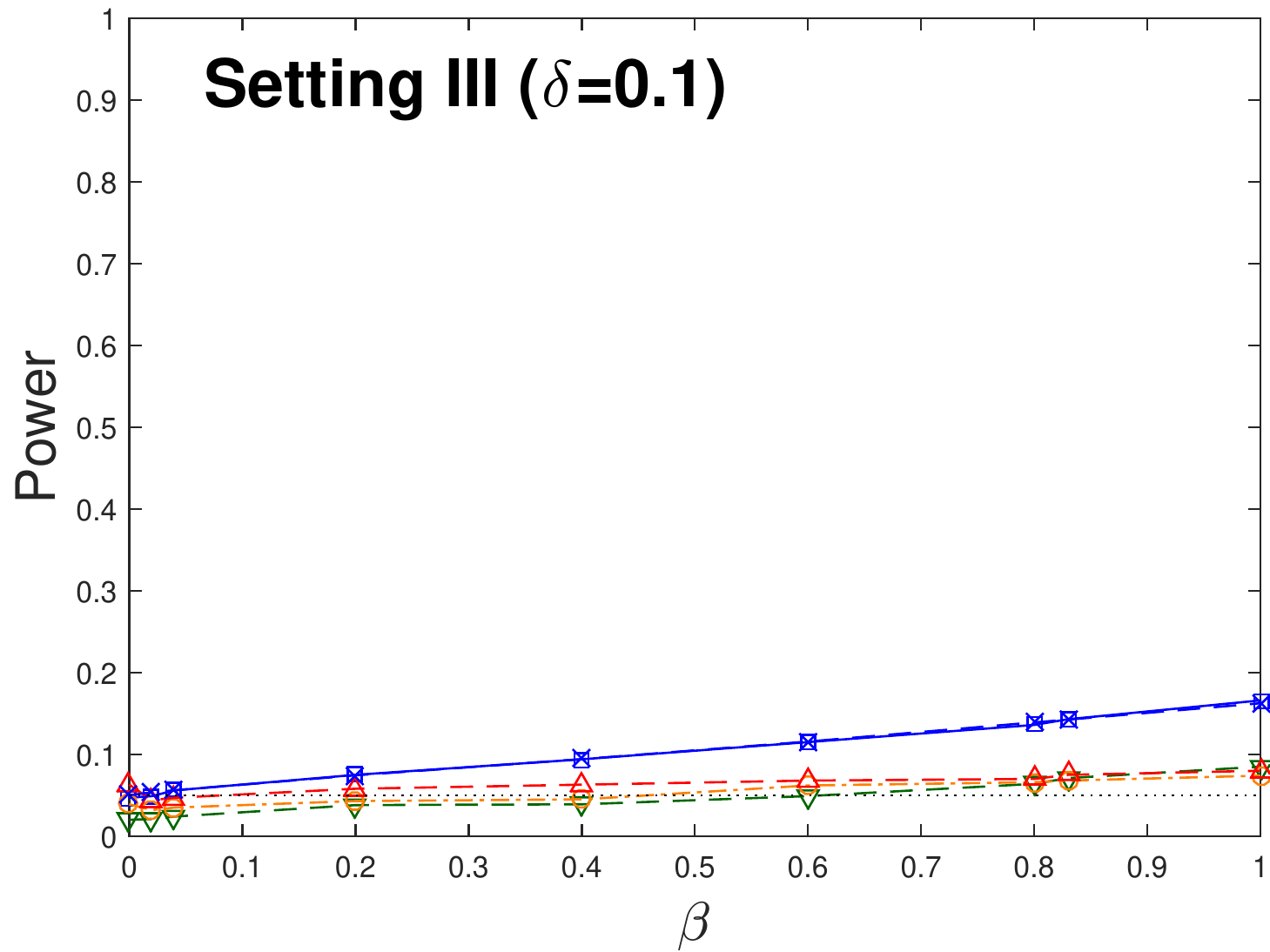}
\end{minipage}%
\begin{minipage}[t]{0.19\textwidth}
\centering
\includegraphics[width=0.95in]{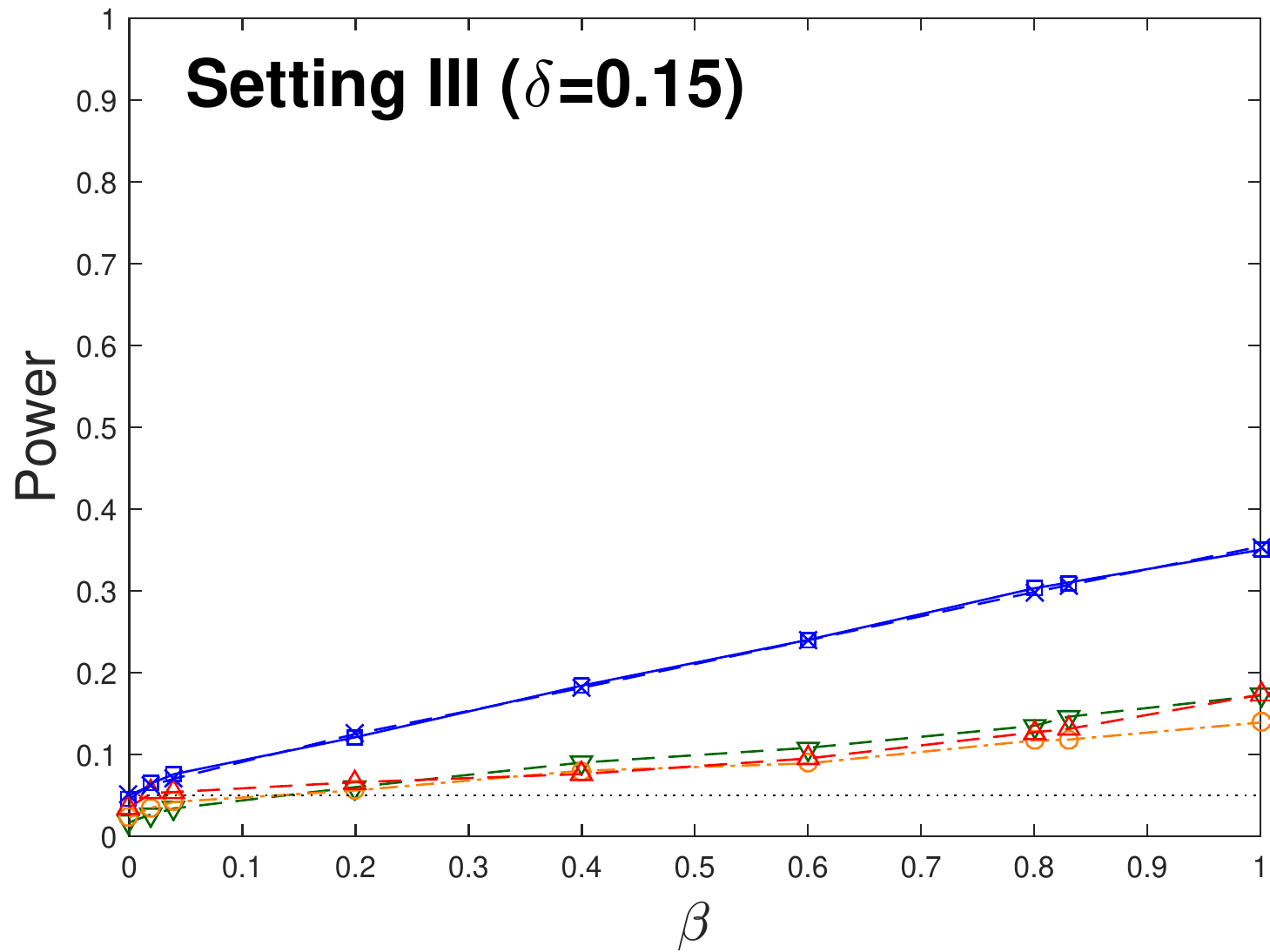}
\end{minipage}%
\begin{minipage}[t]{0.19\textwidth}
\centering
\includegraphics[width=0.95in]{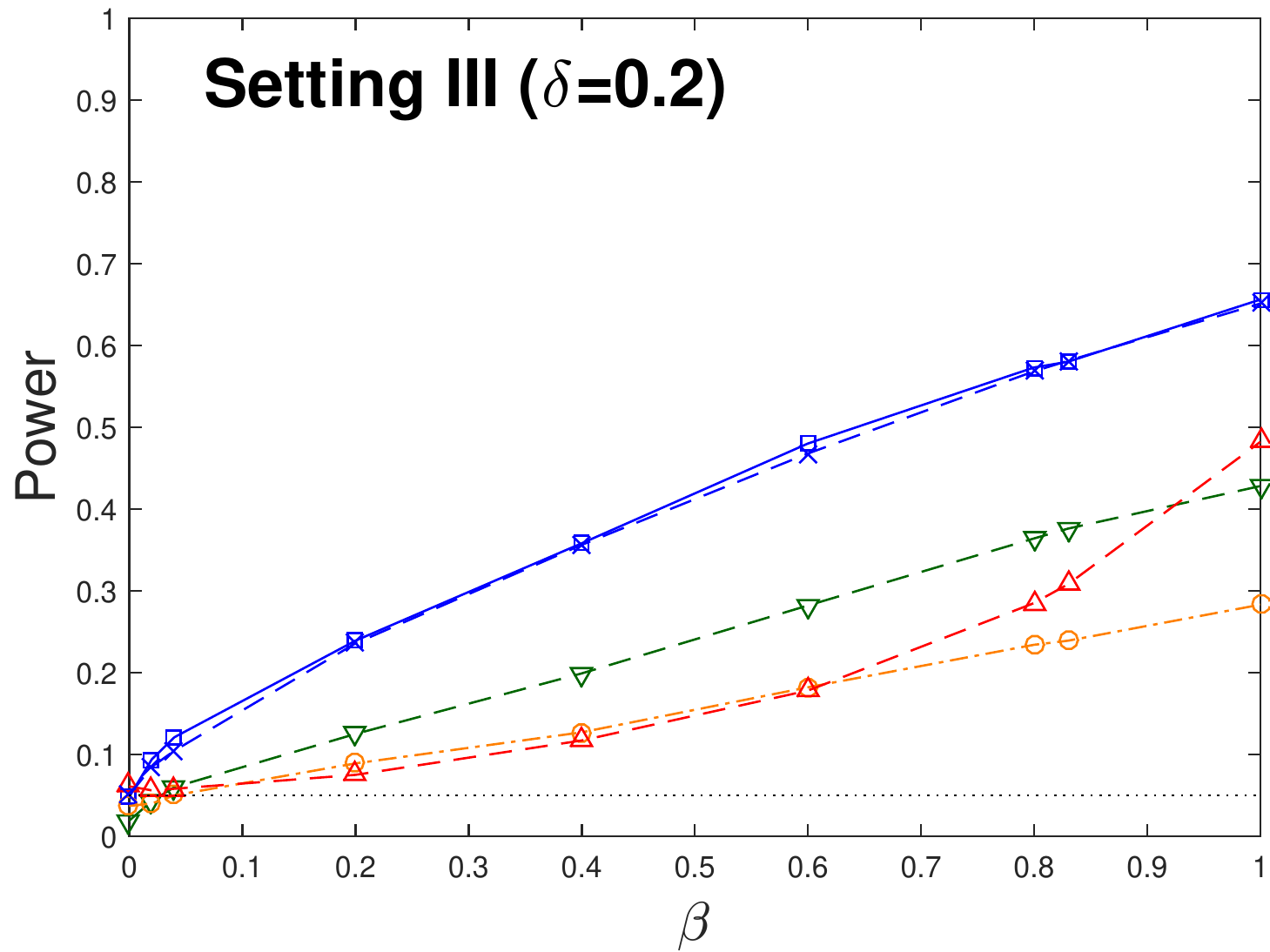}
\end{minipage}%
\begin{minipage}[t]{0.19\textwidth}
\centering
\includegraphics[width=0.95in]{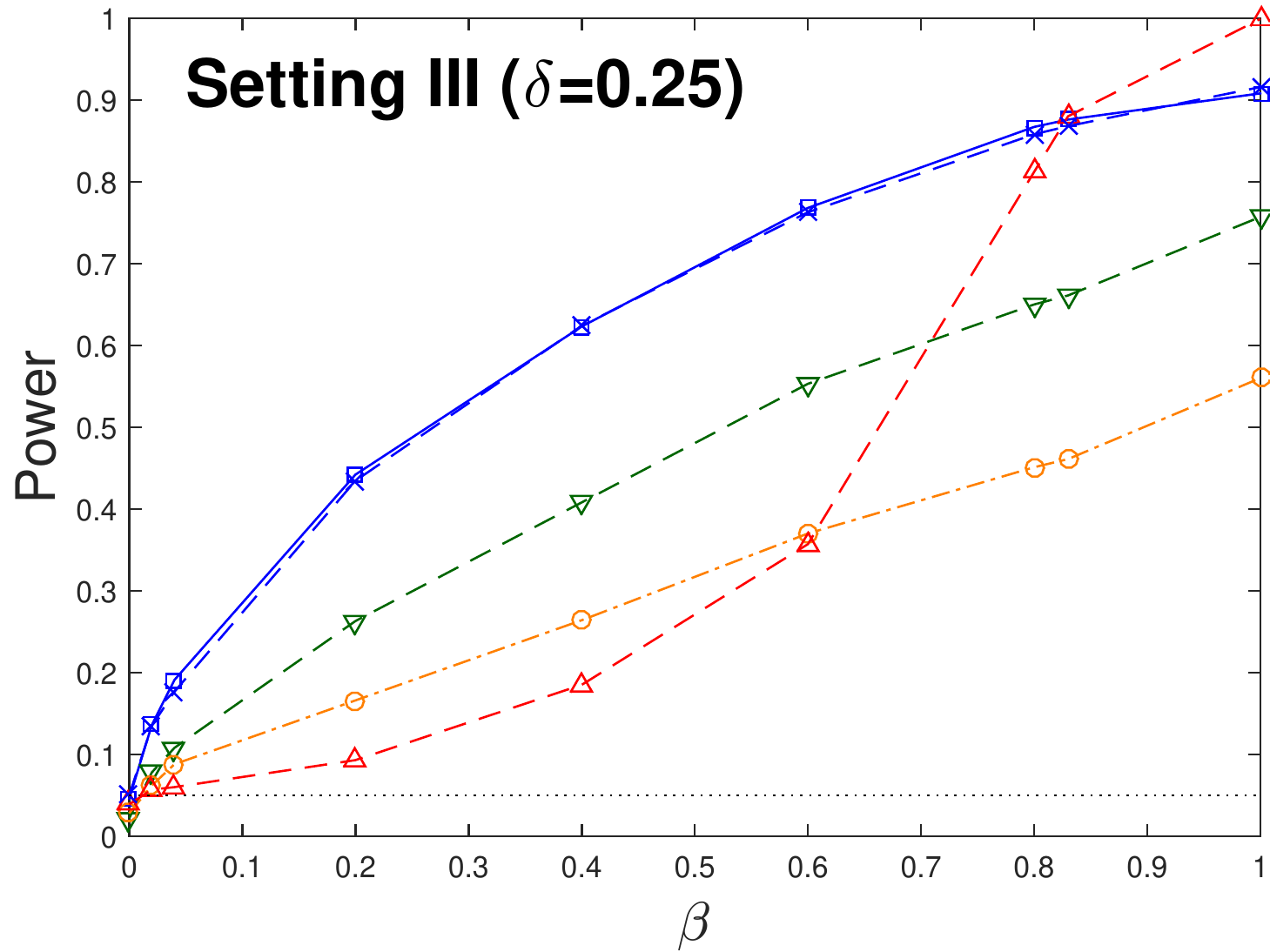}
\end{minipage}%
\begin{minipage}[t]{0.19\textwidth}
\centering
\includegraphics[width=0.95in]{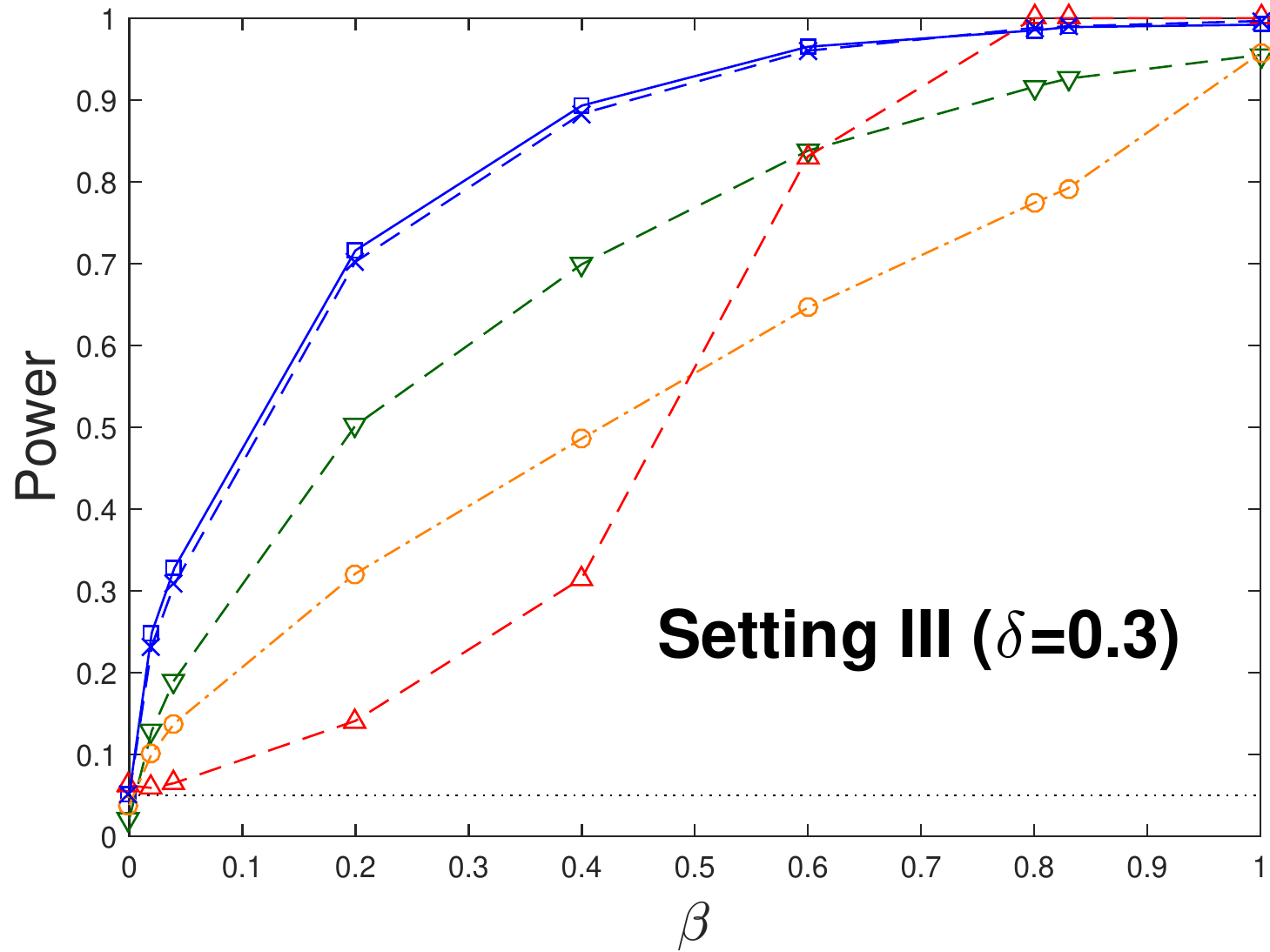}
\end{minipage}%

\begin{minipage}[t]{0.19\textwidth}
\centering
\includegraphics[width=0.95in]{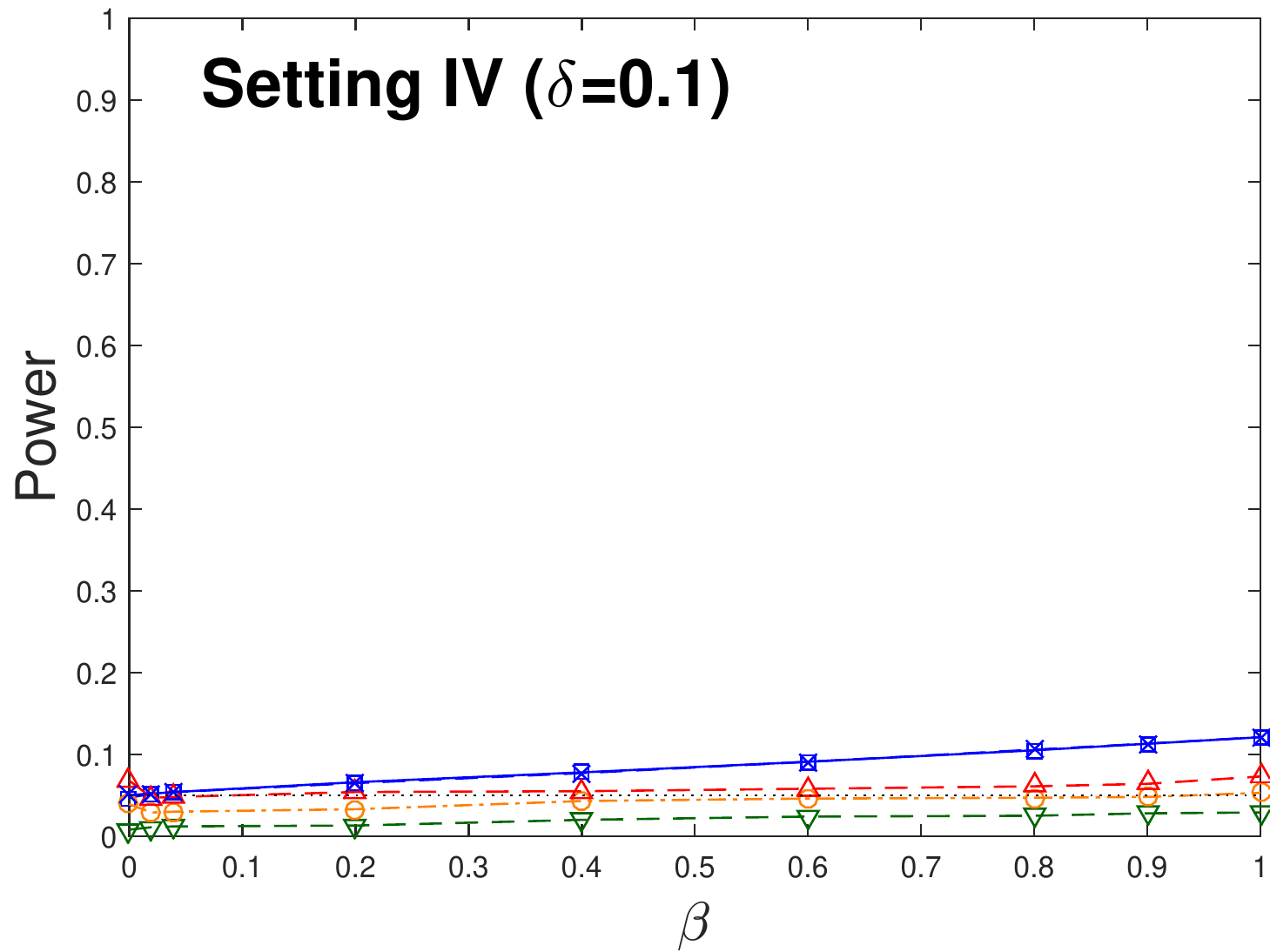}
\end{minipage}%
\begin{minipage}[t]{0.19\textwidth}
\centering
\includegraphics[width=0.95in]{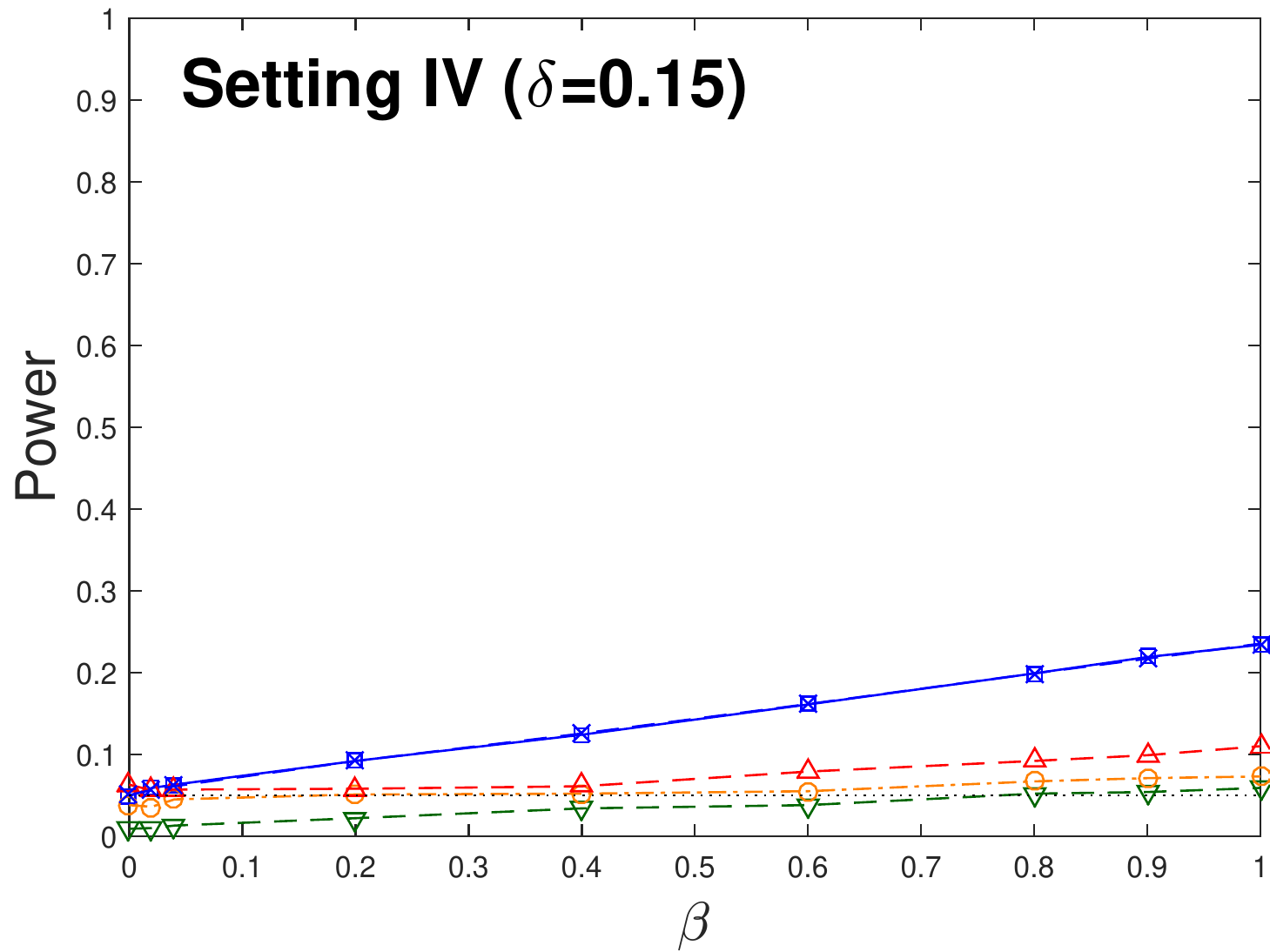}
\end{minipage}
\begin{minipage}[t]{0.19\textwidth}
\centering
\includegraphics[width=0.95in]{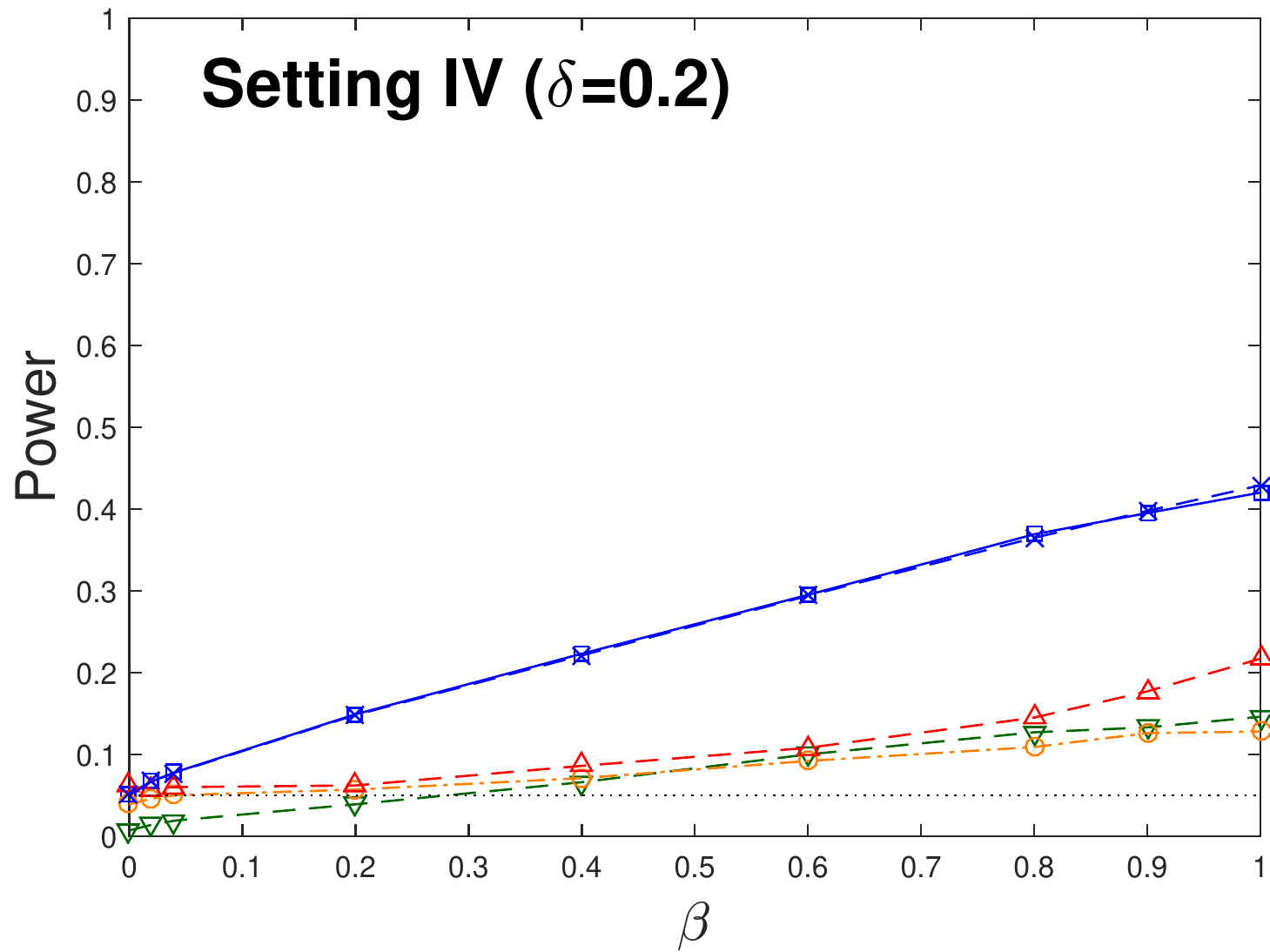}
\end{minipage}%
\begin{minipage}[t]{0.19\textwidth}
\centering
\includegraphics[width=0.95in]{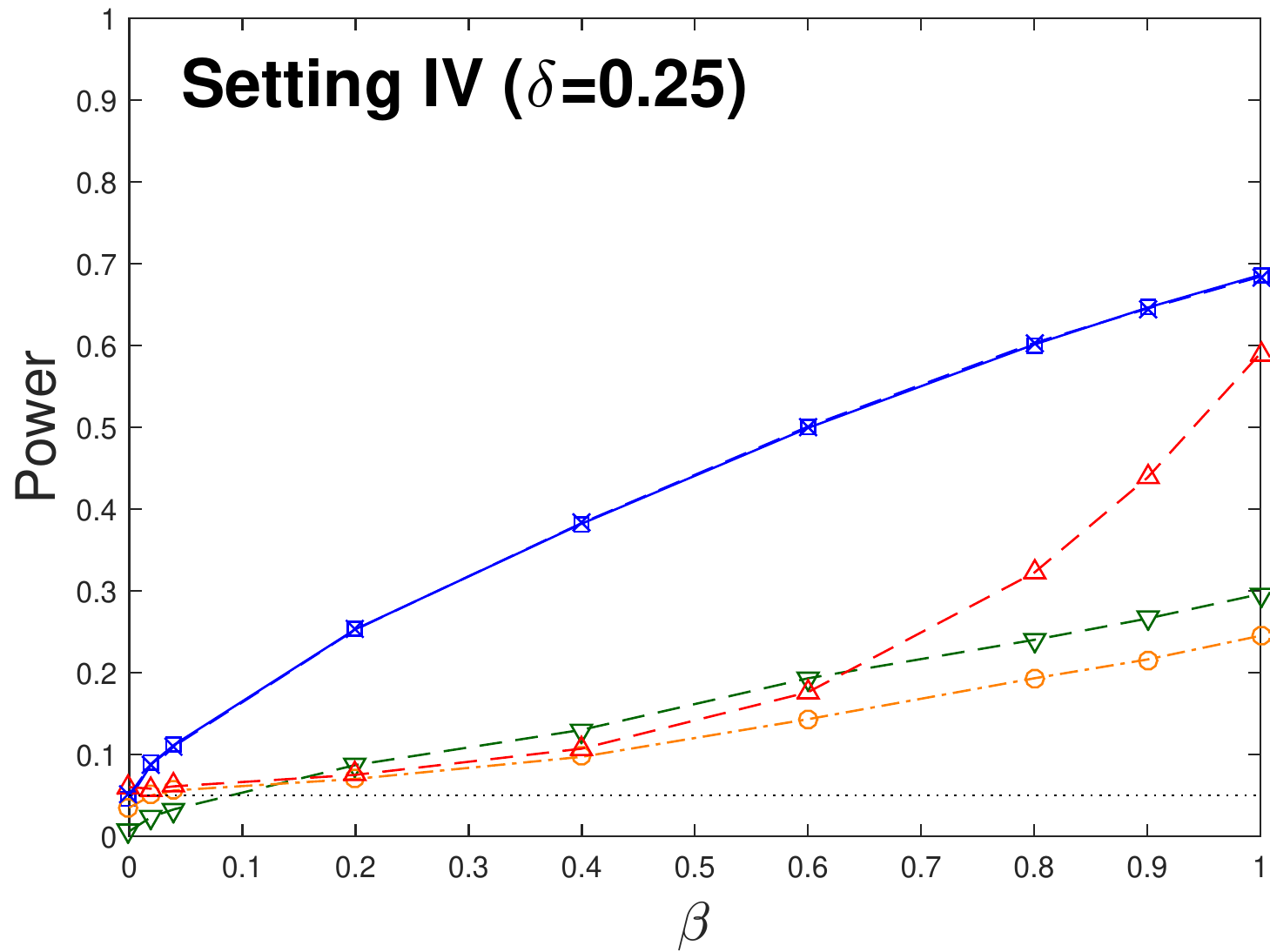}
\end{minipage}
\begin{minipage}[t]{0.19\textwidth}
\centering
\includegraphics[width=0.95in]{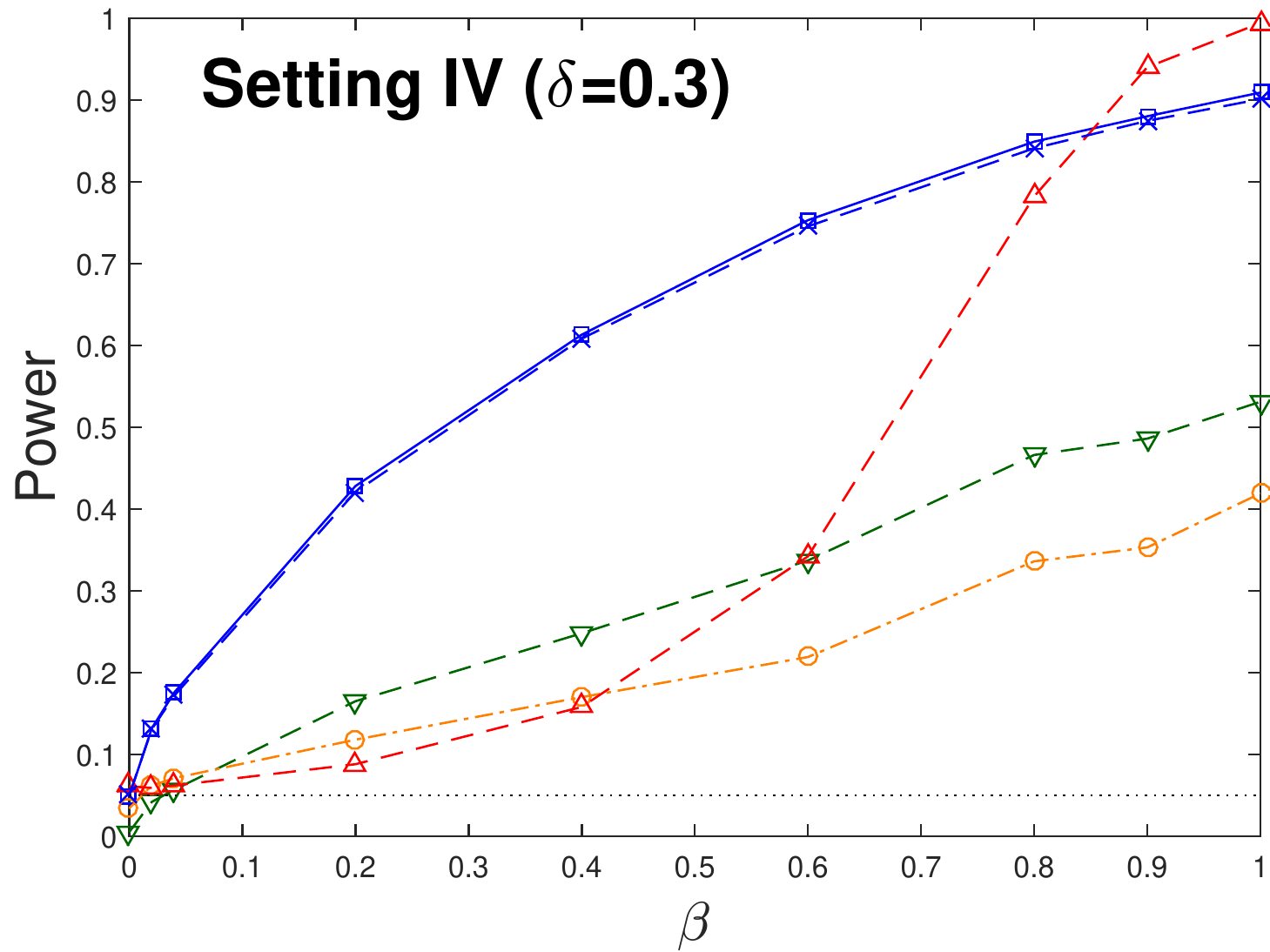}
\end{minipage}%

\begin{minipage}[t]{0.19\textwidth}
\centering
\includegraphics[width=0.95in]{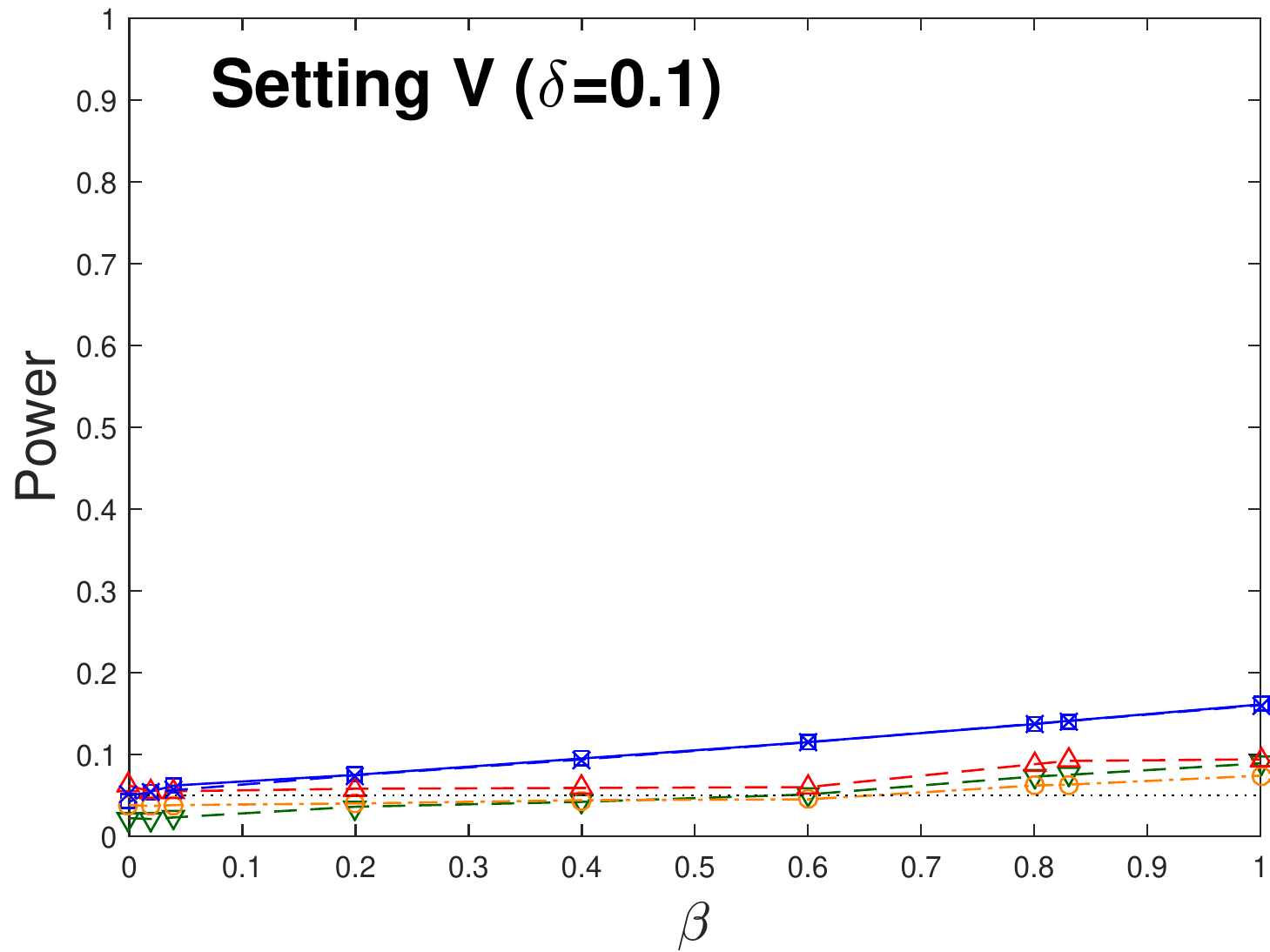}
\end{minipage}%
\begin{minipage}[t]{0.19\textwidth}
\centering
\includegraphics[width=0.95in]{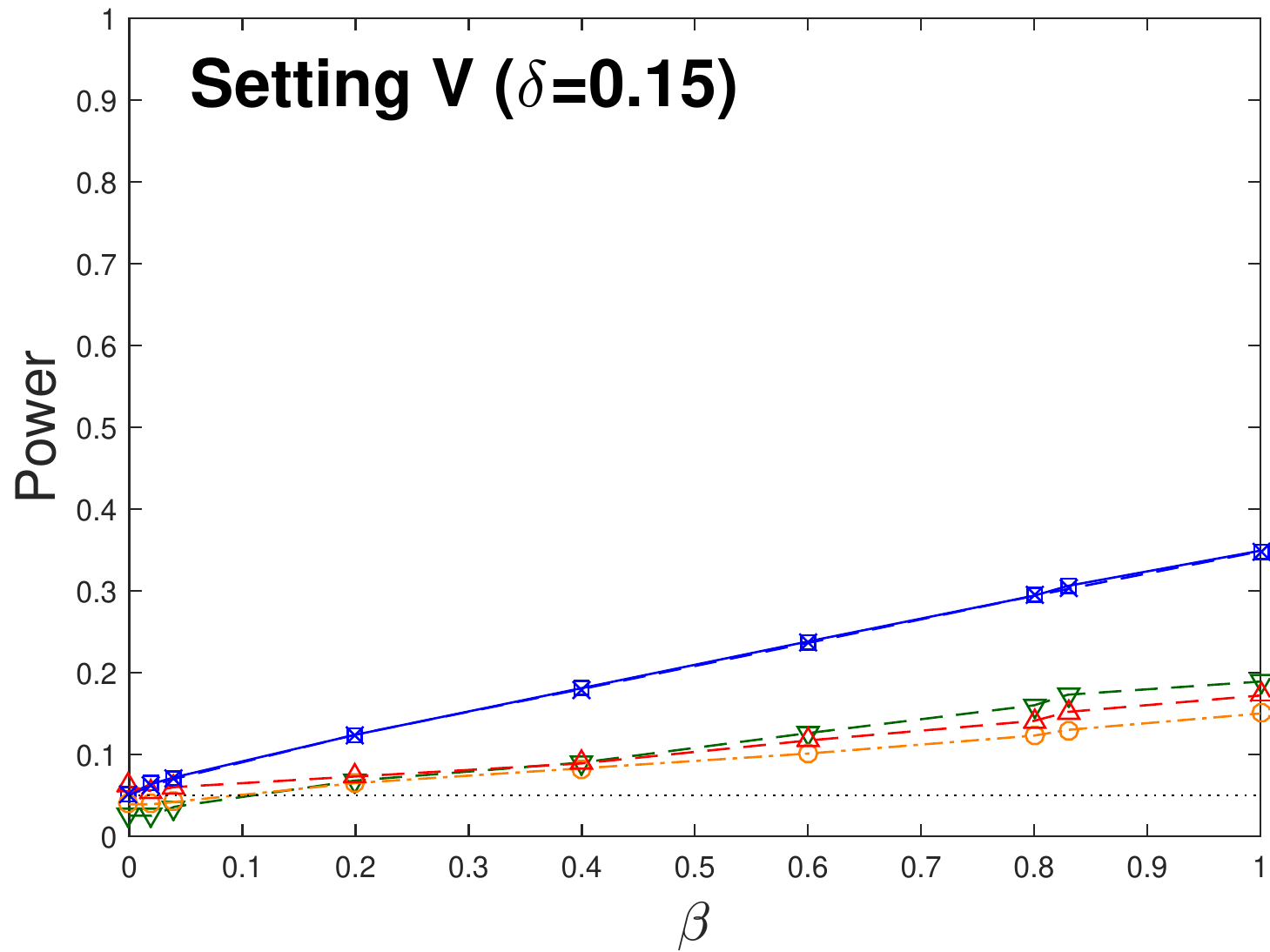}
\end{minipage}%
\begin{minipage}[t]{0.19\textwidth}
\centering
\includegraphics[width=0.95in]{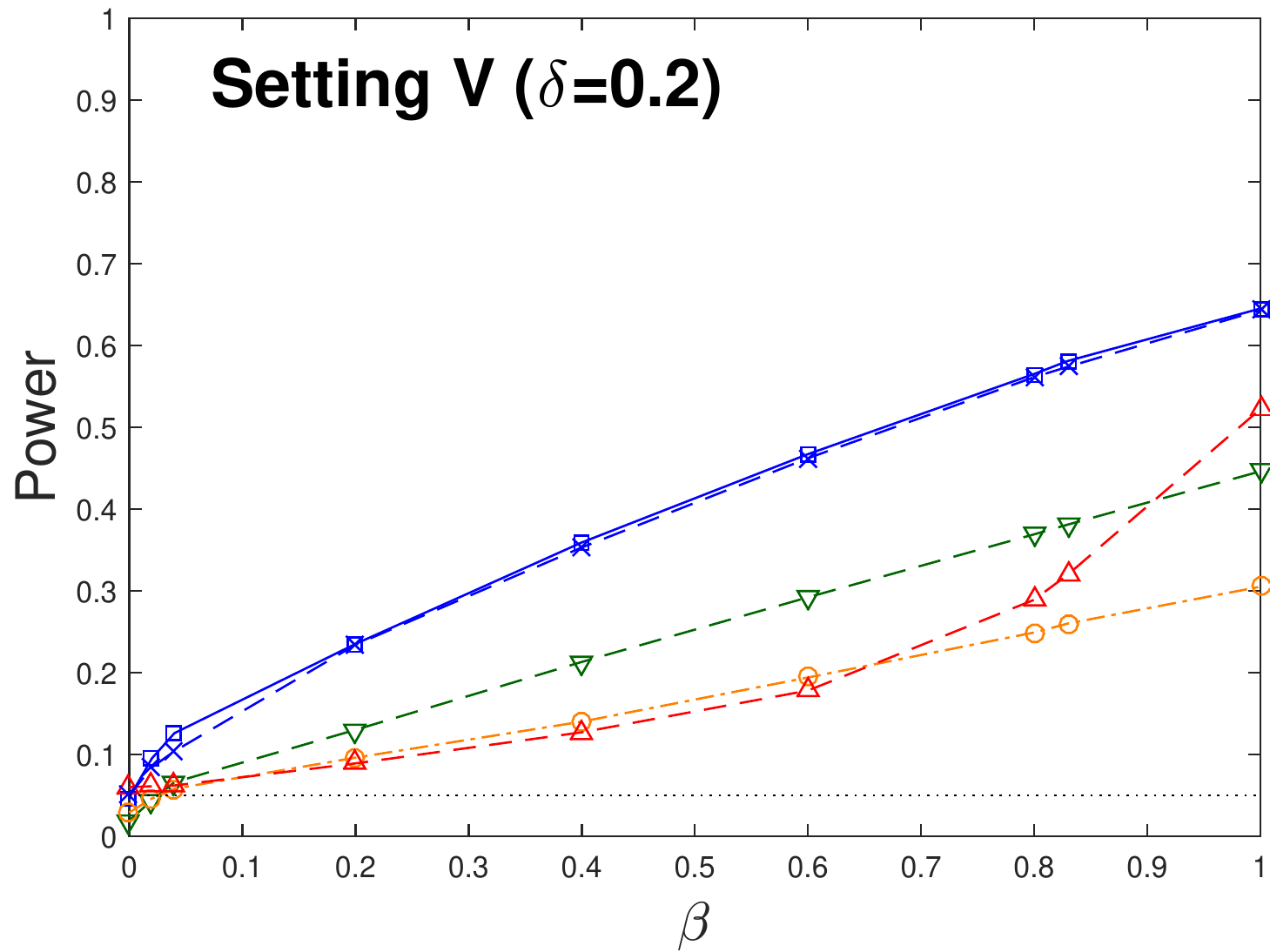}
\end{minipage}%
\begin{minipage}[t]{0.19\textwidth}
\centering
\includegraphics[width=0.95in]{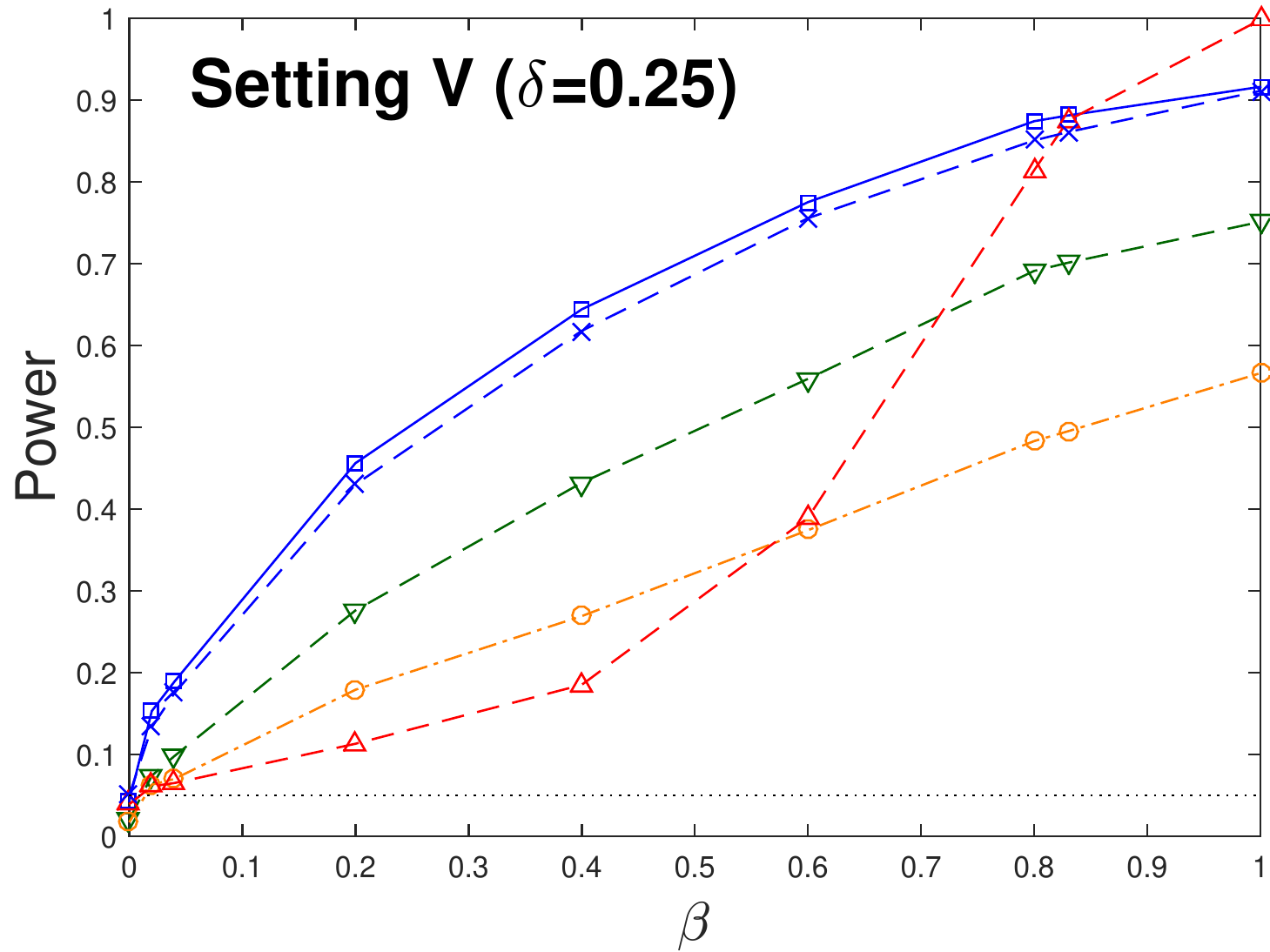}
\end{minipage}%
\begin{minipage}[t]{0.19\textwidth}
\centering
\includegraphics[width=0.95in]{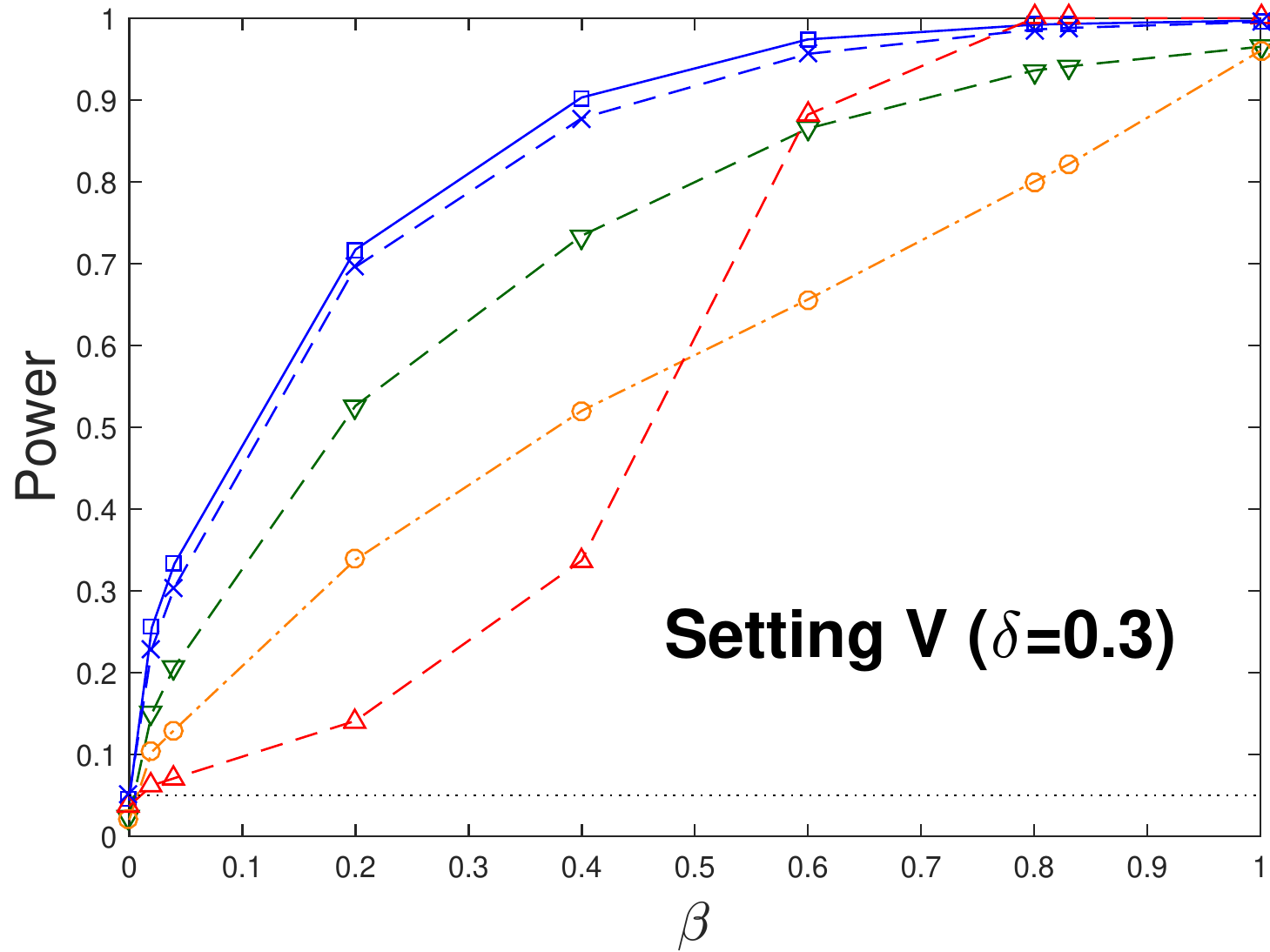}
\end{minipage}%

\begin{minipage}[t]{0.19\textwidth}
\centering
\includegraphics[width=0.95in]{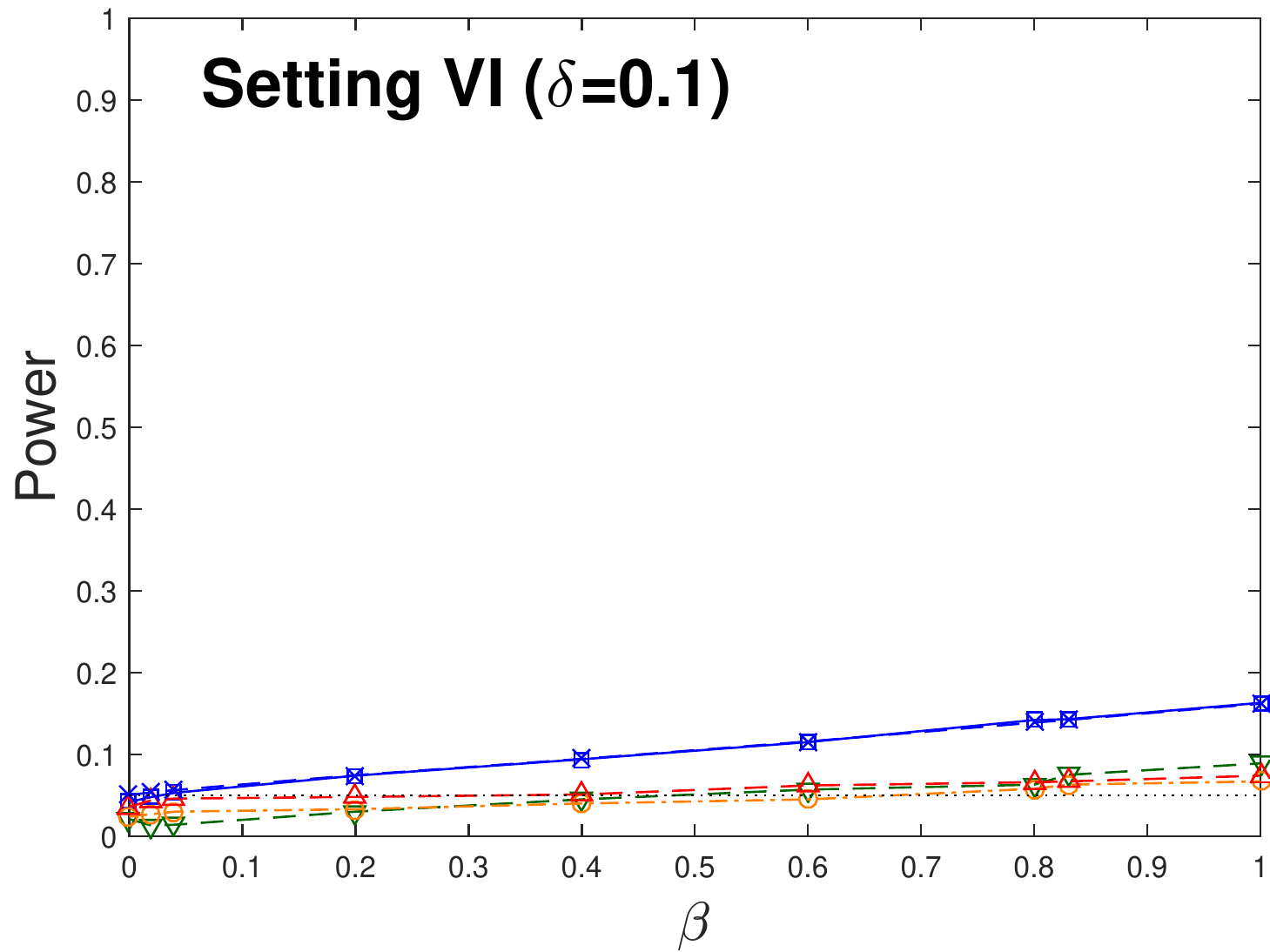}
\end{minipage}%
\begin{minipage}[t]{0.19\textwidth}
\centering
\includegraphics[width=0.95in]{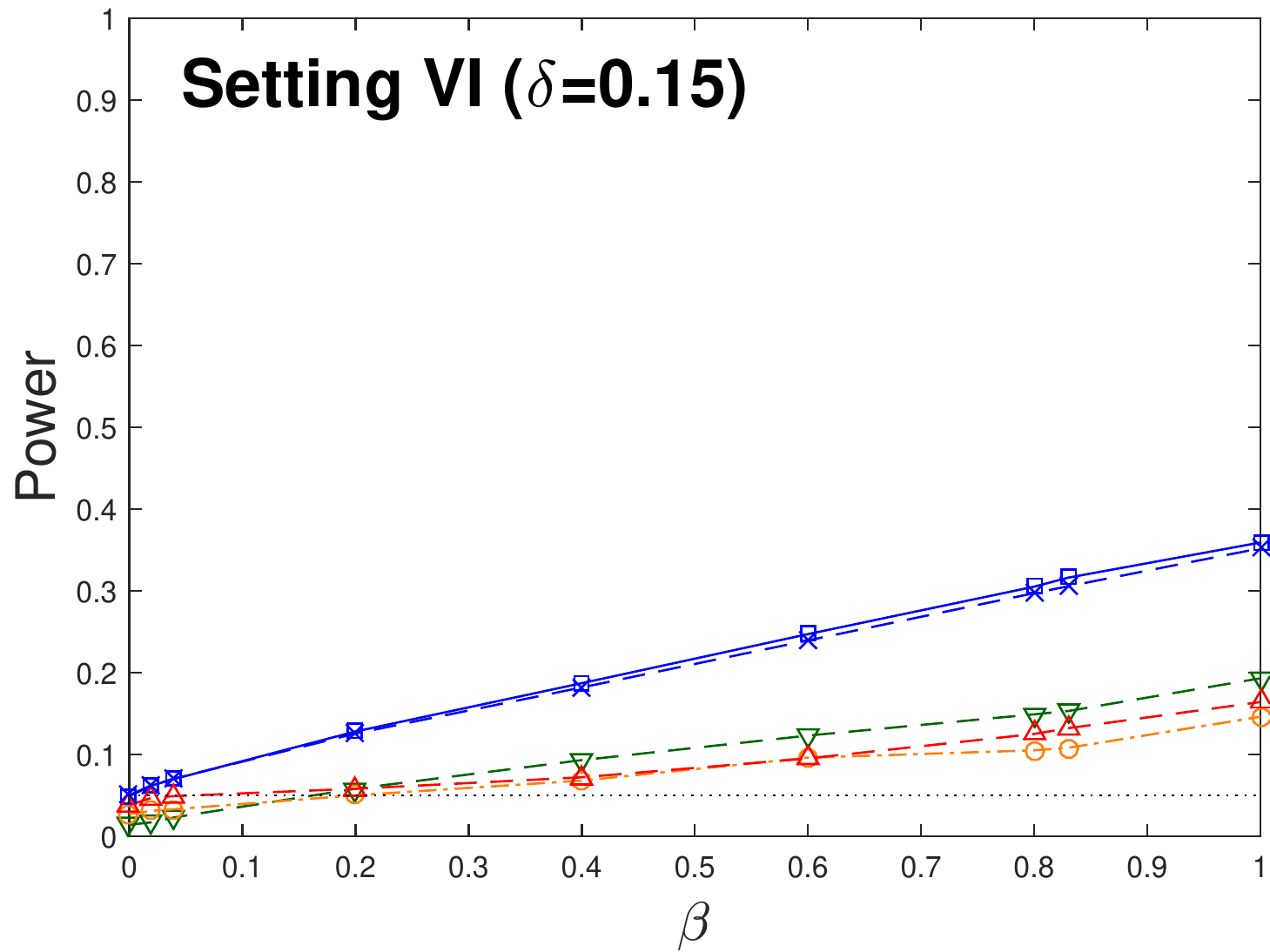}
\end{minipage}%
\begin{minipage}[t]{0.19\textwidth}
\centering
\includegraphics[width=0.95in]{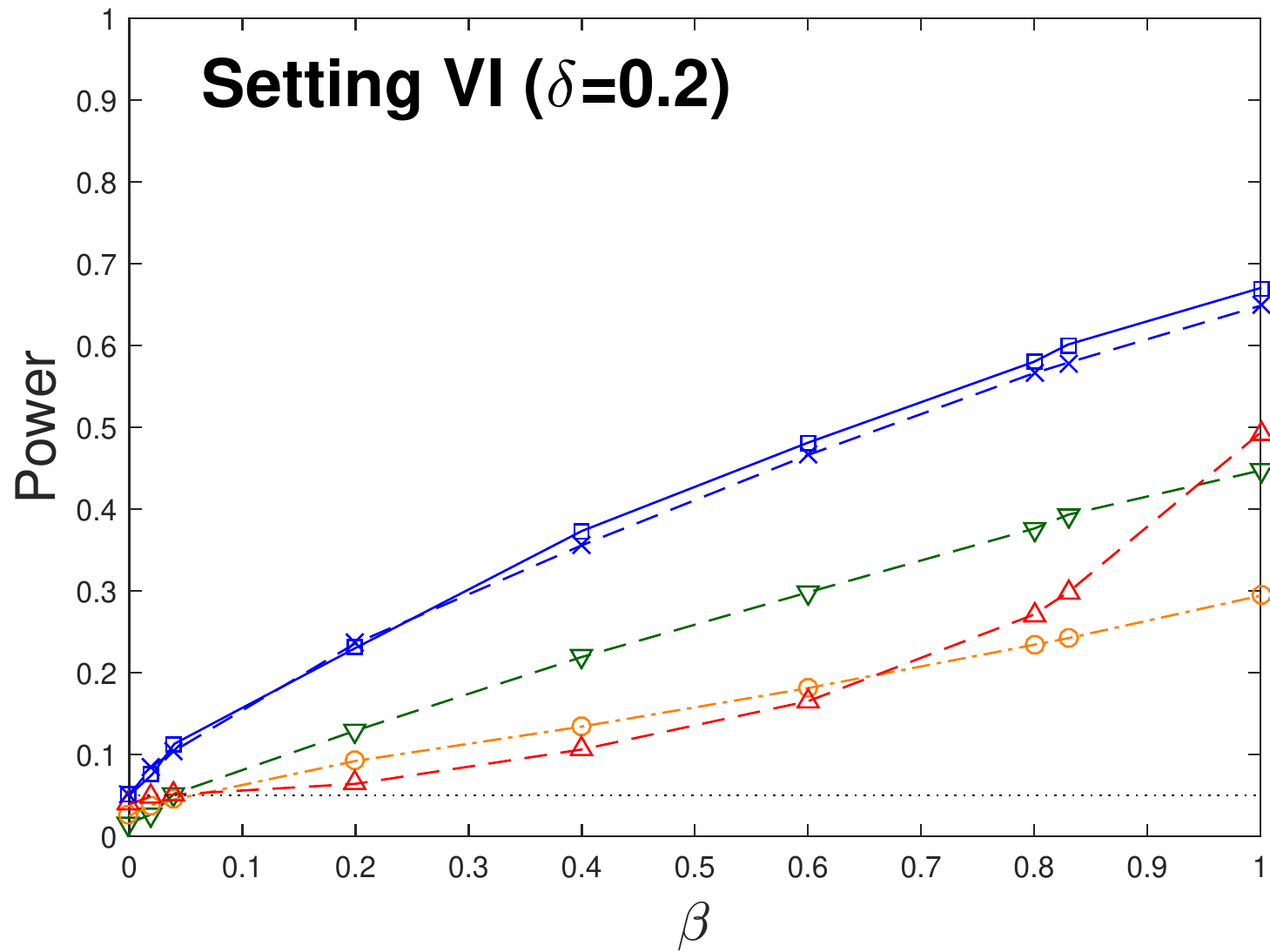}
\end{minipage}%
\begin{minipage}[t]{0.19\textwidth}
\centering
\includegraphics[width=0.95in]{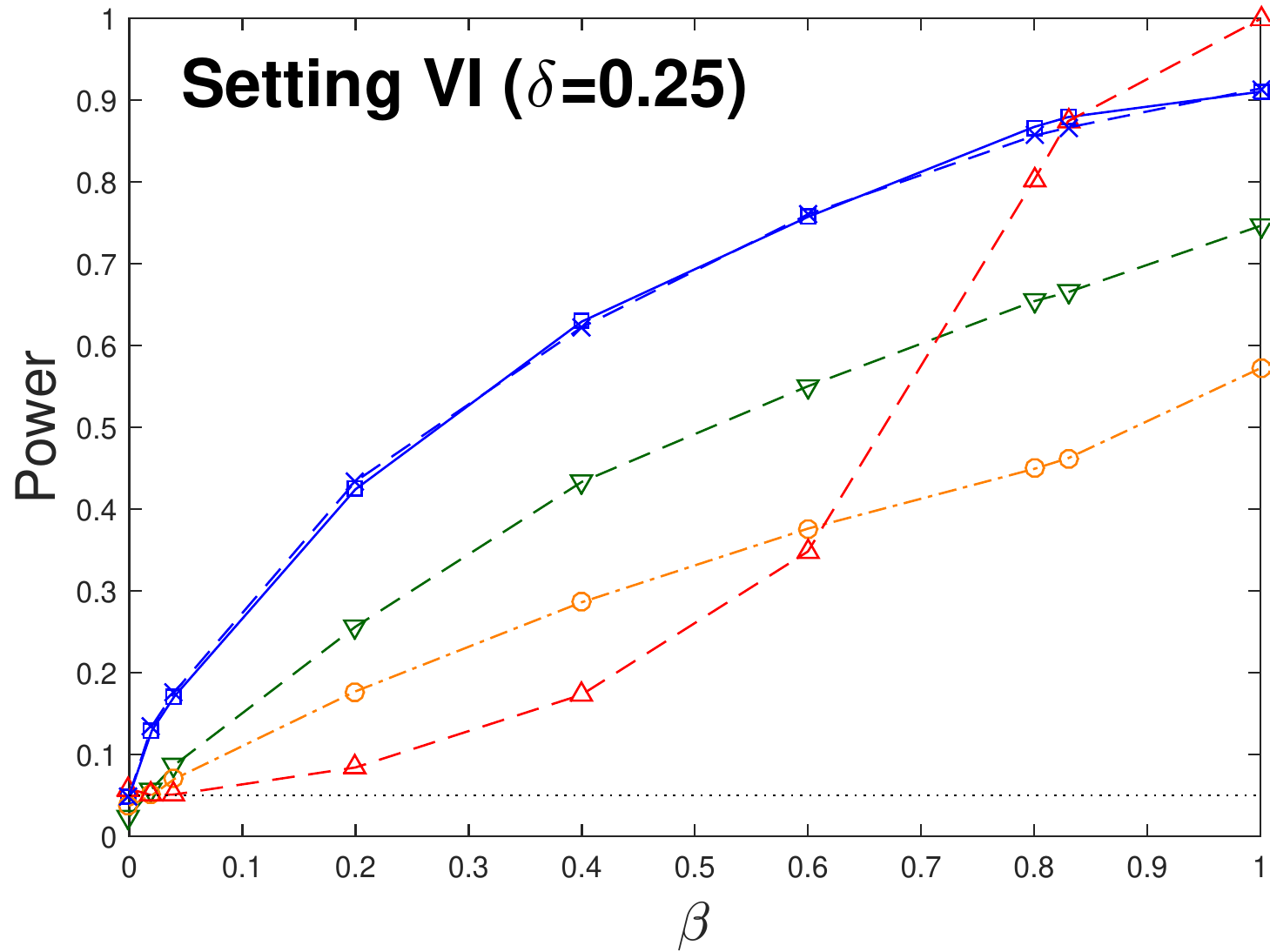}
\end{minipage}%
\begin{minipage}[t]{0.19\textwidth}
\centering
\includegraphics[width=0.95in]{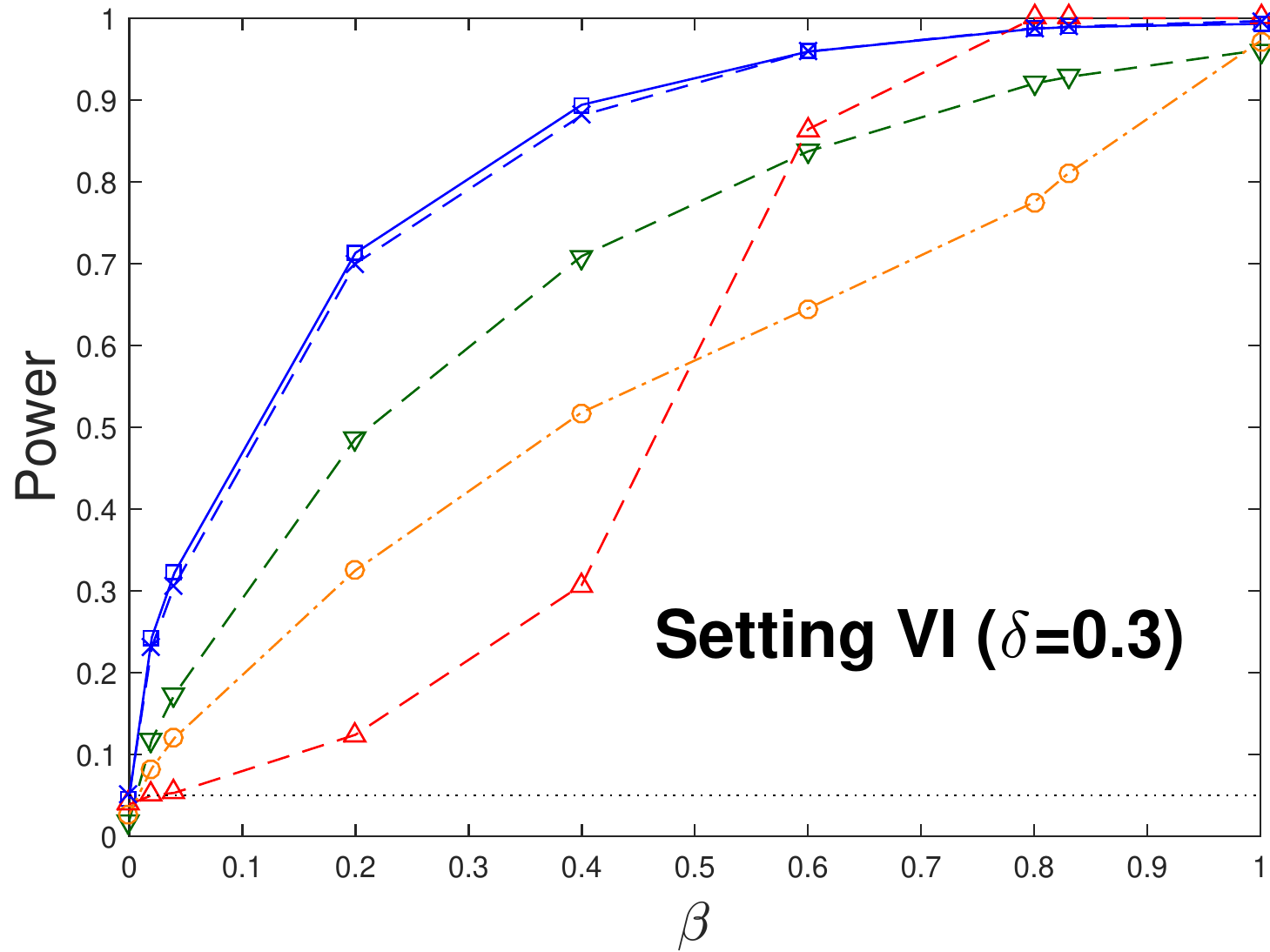}
\end{minipage}%


\begin{minipage}[t]{0.19\textwidth}
\centering
\includegraphics[width=0.95in]{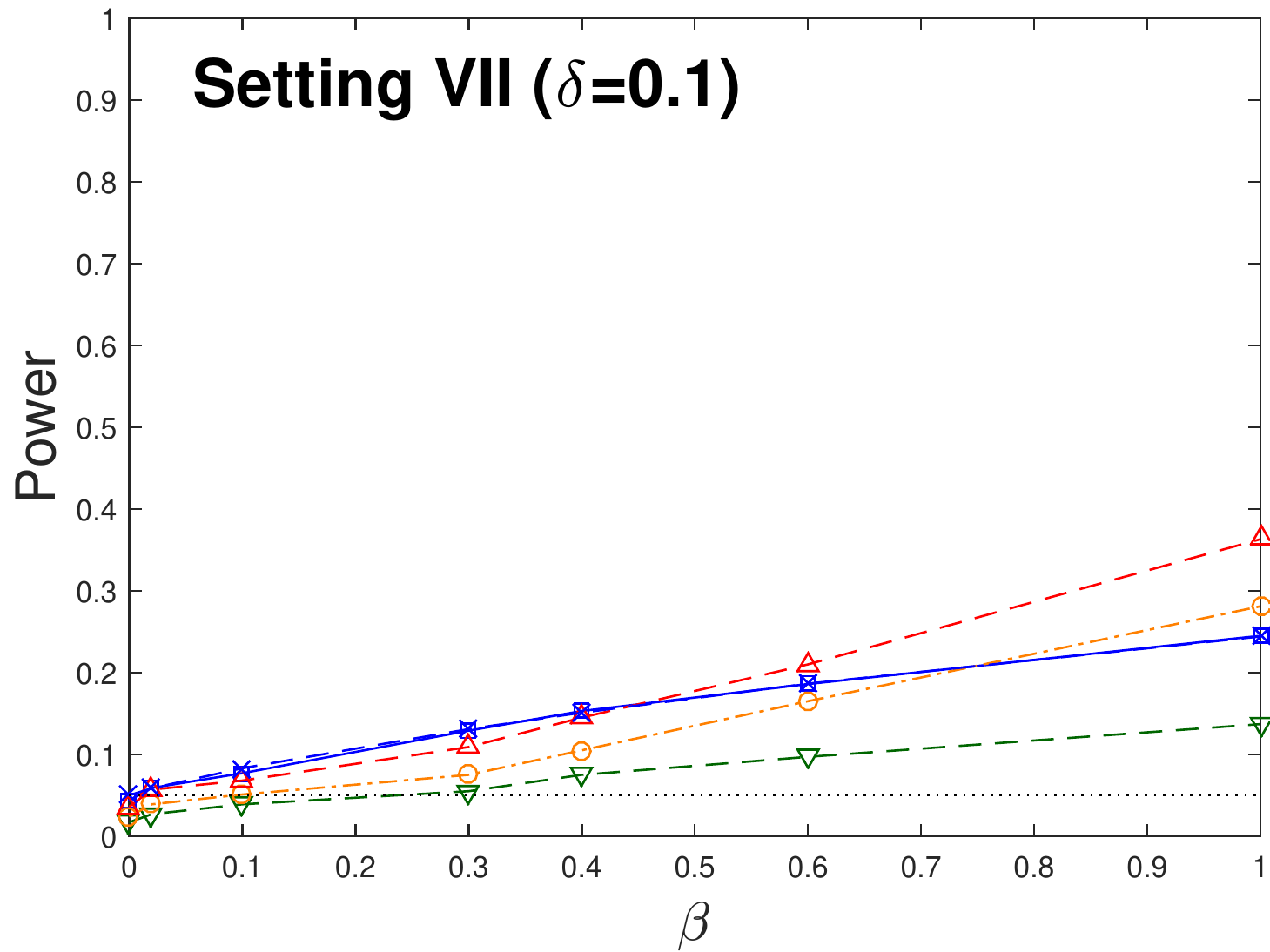}
\end{minipage}%
\begin{minipage}[t]{0.19\textwidth}
\centering
\includegraphics[width=0.95in]{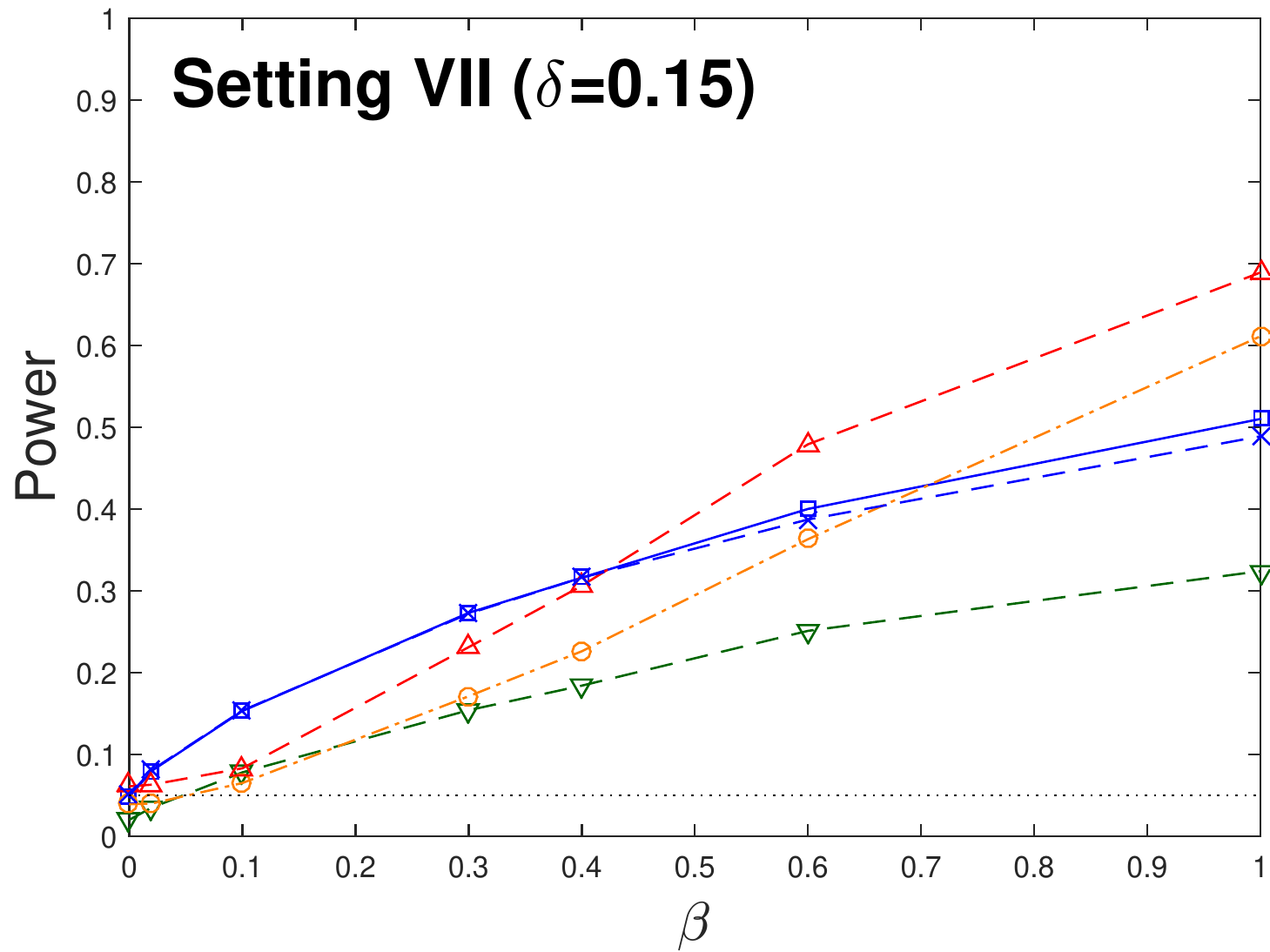}
\end{minipage}%
\begin{minipage}[t]{0.19\textwidth}
\centering
\includegraphics[width=0.95in]{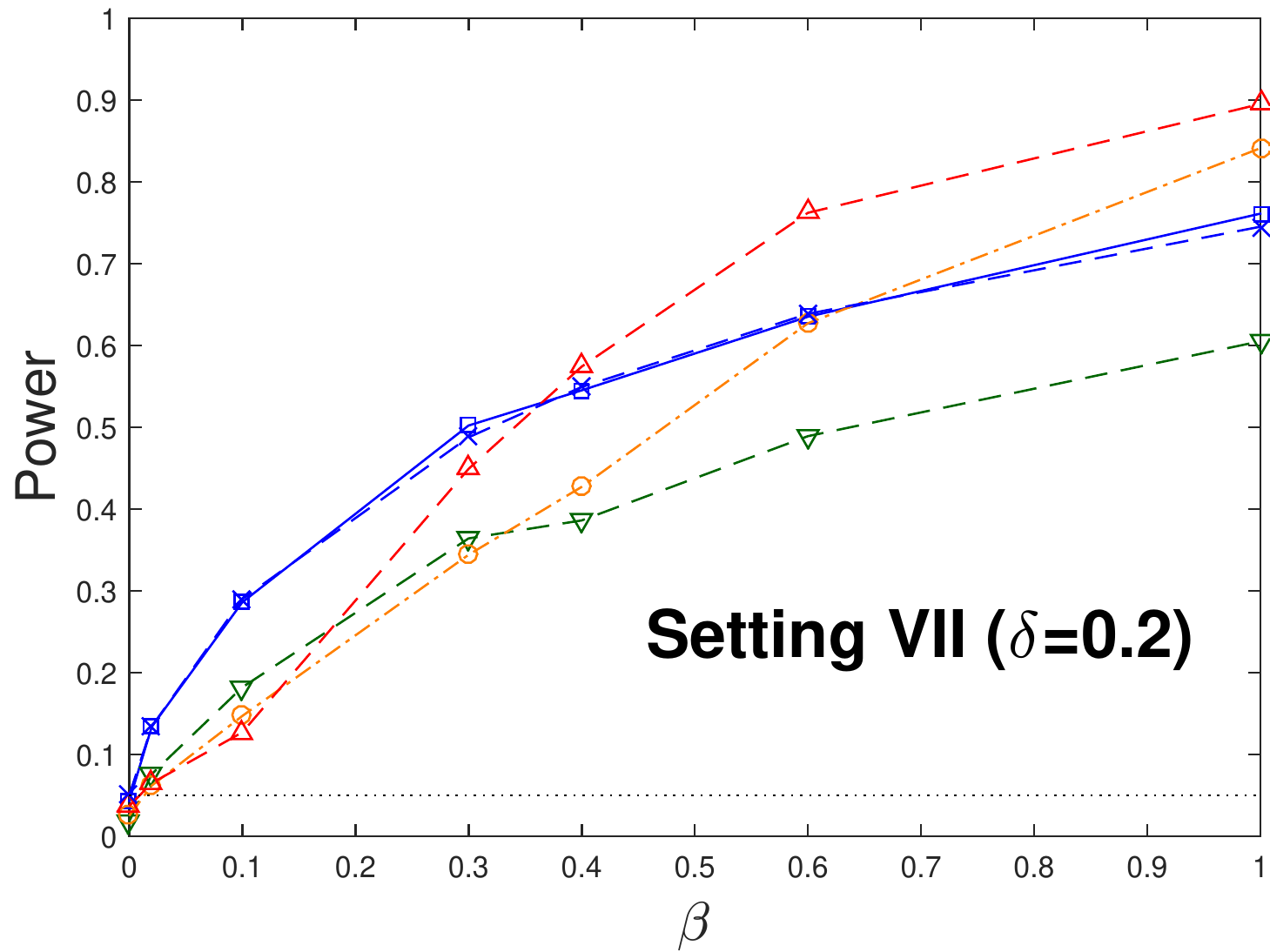}
\end{minipage}%
\begin{minipage}[t]{0.19\textwidth}
\centering
\includegraphics[width=0.95in]{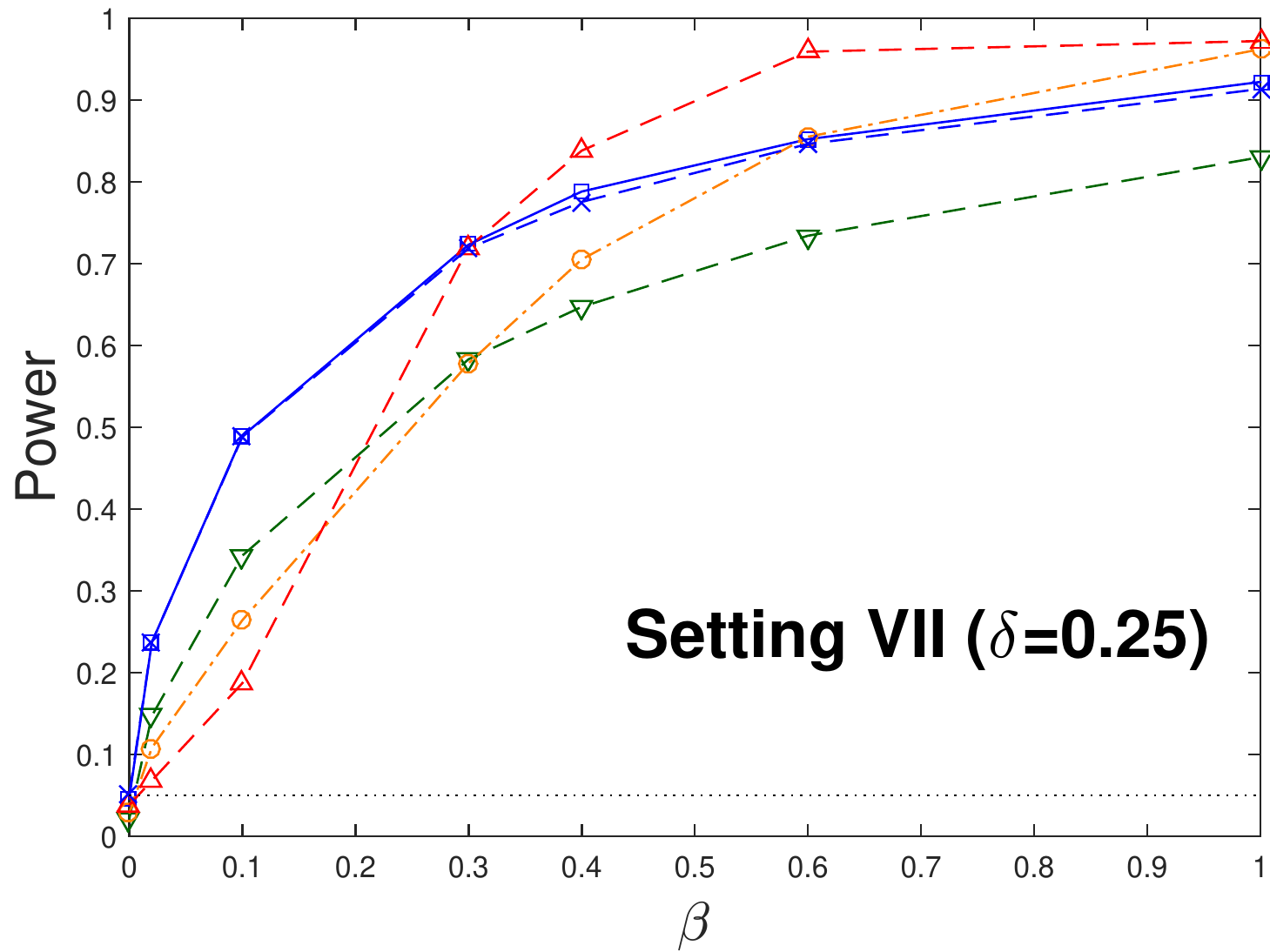}
\end{minipage}%
\begin{minipage}[t]{0.19\textwidth}
\centering
\includegraphics[width=0.95in]{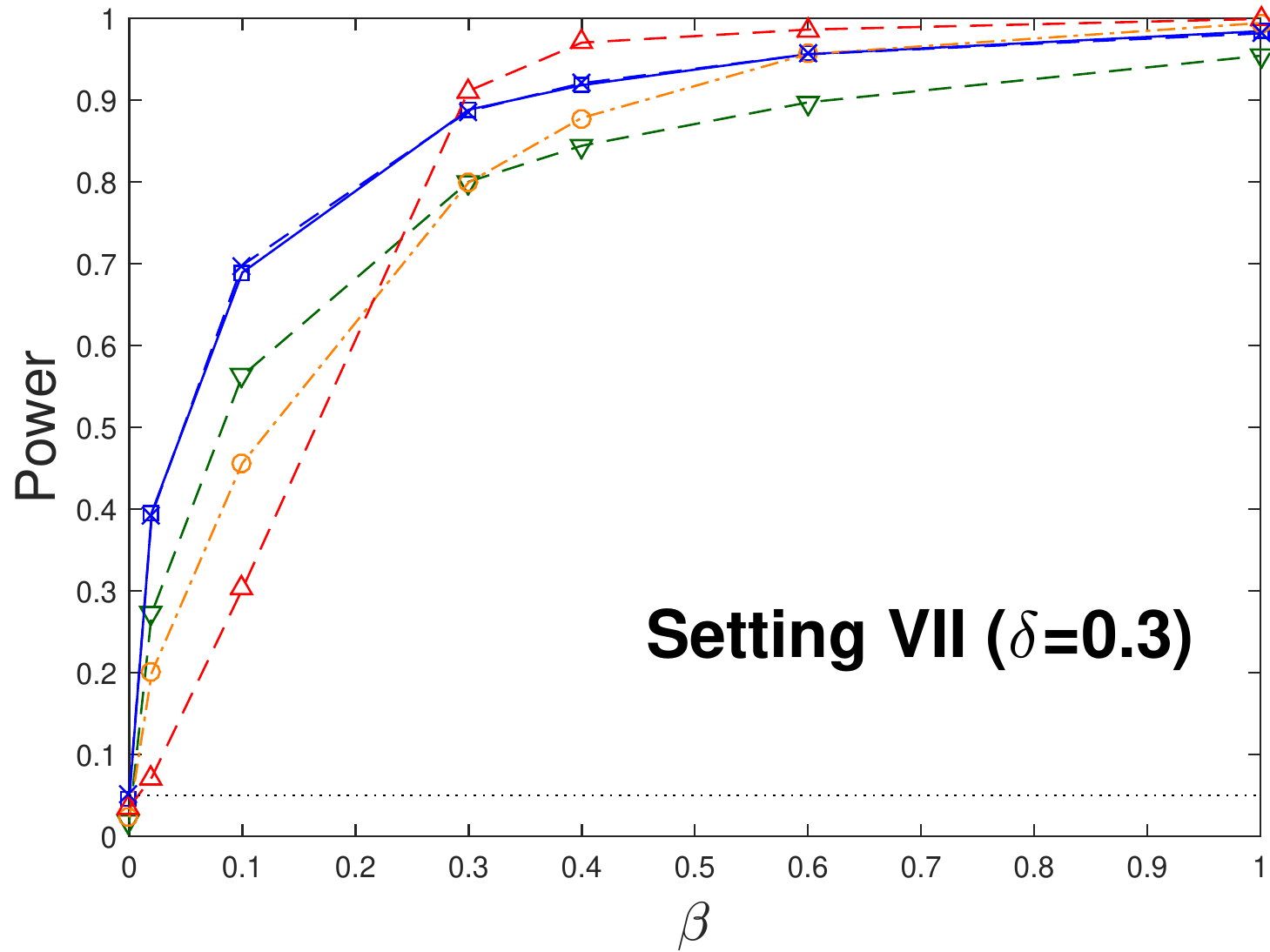}
\end{minipage}%
 \caption{Shown are the bootstrap approximated power curve of the DCF test (crosses), and the empirical power curves of four methods: the DCF test (squares), the CL test (triangles point down), the XL test (circles), and the CQ test (triangles point up) based on $1000$ Monte Carlo runs under Settings I--VII  across different signal  levels of $\delta$ and sparsity levels of $\beta$.}
  \label{fig:1}
\end{figure}

We conduct the four tests and calculate the rejection proportions to assess the empirical power at different signal levels $\delta$ and sparsity levels $\beta$  in each setting as described above, based on $1000$ Monte Carlo runs. {The numerical results of these six settings are shown in {Tables~\ref{tab:1}--\ref{tab:2}}. For visualization, we depict the empirical power plots of all settings in Figure \ref{fig:1}.} We also display the multiplier bootstrap approximation based on another independent set of size $N=10^4$, which agrees well with the empirical size/power of the DCF test and justifies the theoretical assessment in Theorem \ref{theorem:term6}.
We see that the empirical sizes of proposed DCF test agree well with the nominal level $0.05$ in all {six} settings. By comparison, the CQ test is not as stable, and the CL and XL tests show under-estimation of type I error in all settings.

{Regarding power performance under alternatives in these six settings, despite all tests suffering low power for the weak signals $\delta=.1$ and $\delta=.15$, the DCF test still dominates the other tests at all levels of $\beta$. When the signal strength rises to $\delta=.2$, the results in Setting I indicate that the DCF test outperforms the other tests, except for the CQ test when $\beta\ge 80\%$ (a very dense alternative).  Although the power of CQ test increases above that of DCF test at $\beta=80\%$, the gains are not substantial when both tests have high power.
Similar patterns are observed in Settings II, III, V, VI with $\delta=0.25$ for $\beta$ ranging between $80\%$ and $83\%$, and Settings III, IV with $\delta=0.3$ for $\beta$ at $80\%$ and $90\%$, respectively. This phenomenon is visually shown in the power plot in Figure \ref{fig:1}. It is also noted the DCF test dominates the CL ($L_\infty$ type) and XL (combined type) uniformly in these settings over all levels of $\delta$ and $\beta$. To summarize, except for the rapidly increased power of CQ test in very dense alternatives, the DCF test outperforms the other tests over various signal levels of $\delta$ in a broad range of sparsity levels $\beta$, for alternatives with varied magnitudes and signs. Moreover, the gains are sustainable in the situations that the data structures get more complex, e.g., highly unbalanced sizes, heavy-tailed or skewed distributions.
}

We further examine alternatives with common/fixed signal upon reviewer's request under the ``completely relaxed setting'', denoted by Setting VII, where we let  $\mu^Y={\delta}(1,\dots,1_{\lfloor \beta p \rfloor}, 0_{p-\lfloor \beta p \rfloor}')'$. Note that the empirical sizes of four tests in Setting VII are the same as those in Setting III (thus not reported), while the power patterns appear to favor the CQ test when increasing for dense alternatives (DCF still dominates in the range of less dense levels). Here numerical power values are not tabulated for conciseness, given that the visualization in Figure \ref{fig:1} suffices. We conclude this section by pointing out that,
compared to Setting I--VI in which nonzero signals $\theta_k \sim U(-\delta, \delta)$, the alternatives in Setting VII with common/fixed signal are more stringent and easy to be violated in practice.


\section{Real data example}\label{sec:realdata}
We analyze a dataset obtained from the UCI Machine Learning Repository,  \blue{https://archive.ics.uci.edu/ml/datasets/eeg+database}.
The data consist of $122$ individuals, out of which $n=45$ participants belong to the control group, while the remaining $m=77$ are in the alcoholic group.
In the experiment, each subject was shown to a single stimulis (e.g., picture of object) selected from the 1980 Snodgrass and Vanderwart picture set. Then, for each individual, the researchers recorded the EEG measurements which were sampled at 256 Hz$($3.9-msec epoch$)$ for one second from 64 electrodes on that person's scalps, respectively.
As a common practice of data reduction, for each electrode, we pool the 256 records to form 64 measurements by taking the average of the original records on four proximal grid points.
Likewise, we also pool the 64 electrodes by taking the average on every four proximal electrodes, resulting 16 combined electrodes.
For the control group, we let $\mu_{c,  j}=(\mu_{c,  j,  1},\dots,\mu_{c,  j,  64})' \in \mathbb{R}^{64}$ be the common mean vector of the EEG measurements on $j$'th electrode for  $j=1,\dots,16$. For convenience, we write $\mu_{c}=(\mu_{c,  1}',\dots,\mu_{c,  16}')'\in \mathbb{R}^{p}$ with $p=64\times 16=1024$ that is much larger than $n$ and $m$.
Similarly, for the alcoholic group, let $\mu_{a,  j}=(\mu_{a,  j,  1},\dots,\mu_{a,  j,  64})' \in \mathbb{R}^{64}$  be the common mean vector of EEG measurements on $j$'th electrode for $j=1,\dots,16$, and  denote $\mu_{a}=(\mu_{a,  1}',\dots,\mu_{a,  16}')'\in \mathbb{R}^{p}$.
We are interested in the hypothesis test
\begin{align*}
H_0: \mu_{c}=\mu_{a} \quad \text{v.s.} \quad H_a: \mu_{c}\neq \mu_{a}
\end{align*}
to determine whether there is any difference in means of EEG between two groups.
We first carry out the DCF, CL, XL and CQ tests, whose $p$-values are given by $.006$, $.1708$, $.093$ and $.0955$, shown in Table~\ref{tab:3}. In literature \cite{huss:15} provided evidence for the mean difference between two groups, the proposed DCF test indeed detected the difference with statistical significance while the other tests failed to.

For further verification, we carry out random bootstrap with replacement separately within each sample,  and repeat for 500 times. The rejection proportions for the four tests over the 500 bootstrapped datasets are given in  Table~\ref{tab:3}, which shows that the highest rejection proportion among the four tests is achieved by DCF at $82\%$. This is in line with the smallest and significant $p$-value given by the DCF test based on the dataset itself. We also perform 500 random permutations of the whole dataset (i.e., mixing up two groups that eliminate the group difference) and conduct four tests over each permuted dataset.
From Table~\ref{tab:3}, we see that the rejection proportion of the  DCF test $(.046)$ is close to the nominal level $\alpha=.05$, while those of the other tests differ considerably.

\newpage
\global\pdfpageattr\expandafter{\the\pdfpageattr/Rotate 180}
\begin{sidewaystable}[htbp]
\caption{Rejection proportions$(\%)$ calculated for four testing methods at different signal strength levels of $\delta$ and sparsity levels of $\beta$ based on 1000 Monte Carlo runs, where $\beta=0$ corresponds to the null hypothesis  $\beta=1$ to the fully dense alternative, and  $(n, m, p)=(200, 300, 1000)$. }
{\scriptsize
\setlength{\tabcolsep}{1.7mm}{
\begin{tabular*}{19.6cm}{@{\extracolsep{\fill}}l|llll|llll|llll|llll|llll}
\hline \hline
\multicolumn{21}{c}{Setting I: i.i.d equal cov}\\
&\multicolumn{4}{c}{$\delta=.1$}&\multicolumn{4}{c}{$\delta=.15$}&\multicolumn{4}{c}{$\delta=.2$}&\multicolumn{4}{c}{$\delta=.25$}&\multicolumn{4}{c}{$\delta=.3$}\\
\cmidrule(lr){2-5}\cmidrule(lr){6-9}\cmidrule(lr){10-13}\cmidrule(lr){14-17}\cmidrule(lr){18-21}
Test& DCF& CL& XL& CQ & DCF& CL& XL& CQ& DCF& CL& XL& CQ& DCF& CL& XL& CQ& DCF& CL& XL& CQ\\
$\beta=0$&$4.20$& $2.40$& $3.90$& $5.80$&$4.30$& $2.30$& $2.40$& $3.60$&$4.50$& $2.80$& $3.70$& $6.00$&$4.60$& $2.70$& $2.20$& $3.80$&$5.00$& $3.10$& $3.80$& $6.10$\\
$\beta=.02$&$5.00$& $3.20$& $2.50$& $3.40$&$7.50$& $4.80$& $3.70$& $3.50$&$15.4$& $10.5$& $6.50$& $3.90$&$31.7$& $23.3$& $14.6$& $4.40$&$59.0$& $47.9$& $32.6$& $4.90$\\
$\beta=.04$&$5.80$& $3.70$& $2.80$& $3.60$&$10.0$& $6.20$& $4.30$& $3.90$&$20.6$& $14.2$& $8.80$& $4.70$&$40.6$& $30.8$& $20.0$& $5.10$&$72.0$& $58.9$& $41.5$& $5.30$\\
$\beta=.2$&$9.90$& $6.50$& $3.90$& $4.50$&$22.7$& $15.9$& $9.10$& $5.30$&$48.7$& $37.3$& $23.7$& $7.40$&$84.5$& $72.4$& $52.0$& $11.6$&$99.3$& $97.1$& $87.2$& $23.4$\\
$\beta=.4$&$13.9$& $9.40$& $5.30$& $5.20$&$35.3$& $25.4$& $14.4$& $7.80$&$68.8$& $57.1$& $37.9$& $16.5$&$96.8$& $91.1$& $72.7$& $42.5$&$100$& $100$& $97.7$& $96.9$\\
$\beta=.6$&$17.8$& $11.8$& $6.70$& $5.60$&$45.8$& $33.7$& $20.3$& $12.8$&$82.7$& $71.8$& $51.1$& $39.9$&$99.6$& $97.2$& $86.8$& $99.1$&$100$& $100$& $100$& $100$\\
{$\beta=.8$}&$22.4$& $13.8$& $9.00$& $8.30$&$55.5$& $40.1$& $24.4$& $23.1$&$91.3$& $81.7$& $61.5$& $91.7$&$100$& $99.2$& $95.7$& $100$&$100$& $100$& $100$& $100$\\
$\beta=1$&$26.5$& $17.9$& $10.9$& $10.7$&$64.5$& $48.1$& $30.6$& $39.5$&$95.0$& $88.5$& $70.1$& $100$&$100$& $99.6$& $100$& $100$&$100$& $100$& $100$& $100$\\\hline
\multicolumn{21}{c}{Setting II: i.i.d unequal cov}\\
&\multicolumn{4}{c}{$\delta=.1$}&\multicolumn{4}{c}{$\delta=.15$}&\multicolumn{4}{c}{$\delta=.2$}&\multicolumn{4}{c}{$\delta=.25$}&\multicolumn{4}{c}{$\delta=.3$}\\
\cmidrule(lr){2-5}\cmidrule(lr){6-9}\cmidrule(lr){10-13}\cmidrule(lr){14-17}\cmidrule(lr){18-21}
Test& DCF& CL& XL& CQ & DCF& CL& XL& CQ& DCF& CL& XL& CQ& DCF& CL& XL& CQ& DCF& CL& XL& CQ\\
$\beta=0$&$4.90$& $1.80$& $3.70$& $6.10$&$5.20$& $1.30$& $2.20$& $3.80$&$5.00$& $1.60$& $3.60$& $6.00$&$4.80$& $1.20$& $3.50$& $6.30$&$5.00$& $1.90$& $3.90$& $6.20$\\
$\beta=.02$&$4.70$& $1.00$& {$2.40$}& $3.80$&$6.60$& $1.40$& $2.70$& $4.10$&$10.7$& $2.60$& $2.90$& $4.10$&$19.1$& $6.70$& $4.80$& $4.40$&$33.3$& $14.4$& $8.80$& $4.50$\\
$\beta=.04$&$5.80$& $1.30$& $2.50$& $4.10$&$7.90$& $1.80$& $2.80$& $4.30$&$12.5$& $3.50$& $3.40$& {$4.50$}&$24.7$& $9.30$& $6.00$& $4.60$&$42.5$& $20.3$& $12.2$& $5.00$\\
$\beta=.2$&$8.10$& $1.90$& $2.70$& $4.60$&$15.0$& $4.40$& $3.80$& $4.90$&$30.9$& $11.2$& $7.20$& $6.40$&$57.6$& $26.5$& $16.3$& $8.40$&$86.8$& $52.1$& $33.9$& $11.8$\\
$\beta=.4$&$10.6$& $2.80$& $3.10$& $5.70$&$22.4$& $7.20$& $5.70$& $6.50$&$47.3$& $19.6$& $11.6$& $10.0$&$78.7$& $43.2$& $26.6$& $19.1$&$97.5$& $74.1$& $53.2$& $45.7$\\
$\beta=.6$&$13.5$& $3.30$& $3.80$& $6.70$&$29.2$& $9.60$& $6.70$& $8.40$&$59.0$& $26.5$& $17.1$& $18.7$&$90.5$& $56.2$& $36.7$& $54.4$&$99.8$& $88.1$& $70.1$& $99.6$\\
$\beta=.8$&$16.4$& $4.60$& $4.50$& $7.40$&$37.4$& $11.9$& $8.60$& $12.6$&$70.9$& $32.9$& $21.4$& $39.6$&$95.6$& $67.0$& $47.0$& $98.9$&$100$& $94.2$& $90.5$& $100$\\
$\beta=1$&$19.2$& $5.20$& $5.00$& $8.10$&$43.5$& $14.4$& $10.7$& $18.3$&$79.4$& $39.9$& $28.1$& $79.8$&$98.2$& $76.2$& $67.8$& $100$&$100$& $97.7$& $99.9$& $100$\\\hline
\multicolumn{21}{c}{Setting III: completely relaxed}\\
&\multicolumn{4}{c}{$\delta=.1$}&\multicolumn{4}{c}{$\delta=.15$}&\multicolumn{4}{c}{$\delta=.2$}&\multicolumn{4}{c}{$\delta=.25$}&\multicolumn{4}{c}{$\delta=.3$}\\
\cmidrule(lr){2-5}\cmidrule(lr){6-9}\cmidrule(lr){10-13}\cmidrule(lr){14-17}\cmidrule(lr){18-21}
Test& DCF& CL& XL& CQ & DCF& CL& XL& CQ& DCF& CL& XL& CQ& DCF& CL& XL& CQ& DCF& CL& XL& CQ\\
$\beta=0$&$4.70$& $2.00$& $3.90$& $6.30$&$4.50$& $1.70$& $2.30$& $3.50$&$4.80$& $1.90$& $3.70$& $6.10$&$4.60$& $2.20$& $2.80$& $3.90$&$5.10$& $2.10$& $3.80$& $6.20$\\
$\beta=.02$&$4.90$& $2.10$& $3.20$& {$4.40$}&$6.50$& $2.70$& $3.50$& $5.30$&$9.40$& $4.30$& $4.00$& $5.60$&$13.6$& $7.80$& $6.20$& $5.70$&$24.9$& $12.9$& $10.1$& $5.90$\\
$\beta=.04$&$5.60$& $2.40$& $3.50$& $4.70$&$7.60$& $3.40$& $4.20$& $5.40$&$12.1$& $6.00$& $5.00$& $5.80$&$19.1$& $10.8$& $8.80$& $6.00$&$32.8$& $19.1$& $13.8$& $6.50$\\
$\beta=.2$&$7.50$& $3.80$& $4.30$& $5.80$&$12.1$& $6.00$& $5.60$& $6.60$&$23.9$& $12.5$& $8.90$& $7.50$&$44.2$& $26.3$& $16.6$& $9.30$&$71.6$& $50.2$& $32.1$& $14.1$\\
$\beta=.4$&$9.40$& $3.90$& $4.50$& $6.30$&$18.4$& $9.00$& $8.00$& $7.60$&$35.8$& $19.9$& $12.7$& $11.7$&$62.3$& $40.8$& $26.4$& $18.5$&$89.3$& $69.9$& $48.6$& $31.5$\\
$\beta=.6$&$11.5$& $4.90$& $6.20$& $6.80$&$24.0$& $10.8$& $8.90$& $9.50$&$48.0$& $28.2$& $18.2$& $17.8$&$76.8$& $55.3$& $37.0$& $35.7$&$96.5$& $83.8$& $64.6$& $83.1$\\
$\beta=.8$&$13.6$& $6.40$& {$6.60$}& $7.00$&$30.3$& $13.5$& $11.7$& $12.7$&$57.3$& $36.4$& $23.4$& $28.5$&$86.7$& $65.0$& $45.1$& $81.2$&$98.5$& $91.6$& $77.4$& $100$\\
$\beta=.83$&$14.3$& $7.10$& $6.80$& $7.50$&$31.0$& $14.6$& $11.8$& $13.1$&$58.0$& $37.6$& $23.9$& $30.8$&$87.6$& $66.1$& $46.1$& $88.0$&$98.9$& $92.6$& $79.2$& $100$\\
{$\beta=1$}&$16.6$& $8.50$& $7.40$& $8.00$&$35.0$& $17.2$& $13.9$& $17.3$&$65.6$& $42.8$& $28.3$& $48.2$&$90.8$& $75.7$& $56.0$& $99.9$&$99.2$& $95.5$& $95.7$& $100$\\\hline
\end{tabular*}}
}
\label{tab:1}
\end{sidewaystable}
\clearpage
\newpage

\global\pdfpageattr\expandafter{\the\pdfpageattr/Rotate 0}
\begin{sidewaystable}[htbp]
\caption{Rejection proportions$(\%)$ calculated for four testing methods at different signal strength levels of $\delta$ and sparsity levels of $\beta$ based on 1000 Monte Carlo runs, where $\beta=0$ corresponds to the null hypothesis $\beta=1$ to the fully dense alternative,   $(n, m, p)=(100, 400, 1000)$ for Setting IV, and $(n, m, p)=(200, 300, 1000)$ for Settings V and VI. }
{\scriptsize
\setlength{\tabcolsep}{1.7mm}{
\begin{tabular*}{19.6cm}{@{\extracolsep{\fill}}l|llll|llll|llll|llll|llll}
\hline \hline
\multicolumn{21}{c}{Setting IV: completely relaxed and highly unequal sample sizes}\\
&\multicolumn{4}{c}{$\delta=.1$}&\multicolumn{4}{c}{$\delta=.15$}&\multicolumn{4}{c}{$\delta=.2$}&\multicolumn{4}{c}{$\delta=.25$}&\multicolumn{4}{c}{$\delta=.3$}\\
\cmidrule(lr){2-5}\cmidrule(lr){6-9}\cmidrule(lr){10-13}\cmidrule(lr){14-17}\cmidrule(lr){18-21}
Test& DCF& CL& XL& CQ & DCF& CL& XL& CQ& DCF& CL& XL& CQ& DCF& CL& XL& CQ& DCF& CL& XL& CQ\\
$\beta=0$&$4.70$& $.800$& $3.90$& $6.80$&$4.90$& $.900$& $3.80$& $6.30$&$5.20$& $.700$& $3.90$& $6.10$&$4.50$& $.600$& $3.50$& $6.00$&$4.90$& $.500$& $3.40$& $6.10$\\
$\beta=.02$&$5.20$& $1.10$& $2.90$& {$4.70$}&$5.90$& $1.00$& $3.60$& {$5.60$}&$6.70$& $1.40$& $4.60$& $5.80$&$8.90$& $2.40$& $5.00$& $5.80$&$13.2$& $4.20$& $6.20$& $5.90$\\
$\beta=.04$&$5.40$& $1.20$& $3.00$& {$4.80$}&$6.30$& $1.30$& $4.50$& $5.70$&$7.80$& $1.90$& $5.00$& $6.00$&$11.2$& $3.30$& $5.60$& $6.10$&$17.6$& $5.70$& $7.10$& $6.20$\\
$\beta=.2$&$6.60$& $1.30$& $3.30$& $5.40$&$9.20$& $2.20$& $5.10$& $5.80$&$14.9$& $3.90$& $5.70$& $6.20$&$25.3$& $8.70$& $7.00$& $7.50$&$42.8$& $16.5$& $11.8$& $8.80$\\
$\beta=.4$&$7.80$& $2.00$& $4.30$& $5.50$&$12.4$& $3.40$& $5.20$& $6.10$&$22.3$& $6.60$& $7.10$& $8.60$&$38.2$& $13.0$& $9.70$& $10.7$&$61.3$& $24.8$& $17.0$& $15.8$\\
$\beta=.6$&$9.10$& $2.40$& $4.60$& $5.80$&$16.1$& $3.80$& $5.50$& $7.90$&$29.5$& $10.0$& $9.20$& $10.8$&$49.9$& $19.3$& $14.3$& $17.6$&$75.3$& $33.7$& $21.9$& $34.2$\\
$\beta=.8$&$10.5$& $2.50$& $4.70$& $6.10$&$19.9$& $5.20$& $6.70$& $9.20$&$36.9$& $12.7$& $10.9$& $14.5$&$60.1$& $24.0$& $19.3$& $32.2$&$84.9$& $46.6$& $33.6$& $78.2$\\
$\beta=.9$&$11.3$& $2.80$& $4.80$& $6.40$&$21.9$& $5.40$& $7.10$& $9.90$&$39.5$& $13.3$& $12.6$& $17.7$&$64.6$& $26.6$& $21.6$& $43.8$&$88.0$& $48.6$& $35.3$& $94.0$\\
{$\beta=1$}&$12.1$& $2.90$& {$5.30$}& $7.30$&$23.4$& $5.90$& $7.30$& $11.0$&$42.0$& $14.6$& $12.8$& $21.7$&$68.6$& $29.6$& $24.5$& $59.0$&$90.9$& $53.1$& $41.9$& $99.4$\\\hline
\multicolumn{21}{c}{Setting V: completely relaxed and heavy-tailed}\\
&\multicolumn{4}{c}{$\delta=.1$}&\multicolumn{4}{c}{$\delta=.15$}&\multicolumn{4}{c}{$\delta=.2$}&\multicolumn{4}{c}{$\delta=.25$}&\multicolumn{4}{c}{$\delta=.3$}\\
\cmidrule(lr){2-5}\cmidrule(lr){6-9}\cmidrule(lr){10-13}\cmidrule(lr){14-17}\cmidrule(lr){18-21}
Test& DCF& CL& XL& CQ & DCF& CL& XL& CQ& DCF& CL& XL& CQ& DCF& CL& XL& CQ& DCF& CL& XL& CQ\\
$\beta=0$ &$4.20$& $2.20$& $3.80$& $6.20$ &$5.20$& $2.50$& $3.90$& $6.10$&$4.70$& $1.90$& $2.90$& $6.00$ &$4.30$& $2.00$& $1.70$& $3.90$&$4.50$& $2.30$& $2.00$& $3.70$\\
$\beta=.02$&$5.50$& $2.10$& $3.70$& {$5.40$}&$6.40$& $2.50$& {$3.90$}& $5.50$&$9.50$& $4.40$& $4.60$& $6.10$&$15.3$& $7.40$& $6.30$& $6.10$&$25.5$& $15.0$& $10.3$& $6.20$\\
$\beta=.04$&$6.20$& $2.30$& $3.80$& $5.50$&$7.20$& $3.60$& $4.20$& $6.00$&$12.6$& $6.60$& $5.80$& {$6.20$}&$18.9$& $9.80$& $7.00$& $6.50$&$33.3$& $20.7$& $13.0$& $7.10$\\
$\beta=.2$&$7.50$& $3.60$& $4.00$& $5.80$&$12.4$& $6.80$& $6.50$& $7.30$&$23.5$& $13.0$& $9.60$& $8.90$&$45.6$& $27.6$& $17.9$& $11.3$&$71.7$& $52.6$& $33.8$& $14.1$\\
$\beta=.4$&$9.50$& $4.20$& $4.40$& $5.90$&$18.1$& $9.00$& $8.30$& $8.90$&$35.9$& $21.3$& $14.0$& $12.7$&$64.4$& $43.2$& $26.9$& $18.5$&$90.3$& $73.4$& $52.0$& $33.7$\\
$\beta=.6$&$11.5$& $5.10$& $4.50$& $6.00$&$23.8$& $12.6$& $10.1$& $11.7$&$46.7$& $29.2$& $19.4$& $17.8$&$77.5$& $55.9$& $37.4$& $38.9$&$97.4$& $86.5$& $65.6$& $88.2$\\
$\beta=.8$&$13.7$& $7.30$& $6.20$& {$8.80$}&$29.4$& $16.0$& $12.3$& $14.1$&$56.5$& $36.9$& $24.9$& $28.9$&$87.4$& $69.1$& $48.3$& $81.4$&$99.2$& $93.6$& $80.0$& $100$\\
$\beta=.83$&$14.1$& $7.50$& $6.30$& $9.20$&$30.6$& $17.3$& $13.0$& $15.2$&$58.1$& $38.1$& $26.0$& $32.0$&$88.1$& $70.1$& $49.5$& $87.5$&$99.3$& $94.1$& $82.1$& $100$\\
{$\beta=1$}&$16.1$& $8.90$& $7.40$& $9.40$&$34.9$& $18.9$& $15.0$& $17.2$&$64.5$& $44.6$& $30.5$& $52.2$&$91.6$& $75.1$& $56.6$& $99.8$&$99.7$& $96.5$& $96.0$& $100$\\\hline
\multicolumn{21}{c}{Setting VI: completely relaxed and skewed}\\
&\multicolumn{4}{c}{$\delta=.1$}&\multicolumn{4}{c}{$\delta=.15$}&\multicolumn{4}{c}{$\delta=.2$}&\multicolumn{4}{c}{$\delta=.25$}&\multicolumn{4}{c}{$\delta=.3$}\\
\cmidrule(lr){2-5}\cmidrule(lr){6-9}\cmidrule(lr){10-13}\cmidrule(lr){14-17}\cmidrule(lr){18-21}
Test& DCF& CL& XL& CQ & DCF& CL& XL& CQ& DCF& CL& XL& CQ& DCF& CL& XL& CQ& DCF& CL& XL& CQ\\
$\beta=0$&$4.20$& $2.10$& $2.40$& $3.60$&$4.90$& $1.40$& $2.70$& $3.80$&$5.00$& $1.60$& $2.50$& $3.90$&$4.90$& $2.40$& $3.70$& $5.80$&$4.70$& $1.90$& $2.70$& $3.90$\\
$\beta=.02$&$4.80$& $1.30$& $2.70$& {$4.40$}&$6.20$& $1.70$& $3.10$& $4.70$&$7.50$& $2.70$& $3.80$& $4.90$&$12.9$& $5.80$& $5.00$& $5.00$&$24.3$& $11.8$& $8.30$& $5.00$\\
$\beta=.04$&$5.30$& $1.40$& $3.00$& $4.60$&$7.00$& $2.30$& $3.30$& $4.90$&$11.3$& $5.20$& $4.50$& {$5.10$}&$17.1$& $8.70$& $7.00$& $5.10$&$32.2$& $17.3$& $12.0$& $5.30$\\
$\beta=.2$&$7.40$& $3.00$& $3.30$& $4.80$&$12.8$& $5.80$& $5.00$& $5.80$&$23.0$& $12.9$& $9.20$& $6.40$&$42.4$& $25.6$& $17.7$& $8.40$&$71.3$& $48.6$& $32.5$& $12.4$\\
$\beta=.4$&$9.40$& $4.50$& $4.00$& $5.10$&$18.7$& $9.30$& $6.80$& $7.20$&$37.3$& $21.9$& $13.4$& $10.6$&$62.9$& $43.3$& $28.6$& $17.3$&$89.4$& $70.9$& $51.8$& $30.7$\\
$\beta=.6$&$11.5$& $5.70$& $4.50$& $6.20$&$24.7$& $12.3$& $9.60$& $9.50$&$48.1$& $29.8$& $18.1$& $16.5$&$75.7$& $55.0$& $37.6$& $34.8$&$95.9$& $83.7$& $64.5$& $86.4$\\
$\beta=.8$&$14.2$& $6.30$& $5.80$& $6.60$&$30.5$& $14.9$& $10.5$& $12.5$&$58.0$& $37.6$& $23.4$& $27.1$&$86.7$& $65.4$& $44.9$& $80.2$&$98.7$& $92.0$& $77.5$& $100$\\
$\beta=.83$&$14.3$& $7.50$& $6.30$& $6.70$&$31.6$& $15.3$& $10.8$& $13.2$&$60.1$& $39.3$& $24.2$& $29.8$&$87.9$& $66.5$& $46.2$& $87.4$&$98.9$& $92.8$& $81.0$& $100$\\
{$\beta=1$}&$16.3$& $8.90$& $6.70$& $7.40$&$35.9$& $19.3$& $14.6$& $16.4$&$67.0$& $44.7$& $29.4$& $49.3$&$91.0$& $74.6$& $57.2$& $99.9$&$99.3$& $96.1$& $97.2$& $100$\\\hline
\end{tabular*}}
}
\label{tab:2}
\end{sidewaystable}
\clearpage

\begin{table}[htb]
\caption{}
\begin{tabular*}{12cm}{@{\extracolsep{\fill}}lllll}
\hline \hline
\multicolumn{5}{c}{p-values of the four tests based on the dataset.}\\ \hline
Test& DCF& CL& XL& CQ \\
p-value&    $.006$& $.1708$& $.093$& $.0955$\\ \hline\hline
\multicolumn{5}{c}{Rejection proportions$(\%)$ of the four tests over 500 bootstrapped data sets.}\\ \hline
Test& DCF& CL& XL& CQ \\
Rejection proportion& $82$& $65.8$& $65$& $58$ \\\hline\hline
\multicolumn{5}{c}{Rejection proportions$(\%)$  of the four tests over 500 random permutations.}\\ \hline
Test& DCF& CL& XL& CQ \\
Rejection proportion& $4.6$& $1.8$& $3.4$& $7.4$ \\ \hline
\end{tabular*}
\label{tab:3}
\end{table}
%
\section*{Appendix}
We first present some auxiliary lemmas that are key for deriving the main theorems. To introduce  Lemma~\ref{lemma:term1}, for any $\beta>0$ and $y\in \mathbb{R}^p$, we define a function $F_\beta(w)$ as
\begin{align*}
F_\beta(w)=\beta^{-1}\log\big[\sum_{j=1}^p\exp\{\beta(w_j-y_j)\}\big], w\in \mathbb{R}^p,
\end{align*}
which satisfies the property
\begin{align*}
0\leq F_\beta(w)-\max_{1\leq j \leq p}(w_j-y_j)\leq \beta^{-1}\log p,
\end{align*}
for every $w\in \mathbb{R}^p$ by $(1)$ in \cite{chern:14:1}.
In addition, we let $\varphi_0:\mathbb{R}\to [0, 1]$ be a real valued function such that $\varphi_0$ is thrice continuously differentiable and $\varphi_0(z)=1$ for $z\leq 0$ and $\varphi_0(z)=0$ for $z\geq 1$. For any $\phi\geq 1$, define a function $\varphi(z)=\varphi_0(\phi z), z\in \mathbb{R}$. Then, for any $\phi\geq 1$ and $y\in \mathbb{R}^p$, denote $\beta=\phi\log p$ and define a function $\kappa:\mathbb{R}^p\to [0, 1]$ as
 \begin{align}\label{inequality:term4}
\kappa(w)=\varphi_0(\phi F_{\phi\log p}(w))=\varphi(F_{\beta}(w)),  w\in \mathbb{R}^p.
\end{align}
Lemma~\ref{lemma:term1} is devoted to characterize the properties of the function $\kappa$ defined in (\ref{inequality:term4}), which can be also referred to Lemmas A.5 and A.6 in \cite{chern:13}.

\begin{lemma}\label{lemma:term1}
For any $\phi\geq 1$ and $y\in \mathbb{R}^p$, we denote $\beta=\phi \log p$, then the function $\kappa$ defined in (\ref{inequality:term4}) has the following properties, where $\kappa_{jkl}$ denotes $\partial_j\partial_k\partial_l\kappa$.
For any $j, k, l=1,\dots,p$, there exists a nonnegative function $Q_{jkl}$ such that
\begin{enumerate}
\item[1)]$|\kappa_{jkl}(w)|\leq Q_{jkl}(w)$ for all $w\in \mathbb{R}^p$,
\item[2)]$\sum_{j=1}^p\sum_{k=1}^p\sum_{l=1}^pQ_{jkl}(w)\lesssim (\phi^3+\phi^2\beta+\phi \beta^2)\lesssim \phi \beta^2$ for all $w\in \mathbb{R}^p$,
\item[3)]$Q_{jkl}(w)\lesssim Q_{jkl}(w+\tilde{w})\lesssim Q_{jkl}(w)$ for all $w\in \mathbb{R}^p$ \\ and $\tilde{w}\in \{w^*\in \mathbb{R}^p: \max_{1\leq j \leq p}|w_j^*|\beta \leq 1\}$.
\end{enumerate}	
\end{lemma}

To state Lemma~\ref{lemma:term2}, a two-sample extension of Lemma~5.1 in \cite{chern:16:1}, for any sequence of constants $\delta_{n, m}$ that depends on both $n$ and $m$, we denote the quantity $\rho_{n, m}$ by
\begin{eqnarray}
\rho_{n, m}&=&\sup_{v\in [0, 1]}\sup_{y\in \mathbb{R}^p}\big|P\big\{v^{1/2}(S_n^X-n^{1/2}\mu^X+\delta_{n, m}S_m^Y-\delta_{n, m}m^{1/2}\mu^Y)+\nonumber \\
&&(1-v)^{1/2}(S_n^F-n^{1/2}\mu^X+\delta_{n, m}S_m^G-\delta_{n, m}m^{1/2}\mu^Y)\leq y\big\}- \label{eq:rho1} \\
&&P(S_n^F-n^{1/2}\mu^X+\delta_{n, m}S_m^G-\delta_{n, m}m^{1/2}\mu^Y\leq y)\big|. \nonumber
\end{eqnarray}
Lemma~\ref{lemma:term2} provides a bound on $\rho_{n, m}$ under some general conditions.
\begin{lemma}\label{lemma:term2}
For any $\phi_1, \phi_2\geq 1$ and any sequence of constants $\delta_{n, m}$, assume the following condition (a) holds,
\begin{enumerate}
\item[(a)]There exists a universal constant $b>0$ such that $$\min_{1\leq j \leq p}E\{(S_{nj}^X-n^{1/2}\mu_j^X+\delta_{n, m}S_{mj}^Y-\delta_{n, m}m^{1/2}\mu_j^Y)^2\}\geq b.$$
\end{enumerate}	
Then we have
\begin{align*}
\rho_{n,m}\lesssim &n^{-1/2}\phi_1^2(\log p)^2\big\{\phi_1 L_n^X \rho_{n,m}+L_n^X(\log p)^{1/2}+\phi_1 M_n(\phi_1)\big\}+\\
&m^{-1/2}\phi_2^2(\log p)^2|\delta_{n, m}|^3\big\{\phi_2 L_m^Y \rho_{n,m}+L_m^Y(\log p)^{1/2}+\phi_2 M_m^*(\phi_2)\big\}+\\
&(\min\{\phi_1, \phi_2\})^{-1}(\log p)^{1/2},
\end{align*}	
up to a positive universal constant that depends only on $b$, where $\rho_{n, m}$ is defined in (\ref{eq:rho1}).
\end{lemma}

To state Lemma~\ref{lemma:term3} that is a two-sample version of Corollary~5.1 in \cite{chern:16:1}, for any sequence of constants $\delta_{n, m}$ that depends on both $n$ and $m$, we denote the quantity $\rho_{n, m}^*$ by
\begin{eqnarray}
\rho_{n, m}^*&=&\sup_{v\in [0, 1]}\sup_{A\in \mathcal{A}^{Re}}\big|P\big\{v^{1/2}(S_n^X-n^{1/2}\mu^X+\delta_{n, m}S_m^Y-\delta_{n, m}m^{1/2}\mu^Y)+ \nonumber \\
&&(1-v)^{1/2}(S_n^F-n^{1/2}\mu^X+\delta_{n, m}S_m^G-\delta_{n, m}m^{1/2}\mu^Y)\in A\big\}- \label{eq:rho2} \\
&&P(S_n^F-n^{1/2}\mu^X+\delta_{n, m}S_m^G-\delta_{n, m}m^{1/2}\mu^Y\in A)\big|, \nonumber
\end{eqnarray}
which has a similar form to the key quantity $\rho_{n, m}^{**}$ in Theorems~\ref{theorem:term2} and \ref{theorem:term4}.
Lemma~\ref{lemma:term3} gives a bound on $\rho_{n, m}^*$ under some general conditions, and it is important for deriving Lemma~\ref{theorem:term1} and  Theorem~\ref{theorem:term2}.

\begin{lemma}\label{lemma:term3}
For any $\phi_1, \phi_2\geq 1$ and any sequence of constants $\delta_{n, m}$, assume  the following condition (a) holds,
\begin{enumerate}
\item[(a)]There exists a universal constant $b>0$ such that $$\min_{1\leq j \leq p}E\{(S_{nj}^X-n^{1/2}\mu_j^X+\delta_{n, m}S_{mj}^Y-\delta_{n, m}m^{1/2}\mu_j^Y)^2\}\geq b.$$
\end{enumerate}	
Then we have
\begin{align*}
\rho_{n,m}^*\leq &K^*\Big[n^{-1/2}\phi_1^2(\log p)^2\big\{\phi_1 L_n^X \rho_{n,m}^*+L_n^X(\log p)^{1/2}+\phi_1 M_n(\phi_1)\big\}+\\
&m^{-1/2}\phi_2^2(\log p)^2|\delta_{n, m}|^3\big\{\phi_2 L_m^Y \rho_{n,m}^*+L_m^Y(\log p)^{1/2}+\phi_2 M_m^*(\phi_2)\big\}+\\
&(\min\{\phi_1, \phi_2\})^{-1}(\log p)^{1/2}\Big],
\end{align*}	
up to a universal constant $K^*>0$ that depends only on $b$, where $\rho_{n, m}^*$ is defined in (\ref{eq:rho2}).
\end{lemma}
Before stating the next Lemma,  for any $\phi\geq 1$, we  denote $M_n(\phi)=M_n^X(\phi)+M_n^F(\phi)$, where $M_n^X(\phi)$ and $M_n^F(\phi)$ are given as follows, respectively,
\begin{align*}
&n^{-1}\sum_{i=1}^n E\big[\max_{1\leq j \leq p}|X_{ij}-\mu_j^X|^3 1\big\{\max_{1\leq j \leq p}|X_{ij}-\mu_j^X|>n^{1/2}/(4\phi\log p)\big\}\big],\\
&n^{-1}\sum_{i=1}^n E\big[\max_{1\leq j \leq p}|F_{ij}-\mu_j^F|^3 1\big\{\max_{1\leq j \leq p}|F_{ij}-\mu_j^F|>n^{1/2}/(4\phi\log p)\big\}\big],
\end{align*}
similar to those adopted in \cite{chern:16:1}. Likewise, for any $\phi\geq 1$ and any sequence of constants $\delta_{n, m}$ that depends on both $n$ and $m$, we denote $M_m^{*}(\phi)=M_m^Y(\phi)+M_m^G(\phi)$ with $M_m^Y(\phi)$ and $M_m^G(\phi)$ as follows, respectively,
\begin{align*}
&m^{-1}\sum_{i=1}^m E\big[\max_{1\leq j \leq p}|Y_{ij}-\mu_j^Y|^3 1\big\{\max_{1\leq j \leq p}|Y_{ij}-\mu_j^Y|>m^{1/2}/(4|\delta_{n, m}|\phi\log p)\big\}\big],\\
&m^{-1}\sum_{i=1}^m E\big[\max_{1\leq j \leq p}|G_{ij}-\mu_j^G|^3 1\big\{\max_{1\leq j \leq p}|G_{ij}-\mu_j^G|>m^{1/2}/(4|\delta_{n, m}|\phi\log p)\big\}\big].
\end{align*}
Recalling the definition of $\rho_{n, m}^{**}$ in (\ref{eq:rho}), Lemma~\ref{theorem:term1} gives an abstract upper bound on $\rho_{n, m}^{**}$ under mild conditions as follows.
\begin{lemma}\label{theorem:term1} 
For any sequence of constants $\delta_{n, m}$, assume we have the following conditions (a)--(b):
\begin{enumerate}
\item[(a)]There exists a universal constant $b>0$ such that $$\min_{1\leq j \leq p}E\{(S_{nj}^X-n^{1/2}\mu_j^X+\delta_{n, m}S_{mj}^Y-\delta_{n, m}m^{1/2}\mu_j^Y)^2\}\geq b.$$
\item[(b)]There exist two sequences of constants $\bar{L}_n^*$ and $\bar{L}_m^{**}$ such that we have $\bar{L}_n^*\geq L_n^X$ and $\bar{L}_m^{**}\geq L_m^Y$ respectively. Moreover, we also have
\begin{align*}
\phi_n^*&=K_1\{(\bar{L}_n^*)^2(\log p)^4/n\}^{-1/6}\geq 2, \\
\phi_m^{**}&=K_1\{(\bar{L}_m^{**})^2(\log p)^4|\delta_{n, m}|^6/m\}^{-1/6}\geq 2,
\end{align*}
for a universal constant $K_1\in (0, (K^*\vee 2)^{-1}]$, where the positive constant $K^*$ that depends on $n$ as defined in Lemma~\ref{lemma:term3} in the Appendix.
\end{enumerate}	
Then we have the following property, where $\rho_{n, m}^{**}$ is defined in (\ref{eq:rho}),
\begin{align*}
\rho_{n,m}^{**}\leq& K_2\big[\{(\bar{L}_n^*)^2(\log p)^7/n\}^{1/6}+\{M_n(\phi_n^*)/\bar{L}_n^*\}+\\
&\{(\bar{L}_m^{**})^2(\log p)^7|\delta_{n, m}|^6/m\}^{1/6}+\{M_m^*(\phi_m^{**})/\bar{L}_m^{**}\}\big],
\end{align*}	
for a universal constant $K_2>0$ that depends only on $b$.
\end{lemma}

To introduce Lemma~\ref{theorem:term3}, for any sequence of constants $\delta_{n, m}$ that depends on both $n$ and $m$, denote a useful quantity $\hat{\Delta}_{n, m}=\|\hat{\Sigma}^X-{\Sigma}^X+\delta_{n, m}^2(\hat{\Sigma}^Y-{\Sigma}^Y)\|_{\infty}$.
Lemma~\ref{theorem:term3} below gives an abstract upper bound on $\rho_{n, m}^{MB}$ defined in (\ref{eq:rhoMB}).
\begin{lemma}\label{theorem:term3}
For any sequence of constants $\delta_{n, m}$, assume we have the following condition (a):
\begin{enumerate}
\item[(a)]There exists a universal constant $b>0$ such that $$\min_{1\leq j \leq p}E\{(S_{nj}^X-n^{1/2}\mu_j^X+\delta_{n, m}S_{mj}^Y-\delta_{n, m}m^{1/2}\mu_j^Y)^2\}\geq b.$$
\end{enumerate}	
Then for any sequence of constants $\bar{\Delta}_{n, m}>0$, on the event $\{\hat{\Delta}_{n, m}\leq \bar{\Delta}_{n, m}\}$, we have the following property, where $\rho_{n,m}^{MB}$ is defined in (\ref{eq:rhoMB}),
\begin{align*}
\rho_{n,m}^{MB}\lesssim& (\bar{\Delta}_{n, m})^{1/3}(\log p)^{2/3}.
\end{align*}	
\end{lemma}

Lastly, we present two-sample Borel-Cantelli lemma in Lemma~\ref{lemma:term8}.
\begin{lemma}\label{lemma:term8}
Let $\{A_{n, m}:n\geq1, m\geq 1, (n, m)\in A\}$ be a sequence of events in the sample space $\Omega$, where $A$ is the set of all possible combinations $(n, m)$, which has the form $A=\{(n, m): n\geq 1, m\in \sigma(n)\}$ where $\sigma(n)$ is a set of positive integers determined by $n$, possibly the empty set. Assume the following condition (a):
\begin{enumerate}
\item[(a)]$\sum_{n=1}^{\infty}\sum_{m\in \sigma(n)}P(A_{n, m})<\infty. $
\end{enumerate}	
Then we have the following property:
\begin{align*}
P\big(\bigcap_{k_1=1}^{\infty}\bigcap_{k_2=1}^{\infty}\bigcup_{n=k_1}^{\infty}\bigcup_{m\in \varrho(k_2)\cap\sigma(n)}A_{n, m}\big)=0,
\end{align*}
where $\varrho(k_2)=\{k: k\in \mathbb{Z}, k\geq k_2\}$.
\end{lemma}
Note that if $m\in \sigma(n)=\emptyset$, we just delete the roles of those $A_{n, m}$ and $A_{n, m}^c$ during any operations such as union and intersection, and the same applies to $P(A_{n, m})$ and $P(A_{n, m}^c)$ during summation and deduction.

Before preceding, we mention that the derivations of Theorems~\ref{theorem:term2}--\ref{theorem:term4} essentially follow those of their counterparts in \cite{chern:16:1}, but need more technicality to employ the aforesaid  Lemmas~\ref{theorem:term1}--\ref{theorem:term3} to address the challenge arising from unequal sample sizes. The derivation of Corollary~\ref{lemma:term9} is based on Theorem~\ref{theorem:term2} as well as a two-sample Borel-Cantelli lemma (Lemma~\ref{lemma:term8}) that firstly appears in this work as far as we know.

 Theorems~\ref{theorem:term5}--\ref{theorem:term7} regarding the DCF test are newly developed, while no comparable results are present in literature. Thus we present the proofs of Theorems~\ref{theorem:term5}--\ref{theorem:term7} below, while the proofs of Theorems \ref{theorem:term2}--\ref{theorem:term4}, Corollary~\ref{lemma:term9}, and the auxiliary Lemmas are delegated to an online Supplementary material for space economy.
\vspace{0.15in}


\noindent{\em Proof of Theorem~\ref{theorem:term5}}:\ \
First of all, we define a sequence of constants $\delta_{n, m}$ by
\begin{align}\label{inequality:term118}
\delta_{n, m}=-n^{1/2}m^{-1/2}.
\end{align}
Together with condition (a), it can deduced that
\begin{align}\label{inequality:term119}
\delta_2<|\delta_{n, m}|<\delta_1,
\end{align}
with $\delta_1=\{c_2/(1-c_2)\}^{1/2}>0$ and $\delta_2=\{c_1/(1-c_1)\}^{1/2}>0$. Moreover, by combining (\ref{inequality:term118}), (\ref{inequality:term119}) with condition (b), we have
\begin{align}\label{inequality:term120}
&\min_{1\leq j \leq p}E\{(S_{nj}^X-n^{1/2}\mu_j^X+\delta_{n, m}S_{mj}^Y-\delta_{n, m}m^{1/2}\mu_j^Y)^2\}\geq \min\{1, \delta_2^2\}b.
\end{align}
In addition, based on condition (a) and condition (e), one has
\begin{align}\label{inequality:term121}
&B_{n, m}^2\log^7(pm)/m\sim B_{n, m}^2\log^7(pn)/n\to 0.
\end{align}
To this end, by combining (\ref{inequality:term118}), (\ref{inequality:term119}), (\ref{inequality:term120}), (\ref{inequality:term121}), condition (c), condition (d) with Theorem~\ref{theorem:term2}, it can be shown that
\begin{align}\label{inequality:term122}
&\sup_{t\geq 0}\big|P(\|S_n^X-n^{1/2}m^{-1/2}S_m^Y-n^{1/2}(\mu^X-\mu^Y)\|_{\infty}\leq t)-\notag\\
&P(\|S_n^F-n^{1/2}m^{-1/2}S_m^G-n^{1/2}(\mu^X-\mu^Y)\|_{\infty}\leq t)\big|\notag\\
&\leq\rho_{n,m}^{**}\lesssim \{B_{n, m}^2\log^7(pn)/n\}^{1/6}.
\end{align}	
Next, we denote a sequence of constants $\alpha_{n, m}$ by
\begin{align}\label{inequality:term123}
\alpha_{n, m}=(pn)^{-1},
\end{align}
and it is obvious that
\begin{align}\label{inequality:term124}
\alpha_{n, m}\in (0, e^{-1}).
\end{align}
Moreover, by combining condition (a), condition (e) with (\ref{inequality:term123}), we conclude that
\begin{align}\label{inequality:term125}
B_{n, m}^2\log^5(pm)\log^2(1/\alpha_{n, m})/m\sim B_{n, m}^2\log^5(pn)\log^2(1/\alpha_{n, m})/n\to 0.
\end{align}
To this end, by combining (\ref{inequality:term118}), (\ref{inequality:term119}), (\ref{inequality:term120}), (\ref{inequality:term123}), (\ref{inequality:term124}), (\ref{inequality:term125}), condition (c), condition (d) with Theorem~\ref{theorem:term4},  it follows that there exists a universal constant $c^*>0$ such that with probability at least $1-\gamma_{n, m}$,
we have $\rho_{n,m}^{MB}\lesssim\{ B_{n, m}^2\log^7(pn)/n\}^{1/6}$,
 where
$\gamma_{n, m}=(\alpha_{n, m})^{\log(pn)/3}+3(\alpha_{n, m})^{\log^{1/2}(pn)/c_*}+(\alpha_{n, m})^{\log(pm)/3}+\\3(\alpha_{n, m})^{\log^{1/2}(pm)/c_*}+(\alpha_{n, m})^{\log^3(pn)/6}+3(\alpha_{n, m})^{\log^3(pn)/c_*}+(\alpha_{n, m})^{\log^3(pm)/6}+3(\alpha_{n, m})^{\log^3(pm)/c_*}$.
Together with   (a), (\ref{inequality:term123}) and (\ref{inequality:term124}), it is not hard to prove that
\begin{align}\label{inequality:term128}
\sum_{n}\sum_{m}\gamma_{n, m}<\infty.
\end{align}	
Henceforth, by combining (\ref{inequality:term118}), (\ref{inequality:term119}), (\ref{inequality:term120}), (\ref{inequality:term123}), (\ref{inequality:term124}), (\ref{inequality:term125}), (\ref{inequality:term128}), condition (c), condition (d) with Corollary~\ref{lemma:term9}, we reach a conclusion that with probability one,
\begin{align}\label{inequality:term129}
\sup_{t\geq 0}\big|&P_e(\|S_n^{eX}-n^{1/2}m^{-1/2}S_m^{eY}\|_{\infty}\leq t)-P(\|S_n^F-n^{1/2}m^{-1/2}S_m^G-\notag\\
&n^{1/2}(\mu^X-\mu^Y)\|_{\infty}\leq t)\big|\leq\rho_{n,m}^{MB}\lesssim \{B_{n, m}^2\log^7(pn)/n\}^{1/6}.
\end{align}	
Finally, according to  (\ref{inequality:term122}) and (\ref{inequality:term129}), the assertion holds trivially.
\qed


\vspace{0.15in}

\noindent{\em Proof of Theorem~\ref{theorem:term6}}:\ \
Given any $(\mu^X-\mu^Y)$, we have
\begin{align}\label{inequality:term134}
&Power^*(\mu^X-\mu^Y)\notag\\
=&P_{e^*}\{\|S_n^{e^*X}-n^{1/2}m^{-1/2}S_m^{e^*Y}+n^{1/2}(\mu^X-\mu^Y)\|_{\infty}\geq c_B(\alpha)\}\notag\\
=&1-P_{e^*}\{\|S_n^{e^*X}-n^{1/2}m^{-1/2}S_m^{e^*Y}+n^{1/2}(\mu^X-\mu^Y)\|_{\infty}< c_B(\alpha)\}\notag\\
=&1-P_{e^*}\{-n^{1/2}(\mu^X-\mu^Y)-c_B(\alpha)<S_n^{e^*X}-n^{1/2}m^{-1/2}S_m^{e^*Y}<\notag\\
&-n^{1/2}(\mu^X-\mu^Y)+c_B(\alpha)\}\notag\\
=&1-P_{e^*}\{-n^{1/2}(\mu^X-\mu^Y)-c_B(\alpha)<S_n^{e^*X}-n^{1/2}m^{-1/2}S_m^{e^*Y}<\notag\\
&-n^{1/2}(\mu^X-\mu^Y)+c_B(\alpha)\}\notag\\
&+P\{-n^{1/2}(\mu^X-\mu^Y)-c_B(\alpha)<S_n^X-n^{1/2}m^{-1/2}S_m^Y\notag\\
&-n^{1/2}(\mu^X-\mu^Y)<-n^{1/2}(\mu^X-\mu^Y)+c_B(\alpha)\}\notag\\
&-P\{-n^{1/2}(\mu^X-\mu^Y)-c_B(\alpha)<S_n^X-n^{1/2}m^{-1/2}S_m^Y\notag\\
&-n^{1/2}(\mu^X-\mu^Y)<-n^{1/2}(\mu^X-\mu^Y)+c_B(\alpha)\}\notag\\
\geq&1-\sup_{A\in\mathcal{A}^{Re}}\big|P(\|S_n^X-n^{1/2}m^{-1/2}S_m^Y\notag\\
&-n^{1/2}(\mu^X-\mu^Y)\|_{\infty}\in A)-P_{e^*}(\|S_n^{e^*X}-n^{1/2}m^{-1/2}S_m^{e^*Y}\|_{\infty}\in A)\big|\notag\\
&-P\{\|S_n^X-n^{1/2}m^{-1/2}S_m^Y\|_{\infty}<c_B(\alpha)\}\notag\\
=&Power(\mu^X-\mu^Y)-\notag\\
&\sup_{A\in\mathcal{A}^{Re}}\big|P(\|S_n^X-n^{1/2}m^{-1/2}S_m^Y-n^{1/2}(\mu^X-\mu^Y)\|_{\infty}\in A)-\notag\\
&P_{e^*}(\|S_n^{e^*X}-n^{1/2}m^{-1/2}S_m^{e^*Y}\|_{\infty}\in A)\big|.
\end{align}
Likewise, given any $(\mu^X-\mu^Y)$, we have
\begin{align}\label{inequality:term135}
&Power(\mu^X-\mu^Y)\notag\\
=&P\{\|S_n^X-n^{1/2}m^{-1/2}S_m^Y\|_{\infty}\geq c_B(\alpha)\}\notag\\
=&1-P\{\|S_n^X-n^{1/2}m^{-1/2}S_m^Y\|_{\infty}< c_B(\alpha)\}\notag\\
=&1-P\{-c_B(\alpha)<S_n^X-n^{1/2}m^{-1/2}S_m^Y< c_B(\alpha)\}\notag\\
=&1+P_{e^*}\{-n^{1/2}(\mu^X-\mu^Y)-c_B(\alpha)<S_n^{e^*X}-n^{1/2}m^{-1/2}S_m^{e^*Y}<\notag\\
&-n^{1/2}(\mu^X-\mu^Y)+c_B(\alpha)\}-P\{-n^{1/2}(\mu^X-\mu^Y)-c_B(\alpha)<\notag\\
&S_n^X-n^{1/2}m^{-1/2}S_m^Y-n^{1/2}(\mu^X-\mu^Y)<-n^{1/2}(\mu^X-\mu^Y)+c_B(\alpha)\}\notag\\
&-P_{e^*}\{-n^{1/2}(\mu^X-\mu^Y)-c_B(\alpha)<S_n^{e^*X}-n^{1/2}m^{-1/2}S_m^{e^*Y}\notag\\
&<-n^{1/2}(\mu^X-\mu^Y)+c_B(\alpha)\}\notag\\
\geq&1-\sup_{A\in\mathcal{A}^{Re}}\big|P(\|S_n^X-n^{1/2}m^{-1/2}S_m^Y-n^{1/2}(\mu^X-\mu^Y)\|_{\infty}\in A)\notag\\
&-P_{e^*}(\|S_n^{e^*X}-n^{1/2}m^{-1/2}S_m^{e^*Y}\|_{\infty}\in A)\big|\notag\\
&-P_{e^*}\{\|S_n^{e^*X}-n^{1/2}m^{-1/2}S_m^{e^*Y}+n^{1/2}(\mu^X-\mu^Y)\|_{\infty}<c_B(\alpha)\}\notag\\
=&Power^*(\mu^X-\mu^Y)-\notag\\
&\sup_{A\in\mathcal{A}^{Re}}\big|P(\|S_n^X-n^{1/2}m^{-1/2}S_m^Y-n^{1/2}(\mu^X-\mu^Y)\|_{\infty}\in A)-\notag\\
&P_{e^*}(\|S_n^{e^*X}-n^{1/2}m^{-1/2}S_m^{e^*Y}\|_{\infty}\in A)\big|.
\end{align}
Putting (\ref{inequality:term134}) and (\ref{inequality:term135}) together indicates that
\begin{align}\label{inequality:term136}
&\big|Power^*(\mu^X-\mu^Y)-Power(\mu^X-\mu^Y)\big|\notag\\
\leq&\sup_{A\in\mathcal{A}^{Re}}\big|P(\|S_n^X-n^{1/2}m^{-1/2}S_m^Y-n^{1/2}(\mu^X-\mu^Y)\|_{\infty}\in A)-\notag\\
&P_{e^*}(\|S_n^{e^*X}-n^{1/2}m^{-1/2}S_m^{e^*Y}\|_{\infty}\in A)\big|.
\end{align}
Moreover, by similar argument as in the proof of Theorem~\ref{theorem:term5}, one can show that with probability one,
\begin{align}\label{inequality:term137}
&\sup_{A\in\mathcal{A}^{Re}}\big|P(\|S_n^X-n^{1/2}m^{-1/2}S_m^Y-n^{1/2}(\mu^X-\mu^Y)\|_{\infty}\in A)-\notag\\
&P_{e^*}(\|S_n^{e^*X}-n^{1/2}m^{-1/2}S_m^{e^*Y}\|_{\infty}\in A)\big|\notag\\
\lesssim&\{B_{n, m}^2\log^7(pn)/n\}^{1/6}.
\end{align}
Finally, by combining (\ref{inequality:term136}) with (\ref{inequality:term137}), for any $\mu^X-\mu^Y\in \mathbb{R}^p$, we have that with probability one,
\begin{align*}
\big|Power^*(\mu^X-\mu^Y)-Power(\mu^X-\mu^Y)\big|\lesssim\{B_{n, m}^2\log^7(pn)/n\}^{1/6},
\end{align*}
 which completes the proof.
\qed


\vspace{0.15in}

\noindent{\em Proof of Theorem~\ref{theorem:term7}}:\ \
First of all, on the basis of (\ref{inequality:term133}) and the triangle inequality, it is clear that
\begin{align}\label{inequality:aux1}
Power^*(\mu^X-\mu^Y)\geq P_{e^*}\{&\|S_n^{e^*X}-n^{1/2}m^{-1/2}S_m^{e^*Y}\|_{\infty}\\ \notag
&\leq \|n^{1/2}(\mu^X-\mu^Y)\|_\infty-c_B(\alpha)\}.
\end{align}	
At this point, with some abuse of notation, we denote $\{e_j: j\leq p\}$ as the natural basis for $\mathbb{R}^p$. Then it follows from union bound inequality and concentration inequality that for any $t\geq 0$,
\begin{align}\label{inequality:aux2}
&P_{e^*}\{\|S_n^{e^*X}-n^{1/2}m^{-1/2}S_m^{e^*Y}\|_{\infty}\geq t\}\\ \notag
\leq& \sum_{j=1}^pP_{e^*}\{|S_{nj}^{e^*X}-n^{1/2}m^{-1/2}S_{mj}^{e^*Y}|\geq t\}\\ \notag
\leq& \sum_{j=1}^p2\exp \big[-t^2/\{2e_j'(\hat{\Sigma}^X+nm^{-1}\hat{\Sigma}^Y)e_j\}\big]\\ \notag
\leq& 2p\exp \Big(-t^2/\big[2\max_{j\leq p}\{e_j'(\hat{\Sigma}^X+nm^{-1}\hat{\Sigma}^Y)e_j\}\big]\Big).
\end{align}	
By plugging $t=c_B(\alpha)$ into (\ref{inequality:aux2}), it follows from the definition of $c_B(\alpha)$ that
\begin{align}\label{inequality:aux3}
c_B(\alpha)&\leq \big[2\log(2p/\alpha)\max_{j\leq p}\{e_j'(\hat{\Sigma}^X+nm^{-1}\hat{\Sigma}^Y)e_j\}\big]^{1/2}\\ \notag
&\leq \big[4\log(pn)\max_{j\leq p}\{e_j'(\hat{\Sigma}^X+nm^{-1}\hat{\Sigma}^Y)e_j\}\big]^{1/2},
\end{align}	
for sufficiently large $n$. To bound the quantity $\max_{j\leq p}\{e_j'(\hat{\Sigma}^X+nm^{-1}\hat{\Sigma}^Y)e_j\}$, first notice that
\begin{align}\label{inequality:aux4}
&\max_{j\leq p}\{e_j'(\hat{\Sigma}^X+nm^{-1}\hat{\Sigma}^Y)e_j\}=\|\hat{\Sigma}^X+nm^{-1}\hat{\Sigma}^Y\|_\infty\\ \notag
\leq &\|\hat{\Sigma}^X-{\Sigma}^X+nm^{-1}(\hat{\Sigma}^Y-{\Sigma}^Y)\|_\infty+\|\Sigma^X+nm^{-1}\Sigma^Y\|_\infty
\end{align}	
For the term $\|\hat{\Sigma}^X-{\Sigma}^X+nm^{-1}(\hat{\Sigma}^Y-{\Sigma}^Y)\|_\infty$, inequalities (53) and (54) from the Supplementary Material together with (\ref{inequality:term118}), (\ref{inequality:term123}) and condition (a) entails that there exists a universal constant $c_1>0$ such that
\begin{align}\label{inequality:aux5}
&\|\hat{\Sigma}^X-{\Sigma}^X+nm^{-1}(\hat{\Sigma}^Y-{\Sigma}^Y)\|_\infty\leq c_1 \{B_{n, m}^2\log^3 (pn)/n\}^{1/2},
\end{align}	
with probability tending to one.  Regarding the term $\|\Sigma^X+nm^{-1}\Sigma^Y\|_\infty$, one has
\begin{align}\label{inequality:aux6}
&\|\Sigma^X+nm^{-1}\Sigma^Y\|_\infty\leq \|\Sigma^X\|_\infty+nm^{-1}\|\Sigma^Y\|_\infty\leq \|\Sigma^X\|_\infty+c_2\|\Sigma^Y\|_\infty \\ \notag
=&\max_{1\leq j \leq p}\sum_{i=1}^n E\{(X_{ij}-\mu_j^X)^2\}/n+c_2\max_{1\leq j \leq p}\sum_{i=1}^m E\{(Y_{ij}-\mu_j^Y)^2\}/m\\ \notag
\leq&\max_{1\leq j \leq p}\sum_{i=1}^n \big[E\{(X_{ij}-\mu_j^X)^4\}\big]^{1/2}/n+c_2\max_{1\leq j \leq p}\sum_{i=1}^m \big[E\{(Y_{ij}-\mu_j^Y)^4\}\big]^{1/2}/m \\ \notag
\leq&\big[\max_{1\leq j \leq p}\sum_{i=1}^n E\{(X_{ij}-\mu_j^X)^4\}/n\big]^{1/2}+c_2\big[\max_{1\leq j \leq p}\sum_{i=1}^m E\{(Y_{ij}-\mu_j^Y)^4\}/m\big]^{1/2} \\ \notag
\leq& c_3 B_{n, m},
\end{align}
for some universal constants $c_2, c_3>0$, where the second inequality is by condition (a), the third inequality is based on Jensen's inequality, the fourth inequality holds from cauchy schwarz inequality, and the last inequality follows from condition (c).
To this end, by combining (\ref{inequality:aux5}), (\ref{inequality:aux6}), (e) with (\ref{inequality:aux4}), it can be deduced that there exists a universal constant $c_4>0$ such that
\begin{align}\label{inequality:aux7}
&\max_{j\leq p}\{e_j'(\hat{\Sigma}^X+nm^{-1}\hat{\Sigma}^Y)e_j\}\leq c_4 B_{n, m},
\end{align}	
with probability tending to one. Together with (\ref{inequality:aux3}), it can be verified that
\begin{align}\label{inequality:aux8}
c_B(\alpha)\leq \{4c_4 B_{n, m}\log(pn)\}^{1/2},
\end{align}	
with probability tending to one. Now, we set the constant $K_s$ in (f) as
$K_s=4c_4^{1/2}$, and it then follows from (f) and (\ref{inequality:aux8}) that
\begin{align}\label{inequality:aux9}
\|n^{1/2}(\mu^X-\mu^Y)\|_\infty-c_B(\alpha)\geq \{4c_4 B_{n, m}\log(pn)\}^{1/2},
\end{align}
with probability tending to one.  Hence, it can be deduced that with probability tending to one,
\begin{align*}
&Power^*(\mu^X-\mu^Y)\geq P_{e^*}[\|S_n^{e^*X}-n^{1/2}m^{-1/2}S_m^{e^*Y}\|_{\infty}\leq \{4c_4 B_{n, m}\log(pn)\}^{1/2}]\\
=&1-P_{e^*}[\|S_n^{e^*X}-n^{1/2}m^{-1/2}S_m^{e^*Y}\|_{\infty}\geq \{4c_4 B_{n, m}\log(pn)\}^{1/2}]\\
\geq& 1-2p\exp \Big(-4c_4 B_{n, m}\log(pn)/\big[2\max_{j\leq p}\{e_j'(\hat{\Sigma}^X+nm^{-1}\hat{\Sigma}^Y)e_j\}\big]\Big)\\
\geq& 1-2n^{-2} \to 1 \quad \text{as} \quad n\to \infty,
\end{align*}
where the first inequality is based on (\ref{inequality:aux1}) and (\ref{inequality:aux9}), the second inequality holds from (\ref{inequality:aux2}), and the last inequality is by (\ref{inequality:aux7}). This completes the proof.
\qed

\bibliographystyle{imsart-number}
\bibliography{bib-xue,5-31-09-xue}

\end{document}


\begin{frontmatter}
\title{SUPPLEMENTARY MATERIAL FOR \\ Distribution and correlation free two-sample test of high dimensional means}
\runtitle{Distribution and correlation free two-sample mean test}

\begin{aug}
  \author{\fnms{Kaijie}  \snm{Xue}\corref{}\ead[label=e1]{kaijie@nankai.edu.cn}}
  \and
  \author{\fnms{Fang} \snm{Yao}\ead[label=e2]{fyao@math.pku.edu.cn}}

%

  \affiliation{Nankai University and Peking University}

 \address{K. Xue \\ School of Statistics and Data Science \\ Nankai University \\ Tianjin, China\\
\printead{e1}}

\address{F. Yao \\  Corresponding author \\ Department of Probability and Statistics \\ Center for Statistical Science \\ School of Mathematical Sciences \\ Peking University\\ Beijing, China \\
\printead{e2}}



\end{aug}

\end{frontmatter}

\section{Notations}\label{sec:notation}
A brief review of some relevant  notations is given in the following before  the proofs.
For two vectors $x=(x_1,\dots,x_p)\in \mathbb{R}^p$ and $y=(y_1,\dots,y_p)'\in \mathbb{R}^p$, write $x\leq y$ if $x_j\leq y_j$ for all $j=1,\dots,p$. For any $x=(x_1,\dots,x_p)'\in \mathbb{R}^p$ and $a\in \mathbb{R}$,  denote $x+a=(x_1+a,\dots,x_p+a)'$. For any $a, b\in \mathbb{R}$, use the notations $a\vee b=\max\{a, b\}$ and $a\wedge b=\min\{a, b\}$. For any two sequences of constants $a_n$ and $b_n$, write $a_n\lesssim b_n$ if $a_n \leq C b_n$ up to a universal constant $C>0$, and $a_n\sim b_n$ if $a_n \lesssim b_n$ and $b_n \lesssim a_n$.  For any matrix $A=(a_{ij})$, define $\|A\|_{\infty}=\max_{i, j}|a_{ij}|$. For any function $f:\mathbb{R}\to\mathbb{R}$, write $\|f\|_{\infty}=\sup_{z\in \mathbb{R}}|f(z)|$. For a smooth function $g:\mathbb{R}^p\to\mathbb{R}$, we adopt indices to represent the partial derivatives for brevity, for example, $\partial_j\partial_k\partial_l g=g_{jkl}$.
For any $\alpha>0$, define the function $\psi_{\alpha}(x)=\exp(x^\alpha)-1$ for $x\in [0, \infty)$, then for any random variable $X$, define
\begin{align}\label{inequality:term1}
\|X\|_{\psi_{\alpha}}=\inf\{\lambda>0: E\{\psi_{\alpha}(|X|/\lambda)\}\leq 1\},
\end{align}
which is an Orlicz norm for $\alpha\in [1, \infty)$ and a quasi-norm for $\alpha\in (0, 1)$.
For any $\beta>0$ and $y\in \mathbb{R}^p$, we define a function $F_\beta(w)$ as
\begin{align}\label{inequality:term2}
F_\beta(w)=\beta^{-1}\log\big[\sum_{j=1}^p\exp\{\beta(w_j-y_j)\}\big], w\in \mathbb{R}^p,
\end{align}
which satisfies the property
\begin{align}\label{inequality:term3}
0\leq F_\beta(w)-\max_{1\leq j \leq p}(w_j-y_j)\leq \beta^{-1}\log p,
\end{align}
for every $w\in \mathbb{R}^p$ by $(1)$ in \cite{chern:14:1}.
In addition, we let $\varphi_0:\mathbb{R}\to [0, 1]$ be a real valued function such that $\varphi_0$ is thrice continuously differentiable and $\varphi_0(z)=1$ for $z\leq 0$ and $\varphi_0(z)=0$ for $z\geq 1$. For any $\phi\geq 1$, we define a function $\varphi(z)=\varphi_0(\phi z), z\in \mathbb{R}$. Then, for any $\phi\geq 1$ and $y\in \mathbb{R}^p$, we denote $\beta=\phi\log p$ and define a function $\kappa:\mathbb{R}^p\to [0, 1]$ as
 \begin{align}\label{inequality:term4}
\kappa(w)=\varphi_0(\phi F_{\phi\log p}(w))=\varphi(F_{\beta}(w)),  w\in \mathbb{R}^p.
\end{align}

\section{Proofs of Theorems 1--2, and Corollary~1}\label{sec:intro}

\noindent{\em Proof of Theorem~1}:\ \
First, we set $\bar{L}_n^*=\bar{L}_m^{**}=B_{n, m}$. Together with condition (c), it is obvious that
\begin{align}\label{inequality:term60}
\bar{L}_n^*=B_{n, m}\geq L_n^X, \quad \bar{L}_m^{**}=B_{n, m}\geq L_m^Y.
\end{align}	
By combining the definition of the Orlicz norm in (\ref{inequality:term1}) with condition (d), we have
\begin{align}\label{inequality:term61}
\max_{1\leq i \leq n}\max_{1\leq j \leq p}\|X_{ij}-\mu_j^X\|_{\psi_{1}}\leq B_{n, m},  \max_{1\leq i \leq m}\max_{1\leq j \leq p}\|Y_{ij}-\mu_j^Y\|_{\psi_{1}}\leq B_{n, m}.
\end{align}	
Moreover, since $F_{ij}\sim N(\mu_j^X, E(X_{ij}-\mu_j^X)^2)$ for all $i=1,\dots,n$ and $j=1,\dots,p$, it can be deduced that for any $t\geq 0$,
\begin{align}\label{inequality:term62}
P(|F_{ij}-\mu_j^X|>t)\leq 2\exp\big[-t^2/\{2E(X_{ij}-\mu_j^X)^2\}\big],
\end{align}	
for all $i=1,\dots,n$ and $j=1,\dots,p$. Therefore, by combining (\ref{inequality:term62}) with Lemma 2.2.1 in \cite{vaar:96}, we have
\begin{align}\label{inequality:term63}
\|F_{ij}-\mu_j^X\|_{\psi_{2}}\leq \big[(1+2)\{2E(X_{ij}-\mu_j^X)^2\}\big]^{1/2},
\end{align}
for all $i=1,\dots,n$ and $j=1,\dots,p$.	
 Furthermore, by property of the Orlicz norm, it can be shown that for all $i=1,\dots,n$ and $j=1,\dots,p$,
\begin{align}\label{inequality:term64}
\|F_{ij}-\mu_j^X\|_{\psi_{1}}\leq (\log 2)^{2^{-1}-1}\|F_{ij}-\mu_j^X\|_{\psi_{2}}=(\log 2)^{-2^{-1}}\|F_{ij}-\mu_j^X\|_{\psi_{2}}.
\end{align}	
In addition, for all $i=1,\dots,n$ and $j=1,\dots,p$, it can be justified that
\begin{align}\label{inequality:term65}
E(X_{ij}-\mu_j^X)^2\leq 2\|X_{ij}-\mu_j^X\|_{\psi_{1}}^2.
\end{align}	
Henceforth, by combining (\ref{inequality:term61}), (\ref{inequality:term63}), (\ref{inequality:term64}) with (\ref{inequality:term65}), we have
\begin{align}\label{inequality:term66}
\max_{1\leq i \leq n}\max_{1\leq j \leq p}\|F_{ij}-\mu_j^X\|_{\psi_{1}}\leq c_2B_{n, m},
\end{align}	
for a universal constant $c_2>0$. Similar reasoning  leads to
\begin{align}\label{inequality:term67}
\max_{1\leq i \leq m}\max_{1\leq j \leq p}\|G_{ij}-\mu_j^Y\|_{\psi_{1}}\leq c_3B_{n, m},
\end{align}	
for a universal constant $c_3>0$. As a result, by combining (\ref{inequality:term61}), (\ref{inequality:term66}), (\ref{inequality:term67}) with  Lemma 2.2.2 in \cite{vaar:96}, we conclude that
\begin{align}\label{inequality:term68}
&\|\max_{1\leq j \leq p}|X_{ij}-\mu_j^X|\|_{\psi_{1}}\leq c_4B_{n, m}\log p,\quad i=1,\dots,n, \notag\\
&\|\max_{1\leq j \leq p}|F_{ij}-\mu_j^X|\|_{\psi_{1}}\leq c_5B_{n, m}\log p,\quad i=1,\dots,n,\notag\\
&\|\max_{1\leq j \leq p}|Y_{ij}-\mu_j^Y|\|_{\psi_{1}}\leq c_6B_{n, m}\log p,\quad i=1,\dots,m,\notag\\
&\|\max_{1\leq j \leq p}|G_{ij}-\mu_j^Y|\|_{\psi_{1}}\leq c_7B_{n, m}\log p,\quad i=1,\dots,m,
\end{align}	
where $c_4$, $c_5$, $c_6$ and $c_7$ are some positive universal constants. It then follows from (\ref{inequality:term68}) and  Markov's inequality that for any $t\geq 0$,
\begin{align}\label{inequality:term69}
&P(\max_{1\leq j \leq p}|X_{ij}-\mu_j^X|>t)\leq 2\exp\{-t/ (c_4B_{n, m}\log p)\},\quad i=1,\dots,n, \notag\\
&P(\max_{1\leq j \leq p}|F_{ij}-\mu_j^X|>t)\leq 2\exp\{-t/ (c_5B_{n, m}\log p)\},\quad i=1,\dots,n, \notag\\
&P(\max_{1\leq j \leq p}|Y_{ij}-\mu_j^Y|>t)\leq 2\exp\{-t/ (c_6B_{n, m}\log p)\},\quad i=1,\dots,m, \notag\\
&P(\max_{1\leq j \leq p}|G_{ij}-\mu_j^Y|>t)\leq 2\exp\{-t/ (c_7B_{n, m}\log p)\},\quad i=1,\dots,m.
\end{align}
Putting condition (b) and Lemma~3 together yields that for any $\phi_1, \phi_2\geq 1$,
\begin{align*}
\rho_{n,m}^*\leq &K^*\Big[n^{-1/2}\phi_1^2(\log p)^2\big\{\phi_1 L_n^X \rho_{n,m}^*+L_n^X(\log p)^{1/2}+\phi_1 M_n(\phi_1)\big\}+\\
&m^{-1/2}\phi_2^2(\log p)^2|\delta_{n, m}|^3\big\{\phi_2 L_m^Y \rho_{n,m}^*+L_m^Y(\log p)^{1/2}+\phi_2 M_m^*(\phi_2)\big\}+\\
&(\min\{\phi_1, \phi_2\})^{-1}(\log p)^{1/2}\Big],
\end{align*}	
for a universal constant $K^*>0$ that depends only on $b$, where $\rho_{n, m}^*$ is defined in (11) from the main article.
Accordingly, we set a universal constant
\begin{align}\label{inequality:term70}
K_1=\min\{8^{-1}c_4^{-1}c_1^{-1/3}, 8^{-1}c_5^{-1}c_1^{-1/3}, 8^{-1}c_6^{-1}c_1^{-1/3}, 8^{-1}c_7^{-1}c_1^{-1/3}, (K^*\vee 2)^{-1}\},
\end{align}
and it is  trivial to check that
\begin{align}\label{inequality:term71}
K_1\in (0, (K^*\vee 2)^{-1}].
\end{align}
Then, we let
\begin{align}\label{inequality:term72}
\phi_n^*=&K_1\{(B_{n, m})^2(\log p)^4/n\}^{-1/6}, \\ \notag
\phi_m^{**}=&K_1\{(B_{n, m})^2(\log p)^4|\delta_{n, m}|^6/m\}^{-1/6}.
\end{align}
Together with conditions (a) and (e), it can be verified  that
\begin{align}\label{inequality:term73}
\phi_n^*\geq 2, \quad \phi_m^{**}\geq 2.
\end{align}
To this end, by combining (\ref{inequality:term69}), (\ref{inequality:term72}) with Lemma C.1 in \cite{chern:16:1}, we reach a conclusion that
\begin{align}\label{inequality:term74}
M_n^X(\phi_n^*)\lesssim& \big\{n^{1/2}/(\phi_n^*\log p)+B_{n, m}\log p\big\}^3\exp\big[-n^{1/2}/\{4c_4\phi_n^*B_{n, m}(\log p)^2\}\big], \notag\\
M_n^F(\phi_n^*)\lesssim& \big\{n^{1/2}/(\phi_n^*\log p)+B_{n, m}\log p\big\}^3\exp\big[-n^{1/2}/\{4c_5\phi_n^*B_{n, m}(\log p)^2\}\big], \notag\\
M_m^Y(\phi_m^{**})\lesssim& \big\{m^{1/2}/(|\delta_{n, m}|\phi_m^{**}\log p)+B_{n, m}\log p\big\}^3\notag\\
&\exp\big[-m^{1/2}/\{4c_6|\delta_{n, m}|\phi_m^{**}B_{n, m}(\log p)^2\}\big], \notag\\
M_m^G(\phi_m^{**})\lesssim& \big\{m^{1/2}/(|\delta_{n, m}|\phi_m^{**}\log p)+B_{n, m}\log p\big\}^3\notag\\
&\exp\big[-m^{1/2}/\{4c_7|\delta_{n, m}|\phi_m^{**}B_{n, m}(\log p)^2\}\big].
\end{align}
To bound the term $M_n^X(\phi_n^*)$, it can be noted that
\begin{align}\label{inequality:term75}
&n^{1/2}/\{4c_4\phi_n^*B_{n, m}(\log p)^2\}\notag \\
\geq&(4c_4K_1)^{-1}\big[(B_{n, m})^{2}\{\log (pn)\}^{7}/n\big]^{-1/3}\log(pn)\notag \\
\geq&(4c_4K_1)^{-1}c_1^{-1/3}\log(pn)=K_1^{-1}4^{-1}c_4^{-1}c_1^{-1/3}\log(pn)\notag \\
\geq&(8^{-1}c_4^{-1}c_1^{-1/3})^{-1}4^{-1}c_4^{-1}c_1^{-1/3}\log(pn)=2\log(pn),
\end{align}
where the second inequality follows from condition (e), and the last inequality is by (\ref{inequality:term70}). Moreover, we  have that
\begin{align}\label{inequality:term76}
n^{1/2}/(\phi_n^*\log p)\leq n^{1/2}/(2\log p)\lesssim n^{1/2},
\end{align}
where the first inequality holds true due to (\ref{inequality:term73}). It can also be checked that
\begin{align}\label{inequality:term77}
B_{n, m}\log p= \{(B_{n, m})^2(\log p)^2/n\}^{1/2}n^{1/2}\lesssim n^{1/2},
\end{align}
where the last inequality is valid according to condition (e). At this point, by combining (\ref{inequality:term74}), (\ref{inequality:term75}), (\ref{inequality:term76}) with (\ref{inequality:term77}), it can be obtained that
\begin{align}\label{inequality:term78}
M_n^X(\phi_n^*)\lesssim n^{3/2}(pn)^{-2}\lesssim n^{-1/2}.
\end{align}
Similar argument also gives
\begin{align}\label{inequality:term79}
M_n^F(\phi_n^*)\lesssim  n^{-1/2}.
\end{align}
Piecing (\ref{inequality:term78}) and (\ref{inequality:term79}) together leads to
\begin{align}\label{inequality:term80}
M_n(\phi_n^*)= M_n^X(\phi_n^*)+M_n^F(\phi_n^*)\lesssim n^{-1/2}.
\end{align}
To bound the term $M_m^Y(\phi_m^{**})$, notice that
\begin{align}\label{inequality:term81}
&m^{1/2}/\{4c_6|\delta_{n, m}|\phi_m^{**}B_{n, m}(\log p)^2\}\notag \\
\geq&(4c_6K_1)^{-1}\big[(B_{n, m})^{2}\{\log (pm)\}^{7}/m\big]^{-1/3}\log(pm)\notag \\
\geq&(4c_6K_1)^{-1}c_1^{-1/3}\log(pm)=K_1^{-1}4^{-1}c_6^{-1}c_1^{-1/3}\log(pm)\notag \\
\geq&(8^{-1}c_6^{-1}c_1^{-1/3})^{-1}4^{-1}c_6^{-1}c_1^{-1/3}\log(pm)=2\log(pm),
\end{align}
where the second inequality is based on  condition (e), and the last inequality is owing to (\ref{inequality:term70}). Also notice that
\begin{align}\label{inequality:term82}
m^{1/2}/(|\delta_{n, m}|\phi_m^{**}\log p)\lesssim m^{1/2}/(\phi_m^{**}\log p)\lesssim m^{1/2}/(2\log p)\lesssim m^{1/2},
\end{align}
where the first inequality follows from condition (a), and the second inequality is based on (\ref{inequality:term73}). In addition, one has
\begin{align}\label{inequality:term83}
B_{n, m}\log p= \{(B_{n, m})^2(\log p)^2/m\}^{1/2}m^{1/2}\lesssim m^{1/2},
\end{align}
where the last inequality follows from condition (e). Therefore, by combining (\ref{inequality:term74}), (\ref{inequality:term81}), (\ref{inequality:term82}) with (\ref{inequality:term83}), we have
\begin{align}\label{inequality:term84}
M_m^Y(\phi_m^{**})\lesssim m^{3/2}(pm)^{-2}\lesssim m^{-1/2}.
\end{align}
Similar argument yields that
\begin{align}\label{inequality:term85}
M_m^G(\phi_m^{**})\lesssim  m^{-1/2}.
\end{align}
Under (\ref{inequality:term84}) and (\ref{inequality:term85}), we have
\begin{align}\label{inequality:term86}
M_m^*(\phi_m^{**})= M_m^Y(\phi_m^{**})+M_m^G(\phi_m^{**})\lesssim m^{-1/2}.
\end{align}
To this end, by combining condition (a), condition (b), (\ref{inequality:term60}), (\ref{inequality:term71}), (\ref{inequality:term72}), (\ref{inequality:term73}), (\ref{inequality:term80}), (\ref{inequality:term86}) with Lemma~4, we conclude that
\begin{align}\label{inequality:term87}
\rho_{n,m}^{**}\lesssim& \{(B_{n, m})^2(\log p)^7/n\}^{1/6}+n^{-1/2}(B_{n, m})^{-1}+\notag\\
&\{(B_{n, m})^2(\log p)^7|\delta_{n, m}|^6/m\}^{1/6}+m^{-1/2}(B_{n, m})^{-1}\notag\\
\lesssim& \{(B_{n, m})^2(\log p)^7/n\}^{1/6}+n^{-1/2}(B_{n, m})^{-1}+\{(B_{n, m})^2(\log p)^7/m\}^{1/6} \notag\\
&+m^{-1/2}(B_{n, m})^{-1}.
\end{align}
For the term $n^{-1/2}(B_{n, m})^{-1}$ in (\ref{inequality:term87}), it can be shown that
\begin{align}\label{inequality:term88}
n^{-1/2}(B_{n, m})^{-1}&=\big[(B_{n, m})^2\{\log (pn)\}^7/n\big]^{1/6}(B_{n, m})^{-4/3}\{\log (pn)\}^{-7/6}n^{-1/3}\notag\\
&\lesssim \big[(B_{n, m})^2\{\log (pn)\}^7/n\big]^{1/6}.
\end{align}
Likewise, we have
\begin{align}\label{inequality:term89}
m^{-1/2}(B_{n, m})^{-1}\lesssim \big[(B_{n, m})^2\{\log (pm)\}^7/m\big]^{1/6}.
\end{align}
Finally, based on (\ref{inequality:term87}), (\ref{inequality:term88}) and (\ref{inequality:term89}), one has
\begin{align*}
\rho_{n,m}^{**}\lesssim \big[(B_{n, m})^2\{\log (pn)\}^7/n\big]^{1/6}+\big[(B_{n, m})^2\{\log (pm)\}^7/m\big]^{1/6},
\end{align*}
which completes the proof.
\qed

\noindent{\em Proof of Theorem~2}:\ \
First, by combining  (b) with Lemma~5, it is clear that for any sequence of constants $\bar{\Delta}_{n, m}>0$, on the event $\{\hat{\Delta}_{n, m}\leq \bar{\Delta}_{n, m}\}$, we have
\begin{align}\label{inequality:term95}
\rho_{n,m}^{MB}\lesssim& (\bar{\Delta}_{n, m})^{1/3}(\log p)^{2/3}.
\end{align}
Then, our goal is to find a proper bound for the quantity $\hat{\Delta}_{n, m}=\|\hat{\Sigma}^X-{\Sigma}^X+\delta_{n, m}^2(\hat{\Sigma}^Y-{\Sigma}^Y)\|_{\infty}$. To achieve this, we  first note that
\begin{align}\label{inequality:term96}
\hat{\Delta}_{n, m}\leq \hat{\Delta}_{n, m}^{(1)}+(\hat{\Delta}_{n, m}^{(2)})^2+\delta_{n, m}^2\hat{\Delta}_{n, m}^{(3)}+\delta_{n, m}^2(\hat{\Delta}_{n, m}^{(4)})^2,
\end{align}	
where
\begin{align*}
\hat{\Delta}_{n, m}^{(1)}=&\max_{1\leq j, k \leq p}|n^{-1}\sum_{i=1}^n\{(X_{ij}-\mu_j^X)(X_{ik}-\mu_k^X)-E(X_{ij}-\mu_j^X)(X_{ik}-\mu_k^X)\}|,\\
\hat{\Delta}_{n, m}^{(2)}=&\|(\bar{X}-\mu^X)(\bar{X}-\mu^X)'\|_{\infty}^{1/2}=\|\bar{X}-\mu^X\|_{\infty}=\max_{1\leq j \leq p}|n^{-1}\sum_{i=1}^n({X}_{ij}-\mu_j^X)|,\\
\hat{\Delta}_{n, m}^{(3)}=&\max_{1\leq j, k \leq p}|m^{-1}\sum_{i=1}^m\{(Y_{ij}-\mu_j^Y)(Y_{ik}-\mu_k^Y)-E(Y_{ij}-\mu_j^Y)(Y_{ik}-\mu_k^Y)\}|,\\
\hat{\Delta}_{n, m}^{(4)}=&\|(\bar{Y}-\mu^Y)(\bar{Y}-\mu^Y)'\|_{\infty}^{1/2}=\|\bar{Y}-\mu^Y\|_{\infty}=\max_{1\leq j \leq p}|m^{-1}\sum_{i=1}^m({Y}_{ij}-\mu_j^Y)|.
\end{align*}
To bound the term $\hat{\Delta}_{n, m}^{(1)}$, we first denote $$\sigma_{n, m, 1}^2=\max_{1\leq j, k \leq p}\sum_{i=1}^nE\{(X_{ij}-\mu_j^X)(X_{ik}-\mu_k^X)-E(X_{ij}-\mu_j^X)(X_{ik}-\mu_k^X)\}^2,$$ and it can be shown  that
\begin{align}\label{inequality:term97}
\sigma_{n, m, 1}^2\leq& \max_{1\leq j, k \leq p}\sum_{i=1}^nE\{(X_{ij}-\mu_j^X)^2(X_{ik}-\mu_k^X)^2\}\notag \\
\leq& \max_{1\leq j, k \leq p}\big\{\sum_{i=1}^nE(X_{ij}-\mu_j^X)^4\big\}^{1/2}\big\{\sum_{i=1}^nE(X_{ik}-\mu_k^X)^4\big\}^{1/2}\notag \\
=& \max_{1\leq j \leq p}\sum_{i=1}^nE(X_{ij}-\mu_j^X)^4\leq n (B_{n, m})^2,
\end{align}
where the last inequality follows from condition (c). Moreover, we also denote
$$M_{n, m, 1}=\max_{1\leq i \leq n}\max_{1\leq j \leq p}\max_{1\leq k \leq p}|(X_{ij}-\mu_j^X)(X_{ik}-\mu_k^X)-E(X_{ij}-\mu_j^X)(X_{ik}-\mu_k^X)|,$$ and one can show that
\begin{align}\label{inequality:term98}
\|M_{n, m, 1}\|_{\psi_{1/2}}
\leq& \|\max_{1\leq i \leq n}\max_{1\leq j, k \leq p}|(X_{ij}-\mu_j^X)(X_{ik}-\mu_k^X)|+\notag \\
&\max_{1\leq i \leq n}\max_{1\leq j , k\leq p}E\{|(X_{ij}-\mu_j^X)(X_{ik}-\mu_k^X)|\}\|_{\psi_{1/2}}\notag \\
\leq& c_1\|\max_{1\leq i \leq n}\max_{1\leq j , k\leq p}|(X_{ij}-\mu_j^X)(X_{ik}-\mu_k^X)|\|_{\psi_{1/2}}+\notag \\
&c_1\|\max_{1\leq i \leq n}\max_{1\leq j , k\leq p}E\{|(X_{ij}-\mu_j^X)(X_{ik}-\mu_k^X)|\}\|_{\psi_{1/2}}\notag \\
\leq& c_1\|\max_{1\leq i \leq n}\max_{1\leq j , k\leq p}|(X_{ij}-\mu_j^X)(X_{ik}-\mu_k^X)|\|_{\psi_{1/2}}+\notag \\
&c_2\max_{1\leq i \leq n}\max_{1\leq j , k\leq p}E\{|(X_{ij}-\mu_j^X)(X_{ik}-\mu_k^X)|\}\notag \\
=& c_1\|\max_{1\leq i \leq n}\max_{1\leq j \leq p}|X_{ij}-\mu_j^X|\|_{\psi_{1}}^2+\notag \\
&c_2\max_{1\leq i \leq n}\max_{1\leq j , k\leq p}E\{|(X_{ij}-\mu_j^X)(X_{ik}-\mu_k^X)|\}\notag \\
\leq& c_3\big\{\log(pn)\max_{1\leq i \leq n}\max_{1\leq j \leq p}\|X_{ij}-\mu_j^X\|_{\psi_{1}}\big\}^2+\notag \\
&c_2\max_{1\leq i \leq n}\max_{1\leq j , k\leq p}E\{|(X_{ij}-\mu_j^X)(X_{ik}-\mu_k^X)|\}\notag \\
\leq& c_3\{\log(pn)\}^2(B_{n, m})^2+c_2\max_{1\leq i \leq n}\max_{1\leq j \leq p}E\{(X_{ij}-\mu_j^X)^2\}\notag \\
\leq& c_3\{\log(pn)\}^2(B_{n, m})^2+2c_2\max_{1\leq i \leq n}\max_{1\leq j \leq p}\|X_{ij}-\mu_j^X\|_{\psi_{1}}^2\notag \\
\leq& c_3\{\log(pn)\}^2(B_{n, m})^2+2c_2(B_{n, m})^2\leq  c_4\{\log(pn)\}^2(B_{n, m})^2,
\end{align}
where $c_1$, $c_2$, $c_3$ and $c_4$ are some positive universal constants, the second inequality is based on the fact that $\|\cdot\|_{\psi_{1/2}}$ is a quasi-norm, the fourth inequality follows from Lemma 2.2.2 in \cite{vaar:96}, and the fifth inequality is based on condition (d).  Furthermore, we have
\begin{align}\label{inequality:term99}
\{E(M_{n, m, 1}^2)\}^{1/2}\leq (4!\|M_{n, m, 1}\|_{\psi_{1/2}}^2)^{1/2}\leq c_5\{\log(pn)\}^2(B_{n, m})^2,
\end{align}
where $c_5>0$ is a universal constant, and the last inequality is based on (\ref{inequality:term98}). To this end, by combining (\ref{inequality:term97}), (\ref{inequality:term99}) with Lemma E.1 in \cite{chern:16:1}, we have
\begin{align}\label{inequality:term100}
E(\hat{\Delta}_{n, m}^{(1)})\leq& c_6 n^{-1}\big[(\sigma_{n, m, 1}^2\log p)^{1/2}+\{E(M_{n, m, 1}^2)\}^{1/2}\log p\big]\notag \\
\leq& c_6 n^{-1}\big[(n B_{n, m}^2\log p)^{1/2}+c_5\{\log(pn)\}^2(B_{n, m})^2\log p\big]\notag \\
\leq& c_6 n^{-1}\big[\{n B_{n, m}^2\log(pn)\}^{1/2}+c_5\{\log(pn)\}^{3}(B_{n, m})^2\big]\notag \\
\leq& c_6(1+c_5) \{n^{-1} B_{n, m}^2\log(pn)\}^{1/2}\leq c_7 \{n^{-1} B_{n, m}^2\log(pn)\}^{1/2},
\end{align}
where $c_6$ and $c_7$ are some positive universal constants, and the fourth inequality is based on condition (e). Therefore, by combining (\ref{inequality:term97}), (\ref{inequality:term98}), (\ref{inequality:term100}) with Lemma E.2 in \cite{chern:16:1}, it can be deduced that  for any $t>0$,
\begin{align}\label{inequality:term101}
&P\big[\hat{\Delta}_{n, m}^{(1)}>2c_7 \{n^{-1} B_{n, m}^2\log(pn)\}^{1/2}+t\big]\notag \\ \leq& P\big\{\hat{\Delta}_{n, m}^{(1)}\geq 2E(\hat{\Delta}_{n, m}^{(1)})+t\big\}\notag \\
\leq&\exp\big\{-n^2t^2/(3\sigma_{n, m, 1}^2)\big\}+3\exp\big[-\big\{nt/(c_8\|M_{n, m, 1}\|_{\psi_{1/2}})\big\}^{1/2}\big]\notag \\
\leq&\exp\big\{-nt^2/(3B_{n, m}^2)\big\}+3\exp\big[-\big\{nt/(c_8\|M_{n, m, 1}\|_{\psi_{1/2}})\big\}^{1/2}\big]\notag \\
\leq&\exp\big\{-nt^2/(3B_{n, m}^2)\big\}+3\exp\big[-(nt)^{1/2}/\{c_9B_{n, m}\log(pn)\}\big],
\end{align}
where $c_8$ and $c_9$ are some positive universal constants, the first inequality is by (\ref{inequality:term100}), the second inequality is based on Lemma E.2 in \cite{chern:16:1}, the third inequality follows from (\ref{inequality:term97}), and the fourth inequality is due to (\ref{inequality:term98}). By substituting $t=\{n^{-1} B_{n, m}^2\log(pn)\}^{1/2}\log(1/\alpha_{n, m})$ into (\ref{inequality:term101}), it can be obtained that
\begin{align*}
&P\big[\hat{\Delta}_{n, m}^{(1)}>2c_7 \{n^{-1} B_{n, m}^2\log(pn)\}^{1/2}+\{n^{-1} B_{n, m}^2\log(pn)\}^{1/2}\log(1/\alpha_{n, m})\big]\notag \\
\leq&\exp\big\{-\log(pn)\log^2(1/\alpha_{n, m})/3\big\}+\notag \\
&3\exp\big[-\{B_{n, m}^2\log^3(pn)\log^{-2}(1/\alpha_{n, m})/n\}^{-1/4}/c_9\big]\notag \\
\leq&(\alpha_{n, m})^{\log(pn)/3}+3(\alpha_{n, m})^{\log^{1/2}(pn)/c_9}.
\end{align*}
Together with the fact that
\begin{align*}
&2c_7 \{n^{-1} B_{n, m}^2\log(pn)\}^{1/2}+\{n^{-1} B_{n, m}^2\log(pn)\}^{1/2}\log(1/\alpha_{n, m})\\
\leq& c_{10}\{n^{-1} B_{n, m}^2\log(pn)\}^{1/2}\log(1/\alpha_{n, m})
\end{align*}
for a universal constant $c_{10}>0$, we conclude that
\begin{align}\label{inequality:term102}
&P\big[\hat{\Delta}_{n, m}^{(1)}>c_{10}\{B_{n, m}^2\log(pn)\log^2(1/\alpha_{n, m})/n\}^{1/2}\big]\\ \notag
\leq& (\alpha_{n, m})^{\log(pn)/3}+3(\alpha_{n, m})^{\log^{1/2}(pn)/c_9}.
\end{align}
Similar reasoning leads to
\begin{align}\label{inequality:term103}
&P\big[\hat{\Delta}_{n, m}^{(3)}>c_{11}\{B_{n, m}^2\log(pm)\log^2(1/\alpha_{n, m})/m\}^{1/2}\big]\\ \notag
\leq& (\alpha_{n, m})^{\log(pm)/3}+3(\alpha_{n, m})^{\log^{1/2}(pm)/c_{12}},
\end{align}
for some universal constants $c_{11}, c_{12}>0$.
Next, to bound the term $\hat{\Delta}_{n, m}^{(2)}$, we first denote
$\sigma_{n, m, 2}^2=\max_{1\leq j \leq p}\sum_{i=1}^nE\{(X_{ij}-\mu_j^X)^2\}$, and one can deduce that
\begin{align}\label{inequality:term104}
\sigma_{n, m, 2}^2\leq& 2\max_{1\leq j \leq p}\sum_{i=1}^n\|X_{ij}-\mu_j^X\|_{\psi_{1}}^2\leq 2nB_{n, m}^2,
\end{align}
where the last inequality is by condition (d). In addition, we also denote
$M_{n, m, 2}=\max_{1\leq i \leq n}\max_{1\leq j \leq p}|X_{ij}-\mu_j^X|$, and one can show  that
\begin{align}\label{inequality:term105}
\|M_{n, m, 2}\|_{\psi_{1}}\leq c_{13}\log(pn)\max_{1\leq i \leq n}\max_{1\leq j \leq p}\|X_{ij}-\mu_j^X\|_{\psi_{1}}\leq c_{13}\log(pn)B_{n, m},
\end{align}
for a universal constant $c_{13}>0$, which further implies that
\begin{align}\label{inequality:term106}
\{E(M_{n, m, 2}^2)\}^{1/2}\leq (2!\|M_{n, m, 2}\|_{\psi_{1}}^2)^{1/2}\leq c_{14}\log(pn)B_{n, m},
\end{align}
where $c_{14}>0$ is a universal constant. Hence, by combining (\ref{inequality:term104}), (\ref{inequality:term106}) with Lemma E.1 in \cite{chern:16:1}, we have
\begin{align}\label{inequality:term107}
E(\hat{\Delta}_{n, m}^{(2)})\leq& c_{15}n^{-1}\big[(\sigma_{n, m, 2}^2\log p)^{1/2}+\{E(M_{n, m, 2}^2)\}^{1/2}\log p\big]\notag\\
\leq& c_{15}n^{-1}\big[\{2nB_{n, m}^2\log (pn)\}^{1/2}+c_{14}B_{n, m}\log^2 (pn)\big]\notag\\
\leq& c_{15}\{B_{n, m}^2\log (pn)/n\}^{1/2}\big[2^{1/2}+c_{14}\{\log^3 (pn)/n\}^{1/2}\big]\notag\\
\leq& c_{16}\{B_{n, m}^2\log (pn)/n\}^{1/2},
\end{align}
where $c_{15}$ and $c_{16}$ are positive universal constants.
Then, by combining (\ref{inequality:term104}), (\ref{inequality:term105}), (\ref{inequality:term107}) with Lemma E.2 in \cite{chern:16:1}, we have
that for any $t>0$,
\begin{align}\label{inequality:term108}
&P\big[\hat{\Delta}_{n, m}^{(2)}>2c_{16} \{B_{n, m}^2\log(pn)/n\}^{1/2}+t\big]\leq P\big\{\hat{\Delta}_{n, m}^{(2)}\geq 2E(\hat{\Delta}_{n, m}^{(2)})+t\big\}\notag \\
&\leq\exp\big\{-n^2t^2/(3\sigma_{n, m, 2}^2)\big\}+3\exp\big\{-nt/(c_{17}\|M_{n, m, 2}\|_{\psi_{1}})\big\}\notag \\
&\leq\exp\big\{-nt^2/(6B_{n, m}^2)\big\}+3\exp\big[-nt/\{c_{18}B_{n, m}\log(pn)\}\big],
\end{align}
where $c_{17}$ and $c_{18}$ are some positive universal constants.
Likewise, we substitute $t=\{ B_{n, m}^2\log(pn)\log^2(1/\alpha_{n, m})/n\}^{1/4}$ into (\ref{inequality:term108}) to obtain
\begin{align}\label{inequality:term109}
&P\big[\hat{\Delta}_{n, m}^{(2)}>2c_{16} \{B_{n, m}^2\log(pn)/n\}^{1/2}+\{ B_{n, m}^2\log(pn)\log^2(1/\alpha_{n, m})/n\}^{1/4}\big]\notag \\
\leq&\exp\big\{-n^{1/2}\log^{1/2}(pn)\log(1/\alpha_{n, m})/(6 B_{n, m})\big\}+\notag\\
&3\exp\big[-n^{3/4}\log^{1/2}(1/\alpha_{n, m})/\{c_{18}B_{n, m}^{1/2}\log^{3/4}(pn)\}\big]\notag \\
\leq&\exp\big\{-\log^3(pn)\log(1/\alpha_{n, m})/6\big\}+3\exp\big\{-\log^3(pn)\log(1/\alpha_{n, m})/c_{18}\big\}\notag \\
=&(\alpha_{n, m})^{\log^3(pn)/6}+3(\alpha_{n, m})^{\log^3(pn)/c_{18}}.
\end{align}
Moreover, we have
\begin{align}\label{inequality:term110}
&2c_{16} \{B_{n, m}^2\log(pn)/n\}^{1/2}+\{ B_{n, m}^2\log(pn)\log^2(1/\alpha_{n, m})/n\}^{1/4}\notag \\
\leq&\{ B_{n, m}^2\log(pn)\log^2(1/\alpha_{n, m})/n\}^{1/4}\{1+2c_{16}\log^{-1}(pn)\}\notag \\
\leq&c_{19}\{ B_{n, m}^2\log(pn)\log^2(1/\alpha_{n, m})/n\}^{1/4},
\end{align}
for a universal constant $c_{19}>0$. Therefore, by combining (\ref{inequality:term109}) with (\ref{inequality:term110}), we conclude that
\begin{align}\label{inequality:term111}
&P\big[\hat{\Delta}_{n, m}^{(2)}>c_{19}\{ B_{n, m}^2\log(pn)\log^2(1/\alpha_{n, m})/n\}^{1/4}\big]\notag \\
\leq& (\alpha_{n, m})^{\log^3(pn)/6}+3(\alpha_{n, m})^{\log^3(pn)/c_{18}}.
\end{align}
Similar reasoning leads to
\begin{align}\label{inequality:term112}
&P\big[\hat{\Delta}_{n, m}^{(4)}>c_{20}\{ B_{n, m}^2\log(pm)\log^2(1/\alpha_{n, m})/m\}^{1/4}\big]\notag \\
\leq& (\alpha_{n, m})^{\log^3(pm)/6}+3(\alpha_{n, m})^{\log^3(pm)/c_{21}},
\end{align}
for some universal constants $c_{20}, c_{21}>0$. To this end, we set $\bar{\Delta}_{n, m}$ to be
\begin{align}\label{inequality:term113}
\bar{\Delta}_{n, m}=&(c_{10}+c_{19}^2)\{ B_{n, m}^2\log(pn)\log^2(1/\alpha_{n, m})/n\}^{1/2}+\notag\\
&(c_{11}\delta_1^{2}+c_{20}^2\delta_1^{2})\{ B_{n, m}^2\log(pm)\log^2(1/\alpha_{n, m})/m\}^{1/2},
\end{align}
where $\delta_1$ is defined in condition (a). Henceforth, by combining (\ref{inequality:term96}), (\ref{inequality:term102}), (\ref{inequality:term103}), (\ref{inequality:term111}), (\ref{inequality:term112}) with (\ref{inequality:term113}), we reach a conclusion that
\begin{align}\label{inequality:term114}
&P(\hat{\Delta}_{n, m}\leq \bar{\Delta}_{n, m})=1-P(\hat{\Delta}_{n, m}> \bar{\Delta}_{n, m})\notag\\
=&1-P\{\hat{\Delta}_{n, m}^{(1)}+(\hat{\Delta}_{n, m}^{(2)})^2+\delta_{n, m}^2\hat{\Delta}_{n, m}^{(3)}+\delta_{n, m}^2(\hat{\Delta}_{n, m}^{(4)})^2>\bar{\Delta}_{n, m}\}\notag\\
\geq&1-P\big[\hat{\Delta}_{n, m}^{(1)}>c_{10}\{ B_{n, m}^2\log(pn)\log^2(1/\alpha_{n, m})/n\}^{1/2}\big]\notag\\
&-P\big[\hat{\Delta}_{n, m}^{(2)}>c_{19}\{ B_{n, m}^2\log(pn)\log^2(1/\alpha_{n, m})/n\}^{1/4}\big]\notag\\
&-P\big[\hat{\Delta}_{n, m}^{(3)}>c_{11}\{ B_{n, m}^2\log(pm)\log^2(1/\alpha_{n, m})/m\}^{1/2}\big]\notag\\
&-P\big[\hat{\Delta}_{n, m}^{(4)}>c_{20}\{ B_{n, m}^2\log(pm)\log^2(1/\alpha_{n, m})/m\}^{1/4}\big]\geq 1-\gamma_{n, m},
\end{align}
where
$\gamma_{n, m}=(\alpha_{n, m})^{\log(pn)/3}+3(\alpha_{n, m})^{\log^{1/2}(pn)/c_*}+(\alpha_{n, m})^{\log(pm)/3}+\\3(\alpha_{n, m})^{\log^{1/2}(pm)/c_*}+(\alpha_{n, m})^{\log^3(pn)/6}+3(\alpha_{n, m})^{\log^3(pn)/c_*}+(\alpha_{n, m})^{\log^3(pm)/6}+3(\alpha_{n, m})^{\log^3(pm)/c_*}$
with $c_*=\max\{c_9, c_{12}, c_{18}, c_{21}\}$, and the second inequality is by  (a). Finally, based on (\ref{inequality:term95}), (\ref{inequality:term113}) and (\ref{inequality:term114}), it can be concluded  that with probability at least $1-\gamma_{n, m}$, we have $\rho_{n,m}^{MB}\lesssim\{ B_{n, m}^2\log^5(pn)\log^2(1/\alpha_{n, m})/n\}^{1/6}+\{ B_{n, m}^2\log^5(pm)\log^2(1/\alpha_{n, m})/m\}^{1/6}$, which completes the proof.
\qed

\noindent{\em Proof of Corollary~1}:\ \
First, based on Theorem~2, we know that with probability at least $1-\gamma_{n, m}$, one has
$\rho_{n,m}^{MB}\lesssim\{ B_{n, m}^2\log^5(pn)\log^2(1/\alpha_{n, m})/n\}^{1/6}+\{ B_{n, m}^2\log^5(pm)\log^2(1/\alpha_{n, m})/m\}^{1/6}$.
Accordingly, we let $\{A_{n, m}:n\geq1, m\geq 1, m\in\sigma(n)\}$ be a sequence of events where $A_{n, m}$ is the event such that
$\rho_{n,m}^{MB}\lesssim\{ B_{n, m}^2\log^5(pn)\log^2(1/\alpha_{n, m})/n\}^{1/6}+\{ B_{n, m}^2\log^5(pm)\log^2(1/\alpha_{n, m})/m\}^{1/6}$.
 \\ Note that $\sigma(n)$ is a set of positive integers determined by $n$ and the conditions in the Theorem, possibly the empty set.
It then follows that $P(A_{n, m})\geq 1-\gamma_{n, m}$, which further implies that $P(A_{n, m}^c)\leq \gamma_{n, m}$. Together with Lemma~6 and condition (f), it can be concluded that
$P\big(\bigcup_{k_1=1}^{\infty}\bigcup_{k_2=1}^{\infty}\bigcap_{n=k_1}^{\infty}\bigcap_{m\in \varrho(k_2)\cap \sigma(n)}A_{n, m}\big)=1-P\big(\bigcap_{k_1=1}^{\infty}\bigcap_{k_2=1}^{\infty}\bigcup_{n=k_1}^{\infty}\bigcup_{m\in \varrho(k_2)\cap \sigma(n)}A_{n, m}^c\big)=1$,
which completes the proof.
\qed

\section{Proofs of Lemmas}
\noindent{\em Proof of Lemma~1}:\ \
It follows directly from Lemmas A.5 and A.6 in \cite{chern:13}.
\qed

\noindent{\em Proof of Lemma~2}:\ \
Given $\phi_1, \phi_2\geq 1$, we denote $\phi=\min\{\phi_1, \phi_2\}$. Moreover, we denote $\beta=\phi \log p$, and given any $y\in \mathbb{R}^p$, we define the function $F_\beta(w)$ according to (\ref{inequality:term2}) that depends implicitly on $y$. Based on $F_\beta(w)$ and $\phi$, we construct a  function $\kappa(w)=\varphi_0(\phi F_{\beta}(w))$ upon (\ref{inequality:term4}). We also construct a function $h(w, t)$ as
\begin{align*}
h(w, t)=1\big\{-\phi^{-1}-t/\beta<\max_{1\leq j \leq p}(w_j-y_j)\leq \phi^{-1}+t/\beta\big\},
\end{align*}
for all $w \in \mathbb{R}^p$ and $t>0$. Then we let $V^n=\{V_1,\dots,V_n\}$ be an independent copy of $F^n$, while  $W^m=\{W_1,\dots,W_m\}$ is defined to be an independent copy of $G^m$.
Given any $v\in [0, 1]$, we denote the Slepian interpolant as
\begin{align*}
Z(t)=Z^*(t)+Z^{**}(t), t\in [0, 1],
\end{align*}
where
\begin{align*}
&Z^*(t)=\sum_{i=1}^n Z_i^*(t), t\in [0, 1],\\
&Z^{**}(t)=\sum_{i'=1}^m Z_{i'}^{**}(t), t\in [0, 1],
\end{align*}
Note that for all $i=1,\dots,n$, $i'=1,\dots,m$, $t\in [0, 1]$
\begin{align*}
Z_i^*(t)=&(Z_{i1}^*(t),\dots,Z_{ip}^*(t))'=n^{-1/2}\big[t^{1/2}\{v^{1/2}(X_i-\mu^X)+\\
&(1-v)^{1/2}(F_i-\mu^X)\}+(1-t)^{1/2}(V_i-\mu^X)\big],\\
Z_{i'}^{**}(t)=&(Z_{i'1}^{**}(t),\dots,Z_{i'p}^{**}(t))'=m^{-1/2}\delta_{n, m}\big[t^{1/2}\{v^{1/2}(Y_{i'}-\mu^Y)+\\
&(1-v)^{1/2}(G_{i'}-\mu^Y)\}+(1-t)^{1/2}(W_{i'}-\mu^Y)\big].
\end{align*}
Then, we define the leave-one-out terms for $Z^*(t)$ and $Z^{**}(t)$ as
$$Z^{*(i)}(t)=Z^{*}(t)-Z_i^{*}(t), t\in [0, 1], i=1,\dots,n$$
and
$$Z^{**(i)}(t)=Z^{**}(t)-Z_i^{**}(t), t\in [0, 1], i=1,\dots,m$$
respectively. For all $i=1,\dots,n$, $i'=1,\dots,m$, $t\in [0, 1]$, we denote
\begin{align*}
\dot{Z}_i^*(t)=&(\dot{Z}_{i1}^*(t),\dots,\dot{Z}_{ip}^*(t))'=n^{-1/2}\big[t^{-1/2}\{v^{1/2}(X_i-\mu^X)+\\
&(1-v)^{1/2}(F_i-\mu^X)\}-(1-t)^{-1/2}(V_i-\mu^X)\big],\\
\dot{Z}_{i'}^{**}(t)=&(\dot{Z}_{i'1}^{**}(t),\dots,\dot{Z}_{i'p}^{**}(t))'=m^{-1/2}\delta_{n, m}\big[t^{-1/2}\{v^{1/2}(Y_{i'}-\mu^Y)+\\
&(1-v)^{1/2}(G_{i'}-\mu^Y)\}-(1-t)^{-1/2}(W_{i'}-\mu^Y)\big].
\end{align*}
Note that we sometimes omit the $t$ and write $Z=Z(t)$, $\dot{Z}_i^*=\dot{Z}_i^*(t)$, etc, for the sake of brevity. It can be shown that
\begin{align*}
Z(1)=&v^{1/2}(S_n^X-n^{1/2}\mu^X)+(1-v)^{1/2}(S_n^F-n^{1/2}\mu^X)+\\
&v^{1/2}\delta_{n, m}(S_m^Y-m^{1/2}\mu^Y)+(1-v)^{1/2}\delta_{n, m}(S_m^G-m^{1/2}\mu^Y), \\
Z(0)=&S_n^V-n^{1/2}\mu^X+\delta_{n, m}S_m^W-\delta_{n, m}m^{1/2}\mu^Y.
\end{align*}
We further denote a quantity $\mathcal{I}_{n, m}$ as
\begin{align*}
\mathcal{I}_{n, m}=&\kappa(v^{1/2}(S_n^X-n^{1/2}\mu^X)+(1-v)^{1/2}(S_n^F-n^{1/2}\mu^X)+\\
&v^{1/2}\delta_{n, m}(S_m^Y-m^{1/2}\mu^Y)+(1-v)^{1/2}\delta_{n, m}(S_m^G-m^{1/2}\mu^Y))-\\
&\kappa(S_n^V-n^{1/2}\mu^X+\delta_{n, m}S_m^W-\delta_{n, m}m^{1/2}\mu^Y)\\
=&\kappa(Z(1))-\kappa(Z(0))
\end{align*}
Finally, we denote $\lambda_i^*$, $\lambda_{i}^{**}$ as
\begin{align*}
\lambda_i^*=&1\big\{\max_{1\leq j \leq p}|X_{ij}-\mu_j^X|\vee |F_{ij}-\mu_j^X|\vee |V_{ij}-\mu_j^X|\leq \\
&n^{1/2}/(4\beta)\big\}, \quad i=1,\dots,n, \\
\lambda_i^{**}=&1\big\{\max_{1\leq j \leq p}|Y_{ij}-\mu_j^Y|\vee |G_{ij}-\mu_j^Y|\vee |W_{ij}-\mu_j^Y|\leq\\
&m^{1/2}/(4\beta |\delta_{n, m}|)\big\}, \quad i=1,\dots,m.
\end{align*}
Now, we are in a position to start the proof. According to the above notations, it can be seen that
\begin{align}\label{inequality:term5}
\mathcal{I}_{n, m}&=\kappa(Z(1))-\kappa(Z(0))=\int_0^1\frac{d \kappa(Z(t))}{d t} d t \notag \\
&=2^{-1}\sum_{j=1}^p\sum_{i=1}^n\int_0^1\kappa_j(Z)\dot{Z}_{ij}^* d t+2^{-1}\sum_{j'=1}^p\sum_{i'=1}^m\int_0^1\kappa_{j'}(Z)\dot{Z}_{i'j'}^{**} d t \notag \\
&=\mathcal{I}_{n, m}^*+\mathcal{I}_{n, m}^{**},
\end{align}
where
\begin{align}\label{inequality:term6}
\mathcal{I}_{n, m}^*&=2^{-1}\sum_{j=1}^p\sum_{i=1}^n\int_0^1\kappa_j(Z)\dot{Z}_{ij}^* d t
\end{align}
and
\begin{align}\label{inequality:term7}
\mathcal{I}_{n, m}^{**}&=2^{-1}\sum_{j=1}^p\sum_{i=1}^m\int_0^1\kappa_{j}(Z)\dot{Z}_{ij}^{**} d t.
\end{align}
Moreover, by Taylor's theorem, it can be seen that for all $j=1,\dots,p$ and for any $t\in [0,1]$
\begin{align}\label{inequality:term8}
\kappa_{j}(Z)&=\kappa_{j}(Z^{*(i)}+Z^{**})+\sum_{k=1}^p \kappa_{jk}(Z^{*(i)}+Z^{**})Z_{ik}^{*}+ \notag \\
&\sum_{k=1}^p\sum_{l=1}^p \int_0^1(1-\tau)\kappa_{jkl}(Z^{*(i)}+Z^{**}+\tau Z_i^{*})Z_{ik}^{*}Z_{il}^{*}d \tau, \quad i=1,\dots,n
\end{align}
and
\begin{align}\label{inequality:term9}
\kappa_{j}(Z)&=\kappa_{j}(Z^{**(i)}+Z^{*})+\sum_{k=1}^p \kappa_{jk}(Z^{**(i)}+Z^{*})Z_{ik}^{**}+ \notag \\
&\sum_{k=1}^p\sum_{l=1}^p \int_0^1(1-\tau)\kappa_{jkl}(Z^{**(i)}+Z^{*}+\tau Z_i^{**})Z_{ik}^{**}Z_{il}^{**}d \tau, \quad i=1,\dots,m.
\end{align}
By combining (\ref{inequality:term6}) with (\ref{inequality:term8}), it can be seen  that
\begin{align*}
\mathcal{I}_{n, m}^*=&2^{-1}\sum_{j=1}^p\sum_{i=1}^n\int_0^1\kappa_{j}(Z^{*(i)}+Z^{**})\dot{Z}_{ij}^* d t\\
&+2^{-1}\sum_{j=1}^p\sum_{k=1}^p\sum_{i=1}^n\int_0^1\kappa_{jk}(Z^{*(i)}+Z^{**})\dot{Z}_{ij}^*{Z}_{ik}^*d t \\
&+2^{-1}\sum_{j=1}^p\sum_{k=1}^p\sum_{l=1}^p\sum_{i=1}^n\int_0^1\int_0^1(1-\tau)\kappa_{jkl}(Z^{*(i)}+Z^{**}+\tau Z_i^{*})\\
&\cdot \dot{Z}_{ij}^*{Z}_{ik}^*{Z}_{il}^* d \tau d t,
\end{align*}
which implies that
\begin{align}\label{inequality:term10}
E(\mathcal{I}_{n, m}^*)&=2^{-1}(I^*+II^*+III^*),
\end{align}
where
\begin{align*}
I^*=&\sum_{j=1}^p\sum_{i=1}^n\int_0^1E\{\kappa_{j}(Z^{*(i)}+Z^{**})\dot{Z}_{ij}^* \}d t, \\
II^*=&\sum_{j=1}^p\sum_{k=1}^p\sum_{i=1}^n\int_0^1E\{\kappa_{jk}(Z^{*(i)}+Z^{**})\dot{Z}_{ij}^*{Z}_{ik}^*\}d t, \\
III^*=&\sum_{j=1}^p\sum_{k=1}^p\sum_{l=1}^p\sum_{i=1}^n\int_0^1\int_0^1(1-\tau)E\{\kappa_{jkl}(Z^{*(i)}+Z^{**}+\tau Z_i^{*})\\
&\cdot \dot{Z}_{ij}^*{Z}_{ik}^*{Z}_{il}^*\} d \tau d t.
\end{align*}
For $I^*$, we have
\begin{align}\label{inequality:term11}
I^*&=\sum_{j=1}^p\sum_{i=1}^n\int_0^1E\{\kappa_{j}(Z^{*(i)}+Z^{**})\dot{Z}_{ij}^* \}d t \notag \\
&=\sum_{j=1}^p\sum_{i=1}^n\int_0^1E\{\kappa_{j}(Z^{*(i)}+Z^{**})\}E(\dot{Z}_{ij}^* )d t \notag \\
&=0,
\end{align}
where the second equality follows from the independence of $Z^{*(i)}+Z^{**}$ from $\dot{Z}_{ij}^*$,  and the last equality is by $E(\dot{Z}_{ij}^*)=0$. For $II^*$, we have
\begin{align}\label{inequality:term12}
II^*&=\sum_{j=1}^p\sum_{k=1}^p\sum_{i=1}^n\int_0^1E\{\kappa_{jk}(Z^{*(i)}+Z^{**})\dot{Z}_{ij}^*{Z}_{ik}^*\}d t \notag \\
&=\sum_{j=1}^p\sum_{k=1}^p\sum_{i=1}^n\int_0^1E\{\kappa_{jk}(Z^{*(i)}+Z^{**})\}E(\dot{Z}_{ij}^*{Z}_{ik}^*)d t,
\end{align}
where the last equality follows from the independence of $Z^{*(i)}+Z^{**}$ from $\dot{Z}_{ij}^*{Z}_{ik}^*$. Moreover, for every $i=1,\dots,n$; $j, k=1,\dots,p$ and $t\in [0, 1]$, we have
\begin{align}\label{inequality:term13}
E(\dot{Z}_{ij}^*{Z}_{ik}^*)=&E\Big(n^{-1}\big[t^{-1/2}\{v^{1/2}(X_{ij}-\mu_j^X)+(1-v)^{1/2}(F_{ij}-\mu_j^X)\}-\notag \\
&(1-t)^{-1/2}(V_{ij}-\mu_j^X)\big] \cdot \big[t^{1/2}\{v^{1/2}(X_{ik}-\mu_k^X)+\notag \\
&(1-v)^{1/2}(F_{ik}-\mu_k^X)\}+(1-t)^{1/2}(V_{ik}-\mu_k^X)\big]\Big)\notag \\
=&n^{-1}E\Big[\{v^{1/2}(X_{ij}-\mu_j^X)+(1-v)^{1/2}(F_{ij}-\mu_j^X)\}\notag \\
&\cdot \{v^{1/2}(X_{ik}-\mu_k^X)+(1-v)^{1/2}(F_{ik}-\mu_k^X)\}-\notag \\
&E\{(V_{ij}-\mu_j^X)(V_{ik}-\mu_k^X)\}\Big]\notag \\
=&n^{-1}E\{v(X_{ij}-\mu_j^X)(X_{ik}-\mu_k^X)+\notag \\
&(1-v)(F_{ij}-\mu_j^X)(F_{ik}-\mu_k^X)-(V_{ij}-\mu_j^X)(V_{ik}-\mu_k^X)\}\notag \\
=&0,
\end{align}
where the second and the third equalities follow from the independence among $X^n$, $F^n$, $V^n$, and the last equality holds since $E\{(X_{ij}-\mu_j^X)(X_{ik}-\mu_k^X)\}=E\{(F_{ij}-\mu_j^X)(F_{ik}-\mu_k^X)\}=E\{(V_{ij}-\mu_j^X)(V_{ik}-\mu_k^X)\}$. Hence, by combining (\ref{inequality:term12}) with (\ref{inequality:term13}), we conclude that
\begin{align}\label{inequality:term14}
II^*=0.
\end{align}
For $III^*$, we can write it as
\begin{align}\label{inequality:term15}
III^*=III_1^*+III_2^*,
\end{align}
where
\begin{align*}
III_1^*=&\sum_{j=1}^p\sum_{k=1}^p\sum_{l=1}^p\sum_{i=1}^n\int_0^1\int_0^1(1-\tau)E\{\lambda_i^*\kappa_{jkl}(Z^{*(i)}+Z^{**}+\tau Z_i^{*})\\
&\cdot \dot{Z}_{ij}^*{Z}_{ik}^*{Z}_{il}^*\} d \tau d t, \\
III_2^*=&\sum_{j=1}^p\sum_{k=1}^p\sum_{l=1}^p\sum_{i=1}^n\int_0^1\int_0^1(1-\tau)E\{(1-\lambda_i^*)\kappa_{jkl}(Z^{*(i)}+Z^{**}+\tau Z_i^{*})\\
&\cdot \dot{Z}_{ij}^*{Z}_{ik}^*{Z}_{il}^*\} d \tau d t.
\end{align*}
For $III_2^*$, we have
\begin{align}\label{inequality:term16}
|III_2^*|\leq& \sum_{j=1}^p\sum_{k=1}^p\sum_{l=1}^p\sum_{i=1}^n\int_0^1\int_0^1(1-\tau)E\{(1-\lambda_i^*)|\kappa_{jkl}(Z^{*(i)}+Z^{**}+\tau Z_i^{*})|\notag \\
&\cdot |\dot{Z}_{ij}^*{Z}_{ik}^*{Z}_{il}^*|\} d \tau d t \notag \\
\leq&  \sum_{j=1}^p\sum_{k=1}^p\sum_{l=1}^p\sum_{i=1}^n\int_0^1\int_0^1(1-\tau)E\{(1-\lambda_i^*)Q_{jkl}(Z^{*(i)}+Z^{**}+\tau Z_i^{*}) \notag \\
&\cdot |\dot{Z}_{ij}^*{Z}_{ik}^*{Z}_{il}^*|\} d \tau d t\notag \\
\lesssim& \phi \beta^2 \sum_{i=1}^n\int_0^1 E\big\{\big(1-\lambda_i^*\big)\big(\max_{1\leq j, k, l \leq p}|\dot{Z}_{ij}^*{Z}_{ik}^*{Z}_{il}^*|\big)\big\} d t,
\end{align}
where the first inequality follows from Jensen's inequality, and the second and the third inequalities follow from Lemma~1. Moreover, note that for all $i=1,\dots,n$, we have
\begin{align}\label{inequality:term17}
1-\lambda_i^*=&1\big\{\max_{1\leq j \leq p}|X_{ij}-\mu_j^X|\vee |F_{ij}-\mu_j^X|\vee |V_{ij}-\mu_j^X|> n^{1/2}/(4\beta)\big\} \notag \\
\leq &  1\big\{\max_{1\leq j \leq p}|X_{ij}-\mu_j^X|> n^{1/2}/(4\beta)\big\}+\notag \\
&1\big\{\max_{1\leq j \leq p}|F_{ij}-\mu_j^X|> n^{1/2}/(4\beta)\big\}+ \notag \\
&1\big\{\max_{1\leq j \leq p}|V_{ij}-\mu_j^X|> n^{1/2}/(4\beta)\big\}.
\end{align}
Also note that for all $i=1,\dots,n$ and for any $t\in [0, 1]$, we have
\begin{align}\label{inequality:term18}
&\max_{1\leq j, k, l \leq p}|\dot{Z}_{ij}^*{Z}_{ik}^*{Z}_{il}^*| \notag \\
=&\max_{1\leq j, k, l \leq p}\Big|n^{-1/2}\big[t^{-1/2}\{v^{1/2}(X_{ij}-\mu_j^X)+(1-v)^{1/2}(F_{ij}-\mu_j^X)\}-\notag \\
&(1-t)^{-1/2}(V_{ij}-\mu_j^X)\big]\Big|\cdot \Big|n^{-1/2}\big[t^{1/2}\{v^{1/2}(X_{ik}-\mu_k^X)+\notag \\
&(1-v)^{1/2}(F_{ik}-\mu_k^X)\}+(1-t)^{1/2}(V_{ik}-\mu_k^X)\big]\Big|\notag \\
&\cdot \Big|n^{-1/2}\big[t^{1/2}\{v^{1/2}(X_{il}-\mu_l^X)+(1-v)^{1/2}(F_{il}-\mu_l^X)\}+\notag \\
&(1-t)^{1/2}(V_{il}-\mu_l^X)\big]\Big| \notag \\
\leq& 100 n^{-3/2}\{t^{1/2}\wedge (1-t)^{1/2}\}^{-1} \notag \\
&\cdot \max_{1\leq j \leq p} |X_{ij}-\mu_j^X|^3\vee |F_{ij}-\mu_j^X|^3 \vee |V_{ij}-\mu_j^X|^3 \notag \\
\leq& 100 n^{-3/2}\{t^{1/2}\wedge (1-t)^{1/2}\}^{-1}\notag \\
&\cdot (\max_{1\leq j \leq p} |X_{ij}-\mu_j^X|^3+\max_{1\leq j \leq p} |F_{ij}-\mu_j^X|^3+\max_{1\leq j \leq p} |V_{ij}-\mu_j^X|^3).
\end{align}
Based on (\ref{inequality:term16}), (\ref{inequality:term17}) and (\ref{inequality:term18}), it can be deduced that
\begin{align}\label{inequality:term19}
|III_2^*| \lesssim & n^{-3/2}\phi \beta^2 \Big[\int_0^1\{t^{1/2}\wedge (1-t)^{1/2}\}^{-1} d t\Big] \cdot \notag \\
&\sum_{i=1}^n E\Big(\Big[1\big\{\max_{1\leq j \leq p}|X_{ij}-\mu_j^X|> n^{1/2}/(4\beta)\big\}+\notag \\
&1\big\{\max_{1\leq j \leq p}|F_{ij}-\mu_j^X|> n^{1/2}/(4\beta)\big\}+\notag \\
&1\big\{\max_{1\leq j \leq p}|V_{ij}-\mu_j^X|> n^{1/2}/(4\beta)\big\}\Big] \notag \\
&\cdot \Big[\max_{1\leq j \leq p} |X_{ij}-\mu_j^X|^3+\max_{1\leq j \leq p} |F_{ij}-\mu_j^X|^3+\max_{1\leq j \leq p} |V_{ij}-\mu_j^X|^3\Big]\Big) \notag \\
\lesssim & n^{-3/2}\phi \beta^2 \sum_{i=1}^n E\Big(\Big[1\big\{\max_{1\leq j \leq p}|X_{ij}-\mu_j^X|> n^{1/2}/(4\beta)\big\}+\notag \\
&1\big\{\max_{1\leq j \leq p}|F_{ij}-\mu_j^X|> n^{1/2}/(4\beta)\big\}+\notag \\
&1\big\{\max_{1\leq j \leq p}|V_{ij}-\mu_j^X|> n^{1/2}/(4\beta)\big\}\Big]\notag\\
&\cdot \Big[\max_{1\leq j \leq p} |X_{ij}-\mu_j^X|^3+\max_{1\leq j \leq p} |F_{ij}-\mu_j^X|^3+\max_{1\leq j \leq p} |V_{ij}-\mu_j^X|^3\Big]\Big)\notag \\
\lesssim & n^{-1/2}\phi \beta^2 \big\{M_n^X(\phi)+M_n^F(\phi)+M_n^V(\phi)\big\}\notag \\
\lesssim & n^{-1/2}\phi \beta^2 \big\{M_n^X(\phi)+M_n^F(\phi)\big\}\notag \\
\lesssim & n^{-1/2}\phi \beta^2 M_n(\phi),
\end{align}
where the third inequality is by Lemma B.1 in \cite{chern:16:1}, and the fourth inequality follows from the fact that $M_n^F(\phi)=M_n^V(\phi)$.
For $III_1^*$, we have
\begin{align}\label{inequality:term20}
|III_1^*|\leq& \sum_{j=1}^p\sum_{k=1}^p\sum_{l=1}^p\sum_{i=1}^n\int_0^1\int_0^1(1-\tau)E\{\lambda_i^*|\kappa_{jkl}(Z^{*(i)}+Z^{**}+\tau Z_i^{*})|\notag \\
&\cdot |\dot{Z}_{ij}^*{Z}_{ik}^*{Z}_{il}^*|\} d \tau d t \notag \\
\leq&  \sum_{j=1}^p\sum_{k=1}^p\sum_{l=1}^p\sum_{i=1}^n\int_0^1\int_0^1 E\{\lambda_i^*|\kappa_{jkl}(Z^{*(i)}+Z^{**}+\tau Z_i^{*})|\notag \\
&\cdot |\dot{Z}_{ij}^*{Z}_{ik}^*{Z}_{il}^*|\} d \tau d t \notag \\
=&  \sum_{j=1}^p\sum_{k=1}^p\sum_{l=1}^p\sum_{i=1}^n\int_0^1\int_0^1 E\{\lambda_i^*h(Z^{*(i)}+Z^{**}, 1)\notag \\
&\cdot |\kappa_{jkl}(Z^{*(i)}+Z^{**}+\tau Z_i^{*})|\cdot|\dot{Z}_{ij}^*{Z}_{ik}^*{Z}_{il}^*|\} d \tau d t \notag \\
\lesssim&  \sum_{j=1}^p\sum_{k=1}^p\sum_{l=1}^p\sum_{i=1}^n\int_0^1\int_0^1 E\{\lambda_i^*h(Z^{*(i)}+Z^{**}, 1)Q_{jkl}(Z^{*(i)}+Z^{**}) \notag \\
&\cdot |\dot{Z}_{ij}^*{Z}_{ik}^*{Z}_{il}^*|\} d \tau d t \notag \\
\lesssim&  \sum_{j=1}^p\sum_{k=1}^p\sum_{l=1}^p\sum_{i=1}^n\int_0^1 E\{h(Z^{*(i)}+Z^{**}, 1)Q_{jkl}(Z^{*(i)}+Z^{**})\notag \\
&\cdot |\dot{Z}_{ij}^*{Z}_{ik}^*{Z}_{il}^*|\} d t \notag \\
\lesssim& \sum_{j=1}^p\sum_{k=1}^p\sum_{l=1}^p\sum_{i=1}^n\int_0^1 E\{h(Z^{*(i)}+Z^{**}, 1)Q_{jkl}(Z^{*(i)}+Z^{**})\} \notag \\
&\cdot E(|\dot{Z}_{ij}^*{Z}_{ik}^*{Z}_{il}^*|) d t \notag \\
\lesssim& \Delta_1^*+\Delta_2^*,
\end{align}
where the equality follows from the fact that
\begin{align*}
&\lambda_i^*h(Z^{*(i)}+Z^{**}, 1)|\kappa_{jkl}(Z^{*(i)}+Z^{**}+\tau Z_i^{*})|\\
=&\lambda_i^*|\kappa_{jkl}(Z^{*(i)}+Z^{**}+\tau Z_i^{*})|, \quad i=1,\dots,n
\end{align*}
whenever $\tau\in [0, 1]$, and the third inequality is by Lemma~1. In addition, the terms $\Delta_1^*$ and $\Delta_2^*$ are given by
\begin{align*}
\Delta_1^*=&\sum_{j=1}^p\sum_{k=1}^p\sum_{l=1}^p\sum_{i=1}^n\int_0^1 E\{(1-\lambda_i^*)h(Z^{*(i)}+Z^{**}, 1)Q_{jkl}(Z^{*(i)}+Z^{**})\} \\
&\cdot E(|\dot{Z}_{ij}^*{Z}_{ik}^*{Z}_{il}^*|) d t, \\
\Delta_2^*=&\sum_{j=1}^p\sum_{k=1}^p\sum_{l=1}^p\sum_{i=1}^n\int_0^1 E\{\lambda_i^*h(Z^{*(i)}+Z^{**}, 1)Q_{jkl}(Z^{*(i)}+Z^{**})\}\\
&\cdot E(|\dot{Z}_{ij}^*{Z}_{ik}^*{Z}_{il}^*|) d t.
\end{align*}
For $\Delta_1^*$, we have
\begin{align}\label{inequality:term21}
\Delta_1^*=& \sum_{i=1}^n\int_0^1 \sum_{j=1}^p\sum_{k=1}^p\sum_{l=1}^p E\{(1-\lambda_i^*)h(Z^{*(i)}+Z^{**}, 1)Q_{jkl}(Z^{*(i)}+Z^{**})\}\notag \\
&\cdot E(|\dot{Z}_{ij}^*{Z}_{ik}^*{Z}_{il}^*|) d t \notag \\
\leq& \sum_{i=1}^n\int_0^1 \big\{\max_{1\leq j, k, l \leq p} E(|\dot{Z}_{ij}^*{Z}_{ik}^*{Z}_{il}^*|)\big\}\sum_{j=1}^p\sum_{k=1}^p\sum_{l=1}^p E\big\{(1-\lambda_i^*)\notag \\
&\cdot h(Z^{*(i)}+Z^{**}, 1)Q_{jkl}(Z^{*(i)}+Z^{**})\big\}  d t \notag \\
\lesssim& \phi \beta^2\sum_{i=1}^n\int_0^1 \big\{\max_{1\leq j, k, l \leq p} E(|\dot{Z}_{ij}^*{Z}_{ik}^*{Z}_{il}^*|)\big\}E\big\{(1-\lambda_i^*)\notag \\
&\cdot h(Z^{*(i)}+Z^{**}, 1)\big\}  d t \notag \\
\lesssim& \phi \beta^2\sum_{i=1}^n\int_0^1 \big\{\max_{1\leq j, k, l \leq p} E(|\dot{Z}_{ij}^*{Z}_{ik}^*{Z}_{il}^*|)\big\}\big\{E(1-\lambda_i^*)\big\}  d t \notag \\
\lesssim& n^{-1/2}\phi \beta^2 \Big[\int_0^1\{t^{1/2}\wedge (1-t)^{1/2}\}^{-1} d t\Big]\notag \\
&\cdot  \big\{M_n^X(\phi)+M_n^F(\phi)+M_n^V(\phi)\big\} \notag \\
\lesssim& n^{-1/2}\phi \beta^2  \big\{M_n^X(\phi)+M_n^F(\phi)\big\} \notag \\
\lesssim& n^{-1/2}\phi \beta^2  M_n(\phi),
\end{align}
where the first inequality is by Holder's inequality, the second inequality is due to Lemma~1, and the fourth inequality is based on (\ref{inequality:term17}), (\ref{inequality:term18}) and Lemma B.1 in \cite{chern:16:1}. For $\Delta_2^*$, we have
\begin{align}\label{inequality:term22}
\Delta_2^*=&\sum_{j=1}^p\sum_{k=1}^p\sum_{l=1}^p\sum_{i=1}^n\int_0^1 E\{\lambda_i^*h(Z^{*(i)}+Z^{**}, 1)Q_{jkl}(Z^{*(i)}+Z^{**})\}\notag \\
 &\cdot E(|\dot{Z}_{ij}^*{Z}_{ik}^*{Z}_{il}^*|) d t \notag \\
\lesssim& \sum_{j=1}^p\sum_{k=1}^p\sum_{l=1}^p\sum_{i=1}^n\int_0^1 E\{\lambda_i^*h(Z^{*(i)}+Z^{**}, 1)Q_{jkl}(Z^{*}+Z^{**})\} \notag \\
&\cdot E(|\dot{Z}_{ij}^*{Z}_{ik}^*{Z}_{il}^*|) d t \notag \\
\lesssim& \sum_{j=1}^p\sum_{k=1}^p\sum_{l=1}^p\sum_{i=1}^n\int_0^1 E\{\lambda_i^*h(Z^{*(i)}+Z^{**}, 1)h(Z^{*}+Z^{**}, 2)\notag \\
&\cdot Q_{jkl}(Z^{*}+Z^{**})\} E(|\dot{Z}_{ij}^*{Z}_{ik}^*{Z}_{il}^*|) d t \notag \\
\lesssim& \sum_{j=1}^p\sum_{k=1}^p\sum_{l=1}^p\sum_{i=1}^n\int_0^1 E\{h(Z^{*}+Z^{**}, 2)Q_{jkl}(Z^{*}+Z^{**})\}\notag \\
&\cdot E(|\dot{Z}_{ij}^*{Z}_{ik}^*{Z}_{il}^*|) d t \notag \\
\lesssim& \phi \beta^2  \int_0^1 E\{h(Z^{*}+Z^{**}, 2)\}\big\{\max_{1\leq j, k, l\leq p}\sum_{i=1}^n E(|\dot{Z}_{ij}^*{Z}_{ik}^*{Z}_{il}^*|)\big\} d t,
\end{align}
where the first inequality is by Lemma~1, the second inequality follows from the fact that
$$\lambda_i^*h(Z^{*(i)}+Z^{**}, 1)=\lambda_i^*h(Z^{*(i)}+Z^{**}, 1)h(Z^{*}+Z^{**}, 2), \quad i=1,\dots,n,$$
and the fourth inequality is based on Holder's inequality and Lemma~1. Furthermore,
\begin{align*}
&\max_{1\leq j, k, l\leq p}\sum_{i=1}^n E(|\dot{Z}_{ij}^*{Z}_{ik}^*{Z}_{il}^*|)\notag \\
\lesssim& n^{-3/2}\{t^{1/2}\wedge (1-t)^{1/2}\}^{-1}\notag \\
&\cdot \max_{1\leq j \leq p}\sum_{i=1}^n E(|X_{ij}-\mu_j^X|^3+|F_{ij}-\mu_j^X|^3+|V_{ij}-\mu_j^X|^3)\notag \\
\lesssim& n^{-3/2}\{t^{1/2}\wedge (1-t)^{1/2}\}^{-1}\notag \\
&\cdot \big\{\max_{1\leq j \leq p}\sum_{i=1}^n E(|X_{ij}-\mu_j^X|^3)+\max_{1\leq j \leq p}\sum_{i=1}^n E(|F_{ij}-\mu_j^X|^3)\big\}.
\end{align*}
Together with the fact that
$E(|F_{ij}-\mu_j^X|^3)\lesssim \{E(|F_{ij}-\mu_j^X|^2)\}^{3/2} \lesssim \{E(|X_{ij}-\mu_j^X|^2)\}^{3/2}\lesssim E(|X_{ij}-\mu_j^X|^3)$, we conclude that
\begin{align}\label{inequality:term23}
\max_{1\leq j, k, l\leq p}\sum_{i=1}^n E(|\dot{Z}_{ij}^*{Z}_{ik}^*{Z}_{il}^*|) \lesssim n^{-1/2}\{t^{1/2}\wedge (1-t)^{1/2}\}^{-1}L_n^X.
\end{align}
Moreover, by definition of the function $h$, we have
\begin{align}\label{inequality:term24}
E\{h(Z^{*}+Z^{**}, 2)\} =& P(Z^{*}+Z^{**}-y\leq \phi^{-1}+2\beta^{-1})-\notag \\
&P(Z^{*}+Z^{**}-y\leq -\phi^{-1}-2\beta^{-1}).
\end{align}
In addition, it can be  verified that
\begin{align}\label{inequality:term25}
Z^{*}+Z^{**}\stackrel{\text{d}}{=}&(tv)^{1/2}(S_n^X+\delta_{n, m}S_m^Y-n^{1/2}\mu^X-\delta_{n, m}m^{1/2}\mu^Y)+\notag\\
&(1-tv)^{1/2}(S_n^F+\delta_{n, m}S_m^G-n^{1/2}\mu^X-\delta_{n, m}m^{1/2}\mu^Y),
\end{align}
where $\stackrel{\text{d}}{=}$ means having the same distribution. Therefore, under (\ref{inequality:term25}), we have
\begin{align}\label{inequality:term26}
&P(Z^{*}+Z^{**}-y\leq \phi^{-1}+2\beta^{-1})\notag \\
=&P\big\{(tv)^{1/2}(S_n^X+\delta_{n, m}S_m^Y-n^{1/2}\mu^X-\delta_{n, m}m^{1/2}\mu^Y)+\notag\\
&(1-tv)^{1/2}(S_n^F+\delta_{n, m}S_m^G-n^{1/2}\mu^X-\delta_{n, m}m^{1/2}\mu^Y)-y\leq \notag \\
&\phi^{-1}+2\beta^{-1}\big\} \notag \\
=&P\big\{(tv)^{1/2}(S_n^X+\delta_{n, m}S_m^Y-n^{1/2}\mu^X-\delta_{n, m}m^{1/2}\mu^Y)+\notag\\
&(1-tv)^{1/2}(S_n^F+\delta_{n, m}S_m^G-n^{1/2}\mu^X-\delta_{n, m}m^{1/2}\mu^Y)\leq\notag \\
& y+\phi^{-1}+2\beta^{-1}\big\} \notag \\
=&P\big\{(tv)^{1/2}(S_n^X+\delta_{n, m}S_m^Y-n^{1/2}\mu^X-\delta_{n, m}m^{1/2}\mu^Y)+\notag\\
&(1-tv)^{1/2}(S_n^F+\delta_{n, m}S_m^G-n^{1/2}\mu^X-\delta_{n, m}m^{1/2}\mu^Y)\leq\notag \\
& y+\phi^{-1}+2\beta^{-1}\big\}\notag \\
&-P(S_n^F+\delta_{n, m}S_m^G-n^{1/2}\mu^X-\delta_{n, m}m^{1/2}\mu^Y\leq\notag \\
& y+\phi^{-1}+2\beta^{-1})\notag \\
&+P(S_n^F+\delta_{n, m}S_m^G-n^{1/2}\mu^X-\delta_{n, m}m^{1/2}\mu^Y\leq\notag \\
& y+\phi^{-1}+2\beta^{-1})\notag \\
\leq &\rho_{n, m}+P(S_n^F+\delta_{n, m}S_m^G-n^{1/2}\mu^X-\delta_{n, m}m^{1/2}\mu^Y\leq\notag \\
& y+\phi^{-1}+2\beta^{-1}).
\end{align}
Likewise, we have
\begin{align}\label{inequality:term27}
&P(Z^{*}+Z^{**}-y\leq -\phi^{-1}-2\beta^{-1})\notag \\
=&P\big\{(tv)^{1/2}(S_n^X+\delta_{n, m}S_m^Y-n^{1/2}\mu^X-\delta_{n, m}m^{1/2}\mu^Y)+\notag\\
&(1-tv)^{1/2}(S_n^F+\delta_{n, m}S_m^G-n^{1/2}\mu^X-\delta_{n, m}m^{1/2}\mu^Y)-y\leq\notag \\
& -\phi^{-1}-2\beta^{-1}\big\} \notag \\
=&P\big\{(tv)^{1/2}(S_n^X+\delta_{n, m}S_m^Y-n^{1/2}\mu^X-\delta_{n, m}m^{1/2}\mu^Y)+\notag\\
&(1-tv)^{1/2}(S_n^F+\delta_{n, m}S_m^G-n^{1/2}\mu^X-\delta_{n, m}m^{1/2}\mu^Y)\leq\notag \\
& y-\phi^{-1}-2\beta^{-1}\big\} \notag \\
=&P\big\{(tv)^{1/2}(S_n^X+\delta_{n, m}S_m^Y-n^{1/2}\mu^X-\delta_{n, m}m^{1/2}\mu^Y)+\notag\\
&(1-tv)^{1/2}(S_n^F+\delta_{n, m}S_m^G-n^{1/2}\mu^X-\delta_{n, m}m^{1/2}\mu^Y)\leq\notag \\
& y-\phi^{-1}-2\beta^{-1}\big\} \notag \\
&-P(S_n^F+\delta_{n, m}S_m^G-n^{1/2}\mu^X-\delta_{n, m}m^{1/2}\mu^Y\leq\notag \\
& y-\phi^{-1}-2\beta^{-1})\notag\\
&+P(S_n^F+\delta_{n, m}S_m^G-n^{1/2}\mu^X-\delta_{n, m}m^{1/2}\mu^Y\leq\notag \\
& y-\phi^{-1}-2\beta^{-1})\notag \\
\geq &-\rho_{n, m}+P(S_n^F+\delta_{n, m}S_m^G-n^{1/2}\mu^X-\delta_{n, m}m^{1/2}\mu^Y\leq\notag \\
& y-\phi^{-1}-2\beta^{-1}).
\end{align}
Based on (\ref{inequality:term24}), (\ref{inequality:term26}) and (\ref{inequality:term27}), we have
\begin{align}\label{inequality:term28}
&E\{h(Z^{*}+Z^{**}, 2)\} \notag \\
\leq & 2\rho_{n, m}+P(S_n^F+\delta_{n, m}S_m^G-n^{1/2}\mu^X-\delta_{n, m}m^{1/2}\mu^Y\leq\notag \\
& y+\phi^{-1}+2\beta^{-1})\notag\\
&-P(S_n^F+\delta_{n, m}S_m^G-n^{1/2}\mu^X-\delta_{n, m}m^{1/2}\mu^Y\leq\notag \\
& y-\phi^{-1}-2\beta^{-1})\notag \\
\lesssim & \rho_{n, m}+(\phi^{-1}+2\beta^{-1})(\log p)^{1/2}\notag \\
\lesssim & \rho_{n, m}+\phi^{-1}(\log p)^{1/2},
\end{align}
where the second inequality follows from combining condition (a) with Lemma A.1 in \cite{chern:16:1}. Therefore, based on (\ref{inequality:term22}), (\ref{inequality:term23}) and (\ref{inequality:term28}), it can be verified  that
\begin{align}\label{inequality:term29}
\Delta_2^*&\lesssim \phi \beta^2  \int_0^1 E\{h(Z^{*}+Z^{**}, 2)\}\big\{\max_{1\leq j, k, l\leq p}\sum_{i=1}^n E(|\dot{Z}_{ij}^*{Z}_{ik}^*{Z}_{il}^*|)\big\} d t \notag \\
&\lesssim n^{-1/2}\phi \beta^2 \big\{L_n^X\rho_{n, m}+\phi^{-1}L_n^X(\log p)^{1/2}\big\}.
\end{align}
Under (\ref{inequality:term15}), (\ref{inequality:term19}), (\ref{inequality:term20}), (\ref{inequality:term21}) and (\ref{inequality:term29}), it can be deduced that
\begin{align}\label{inequality:term30}
|III^*|&\lesssim n^{-1/2}\phi \beta^2  M_n(\phi)+n^{-1/2}\phi \beta^2 \big\{L_n^X\rho_{n, m}+\phi^{-1}L_n^X(\log p)^{1/2}\big\}\notag \\
&\lesssim n^{-1/2} \beta^2 \big\{\phi L_n^X\rho_{n, m}+L_n^X(\log p)^{1/2}+\phi   M_n(\phi)\big\}\notag \\
&\lesssim n^{-1/2} \phi^2 (\log p)^2 \big\{\phi L_n^X\rho_{n, m}+L_n^X(\log p)^{1/2}+\phi   M_n(\phi)\big\}\notag \\
&\lesssim n^{-1/2} \phi_1^2 (\log p)^2 \big\{\phi_1 L_n^X\rho_{n, m}+L_n^X(\log p)^{1/2}+\phi_1   M_n(\phi_1)\big\},
\end{align}
where the last inequality follows from the facts that $\phi_1\geq \phi$ and $M_n(\phi_1)\geq M_n(\phi)$.
Hence by combining (\ref{inequality:term10}), (\ref{inequality:term11}), (\ref{inequality:term14}) with (\ref{inequality:term30}), it is clear that
\begin{align}\label{inequality:term31}
|E(\mathcal{I}_{n, m}^*)|\leq& 2^{-1}|III^*|\notag \\
\lesssim& n^{-1/2} \phi_1^2 (\log p)^2 \big\{\phi_1 L_n^X\rho_{n, m}+L_n^X(\log p)^{1/2}+\phi_1   M_n(\phi_1)\big\}.
\end{align}
Next, we start to bound the quantity $|E(\mathcal{I}_{n, m}^{**})|$. Similarly, by combining (\ref{inequality:term7}) with (\ref{inequality:term9}), we have
\begin{align*}
\mathcal{I}_{n, m}^{**}=&2^{-1}\sum_{j=1}^p\sum_{i=1}^m\int_0^1\kappa_{j}(Z^{**(i)}+Z^{*})\dot{Z}_{ij}^{**} d t+\\
&2^{-1}\sum_{j=1}^p\sum_{k=1}^p\sum_{i=1}^m\int_0^1\kappa_{jk}(Z^{**(i)}+Z^{*})\dot{Z}_{ij}^{**}{Z}_{ik}^{**}d t+ \\
&2^{-1}\sum_{j=1}^p\sum_{k=1}^p\sum_{l=1}^p\sum_{i=1}^m\int_0^1\int_0^1(1-\tau)\kappa_{jkl}(Z^{**(i)}+Z^{*}+\tau Z_i^{**})\\
&\cdot \dot{Z}_{ij}^{**}{Z}_{ik}^{**}{Z}_{il}^{**} d \tau d t,
\end{align*}
which implies that
\begin{align}\label{inequality:term32}
E(\mathcal{I}_{n, m}^{**})&=2^{-1}(I^{**}+II^{**}+III^{**}),
\end{align}
where
\begin{align*}
I^{**}=&\sum_{j=1}^p\sum_{i=1}^m\int_0^1E\{\kappa_{j}(Z^{**(i)}+Z^{*})\dot{Z}_{ij}^{**} \}d t, \\
II^{**}=&\sum_{j=1}^p\sum_{k=1}^p\sum_{i=1}^m\int_0^1E\{\kappa_{jk}(Z^{**(i)}+Z^{*})\dot{Z}_{ij}^{**}{Z}_{ik}^{**}\}d t, \\
III^{**}=&\sum_{j=1}^p\sum_{k=1}^p\sum_{l=1}^p\sum_{i=1}^m\int_0^1\int_0^1(1-\tau)E\{\kappa_{jkl}(Z^{**(i)}+Z^{*}+\tau Z_i^{**})\\
&\cdot \dot{Z}_{ij}^{**}{Z}_{ik}^{**}{Z}_{il}^{**}\} d \tau d t.
\end{align*}
For $I^{**}$, we have
\begin{align}\label{inequality:term33}
I^{**}&=\sum_{j=1}^p\sum_{i=1}^m\int_0^1E\{\kappa_{j}(Z^{**(i)}+Z^{*})\dot{Z}_{ij}^{**} \}d t \notag \\
&=\sum_{j=1}^p\sum_{i=1}^m\int_0^1E\{\kappa_{j}(Z^{**(i)}+Z^{*})\}E(\dot{Z}_{ij}^{**} )d t \notag \\
&=0,
\end{align}
where the second equality follows from the independence of $Z^{**(i)}+Z^{*}$ from $\dot{Z}_{ij}^{**}$, and the last equality is by $E(\dot{Z}_{ij}^{**})=0$.
For $II^{**}$, we have
\begin{align}\label{inequality:term34}
II^{**}&=\sum_{j=1}^p\sum_{k=1}^p\sum_{i=1}^m\int_0^1E\{\kappa_{jk}(Z^{**(i)}+Z^{*})\dot{Z}_{ij}^{**}{Z}_{ik}^{**}\}d t \notag \\
&=\sum_{j=1}^p\sum_{k=1}^p\sum_{i=1}^m\int_0^1E\{\kappa_{jk}(Z^{**(i)}+Z^{*})\}E(\dot{Z}_{ij}^{**}{Z}_{ik}^{**})d t,
\end{align}
where the last equality is based on the independence of $Z^{**(i)}+Z^{*}$ from $\dot{Z}_{ij}^{**}{Z}_{ik}^{**}$. In addition, for all $i=1,\dots,m$, $j, k=1,\dots,p$, we have
\begin{align}\label{inequality:term35}
E(\dot{Z}_{ij}^{**}{Z}_{ik}^{**})=&m^{-1}\delta_{n, m}^2 E\Big(\big[t^{-1/2}\{v^{1/2}(Y_{ij}-\mu_j^Y)+(1-v)^{1/2}(G_{ij}-\mu_j^Y)\}-\notag \\
&(1-t)^{-1/2}(W_{ij}-\mu_j^Y)\big] \cdot \big[t^{1/2}\{v^{1/2}(Y_{ik}-\mu_k^Y)+\notag \\
&(1-v)^{1/2}(G_{ik}-\mu_k^Y)\}+(1-t)^{1/2}(W_{ik}-\mu_k^Y)\big]\Big)\notag \\
=&m^{-1}\delta_{n, m}^2 E\Big[\{v^{1/2}(Y_{ij}-\mu_j^Y)+(1-v)^{1/2}(G_{ij}-\mu_j^Y)\}\notag \\
&\cdot \{v^{1/2}(Y_{ik}-\mu_k^Y)+(1-v)^{1/2}(G_{ik}-\mu_k^Y)\}-\notag \\
&(W_{ij}-\mu_j^Y)(W_{ik}-\mu_k^Y)\Big]\notag \\
=&m^{-1}\delta_{n, m}^2E\{v(Y_{ij}-\mu_j^Y)(Y_{ik}-\mu_k^Y)+\notag \\
&(1-v)(G_{ij}-\mu_j^Y)(G_{ik}-\mu_k^Y)-(W_{ij}-\mu_j^Y)(W_{ik}-\mu_k^Y)\}\notag \\
=&0,
\end{align}
where the second and the third equalities are based on the independence among $Y^m$, $G^m$, $W^m$, and the last equality follows from $E\{(Y_{ij}-\mu_j^Y)(Y_{ik}-\mu_k^Y)\}=E\{(G_{ij}-\mu_j^Y)(G_{ik}-\mu_k^Y)\}=E\{(W_{ij}-\mu_j^Y)(W_{ik}-\mu_k^Y)\}$. Thus, by combining (\ref{inequality:term34}) with (\ref{inequality:term35}), we have
\begin{align}\label{inequality:term36}
II^{**}=0.
\end{align}
For $III^{**}$, it could be decomposed as
\begin{align}\label{inequality:term37}
III^{**}=III_1^{**}+III_2^{**},
\end{align}
where
\begin{align*}
III_1^{**}=&\sum_{j=1}^p\sum_{k=1}^p\sum_{l=1}^p\sum_{i=1}^m\int_0^1\int_0^1(1-\tau)E\{\lambda_i^{**}\kappa_{jkl}(Z^{**(i)}+Z^{*}+\tau Z_i^{**})\\
&\cdot \dot{Z}_{ij}^{**}{Z}_{ik}^{**}{Z}_{il}^{**}\} d \tau d t, \\
III_2^{**}=&\sum_{j=1}^p\sum_{k=1}^p\sum_{l=1}^p\sum_{i=1}^m\int_0^1\int_0^1(1-\tau)E\{(1-\lambda_i^{**})\kappa_{jkl}(Z^{**(i)}+Z^{*}+\tau Z_i^{**})\\
&\cdot \dot{Z}_{ij}^{**}{Z}_{ik}^{**}{Z}_{il}^{**}\} d \tau d t.
\end{align*}
For $III_2^{**}$, it is easy to observe that
\begin{align}\label{inequality:term38}
|III_2^{**}|\leq& \sum_{j=1}^p\sum_{k=1}^p\sum_{l=1}^p\sum_{i=1}^m\int_0^1\int_0^1(1-\tau)E\{(1-\lambda_i^{**})\notag \\
&\cdot|\kappa_{jkl}(Z^{**(i)}+Z^{*}+\tau Z_i^{**})|\cdot |\dot{Z}_{ij}^{**}{Z}_{ik}^{**}{Z}_{il}^{**}|\} d \tau d t \notag \\
\leq&  \sum_{j=1}^p\sum_{k=1}^p\sum_{l=1}^p\sum_{i=1}^m\int_0^1\int_0^1(1-\tau)E\{(1-\lambda_i^{**})\notag \\
&\cdot Q_{jkl}(Z^{**(i)}+Z^{*}+\tau Z_i^{**})\cdot |\dot{Z}_{ij}^{**}{Z}_{ik}^{**}{Z}_{il}^{**}|\} d \tau d t\notag \\
\lesssim& \phi \beta^2 \sum_{i=1}^m\int_0^1 E\big\{\big(1-\lambda_i^{**}\big)\big(\max_{1\leq j, k, l \leq p} |\dot{Z}_{ij}^{**}{Z}_{ik}^{**}{Z}_{il}^{**}|\big)\big\} d t,
\end{align}
where the first inequality is by Jensen's inequality, and the second and the third inequalities are based on Holder's inequality and Lemma~1. Note that for every $i=1,\dots,m$, we have
\begin{align}\label{inequality:term39}
1-\lambda_i^{**}=&1\big\{\max_{1\leq j \leq p}|Y_{ij}-\mu_j^Y|\vee |G_{ij}-\mu_j^Y|\vee |W_{ij}-\mu_j^Y|> m^{1/2}/(4\beta |\delta_{n, m}|)\big\} \notag \\
\leq &  1\big\{\max_{1\leq j \leq p}|Y_{ij}-\mu_j^Y|> m^{1/2}/(4\beta |\delta_{n, m}|)\big\}+\notag \\
&1\big\{\max_{1\leq j \leq p}|G_{ij}-\mu_j^Y|> m^{1/2}/(4\beta |\delta_{n, m}|)\big\}+ \notag \\
&1\big\{\max_{1\leq j \leq p}|W_{ij}-\mu_j^Y|> m^{1/2}/(4\beta |\delta_{n, m}|)\big\}.
\end{align}
We also notice that for every $i=1,\dots,n$,
\begin{align}\label{inequality:term40}
&\max_{1\leq j, k, l \leq p}|\dot{Z}_{ij}^{**}{Z}_{ik}^{**}{Z}_{il}^{**}| \notag \\
\leq& 100 m^{-3/2}|\delta_{n, m}|^3\{t^{1/2}\wedge (1-t)^{1/2}\}^{-1} \notag \\
&\cdot \max_{1\leq j \leq p} |Y_{ij}-\mu_j^Y|^3\vee |G_{ij}-\mu_j^Y|^3 \vee |W_{ij}-\mu_j^Y|^3 \notag \\
\leq& 100 m^{-3/2}|\delta_{n, m}|^3\{t^{1/2}\wedge (1-t)^{1/2}\}^{-1} \notag \\
&\cdot (\max_{1\leq j \leq p} |Y_{ij}-\mu_j^Y|^3+\max_{1\leq j \leq p} |G_{ij}-\mu_j^Y|^3+\max_{1\leq j \leq p} |W_{ij}-\mu_j^Y|^3).
\end{align}
Therefore, under (\ref{inequality:term38}), (\ref{inequality:term39}) and (\ref{inequality:term40}), we have
\begin{align}\label{inequality:term41}
&|III_2^{**}| \notag \\
\lesssim & m^{-3/2}\phi \beta^2 |\delta_{n, m}|^3\sum_{i=1}^m E\Big(\Big[1\big\{\max_{1\leq j \leq p}|Y_{ij}-\mu_j^Y|> m^{1/2}/(4\beta |\delta_{n, m}|)\big\}\notag \\
&+1\big\{\max_{1\leq j \leq p}|G_{ij}-\mu_j^Y|> m^{1/2}/(4\beta |\delta_{n, m}|)\big\}\notag \\
&+1\big\{\max_{1\leq j \leq p}|W_{ij}-\mu_j^Y|> m^{1/2}/(4\beta |\delta_{n, m}|)\big\}\Big]\notag \\
&\cdot \Big[\max_{1\leq j \leq p} |Y_{ij}-\mu_j^Y|^3+\max_{1\leq j \leq p} |G_{ij}-\mu_j^Y|^3+\max_{1\leq j \leq p} |W_{ij}-\mu_j^Y|^3\Big]\Big) \notag \\
\lesssim & m^{-1/2}\phi \beta^2 |\delta_{n, m}|^3 \big\{M_m^Y(\phi)+M_m^G(\phi)+M_m^W(\phi)\big\}\notag \\
\lesssim & m^{-1/2}\phi \beta^2 |\delta_{n, m}|^3 \big\{M_m^Y(\phi)+M_m^G(\phi)\big\}\notag \\
\lesssim & m^{-1/2} \phi \beta^2 |\delta_{n, m}|^3 M_m^*(\phi),
\end{align}
where the second inequality follows from Lemma B.1 in \cite{chern:16:1}, and the third inequality is based on the fact that $M_m^G(\phi)=M_m^W(\phi)$. For $III_1^{**}$, we have
\begin{align}\label{inequality:term42}
&|III_1^{**}|\notag \\
\leq& \sum_{j=1}^p\sum_{k=1}^p\sum_{l=1}^p\sum_{i=1}^m\int_0^1\int_0^1(1-\tau)E\{\lambda_i^{**}|\kappa_{jkl}(Z^{**(i)}+Z^{*}+\tau Z_i^{**})|\notag \\
&\cdot |\dot{Z}_{ij}^{**}{Z}_{ik}^{**}{Z}_{il}^{**}|\} d \tau d t \notag \\
\leq&  \sum_{j=1}^p\sum_{k=1}^p\sum_{l=1}^p\sum_{i=1}^m\int_0^1\int_0^1 E\{\lambda_i^{**}|\kappa_{jkl}(Z^{**(i)}+Z^{*}+\tau Z_i^{**})|\notag \\
&\cdot |\dot{Z}_{ij}^{**}{Z}_{ik}^{**}{Z}_{il}^{**}|\} d \tau d t \notag \\
=&  \sum_{j=1}^p\sum_{k=1}^p\sum_{l=1}^p\sum_{i=1}^m\int_0^1\int_0^1 E\{\lambda_i^{**}h(Z^{**(i)}+Z^{*}, 1)\notag \\
&\cdot |\kappa_{jkl}(Z^{**(i)}+Z^{*}+\tau Z_i^{**})|\cdot |\dot{Z}_{ij}^{**}{Z}_{ik}^{**}{Z}_{il}^{**}|\} d \tau d t \notag \\
\lesssim&  \sum_{j=1}^p\sum_{k=1}^p\sum_{l=1}^p\sum_{i=1}^m\int_0^1\int_0^1 E\{\lambda_i^{**}h(Z^{**(i)}+Z^{*}, 1)Q_{jkl}(Z^{**(i)}+Z^{*}) \notag \\
&\cdot |\dot{Z}_{ij}^{**}{Z}_{ik}^{**}{Z}_{il}^{**}|\} d \tau d t \notag \\
\lesssim&  \sum_{j=1}^p\sum_{k=1}^p\sum_{l=1}^p\sum_{i=1}^m\int_0^1 E\{h(Z^{**(i)}+Z^{*}, 1)Q_{jkl}(Z^{**(i)}+Z^{*}) \notag \\
&\cdot |\dot{Z}_{ij}^{**}{Z}_{ik}^{**}{Z}_{il}^{**}|\}  d t \notag \\
\lesssim& \sum_{j=1}^p\sum_{k=1}^p\sum_{l=1}^p\sum_{i=1}^m\int_0^1 E\{h(Z^{**(i)}+Z^{*}, 1)Q_{jkl}(Z^{**(i)}+Z^{*})\}\notag \\
&\cdot E(|\dot{Z}_{ij}^{**}{Z}_{ik}^{**}{Z}_{il}^{**}|) d t \notag \\
\lesssim& \Delta_1^{**}+\Delta_2^{**},
\end{align}
where the third inequality follows from Lemma~1, and the terms $\Delta_1^{**}$ and $\Delta_2^{**}$ are
\begin{align*}
\Delta_1^{**}=&\sum_{j=1}^p\sum_{k=1}^p\sum_{l=1}^p\sum_{i=1}^m\int_0^1 E\{(1-\lambda_i^{**})h(Z^{**(i)}+Z^{*}, 1)\\
&\cdot Q_{jkl}(Z^{**(i)}+Z^{*})\} E(|\dot{Z}_{ij}^{**}{Z}_{ik}^{**}{Z}_{il}^{**}|) d t, \\
\Delta_2^{**}=&\sum_{j=1}^p\sum_{k=1}^p\sum_{l=1}^p\sum_{i=1}^m\int_0^1 E\{\lambda_i^{**} h(Z^{**(i)}+Z^{*}, 1)\\
&\cdot Q_{jkl}(Z^{**(i)}+Z^{*})\} E(|\dot{Z}_{ij}^{**}{Z}_{ik}^{**}{Z}_{il}^{**}|) d t.
\end{align*}
For $\Delta_1^{**}$, it can be deduced  that
\begin{align}\label{inequality:term43}
&\Delta_1^{**}\nonumber \\
=& \sum_{j=1}^p\sum_{k=1}^p\sum_{l=1}^p\sum_{i=1}^m\int_0^1 E\{(1-\lambda_i^{**})h(Z^{**(i)}+Z^{*}, 1)Q_{jkl}(Z^{**(i)}+Z^{*})\}\notag \\ &\cdot E(|\dot{Z}_{ij}^{**}{Z}_{ik}^{**}{Z}_{il}^{**}|) d t \nonumber \\
\leq& \sum_{i=1}^m\int_0^1 \big\{\max_{1\leq j, k, l \leq p} E(|\dot{Z}_{ij}^{**}{Z}_{ik}^{**}{Z}_{il}^{**}|)\big\}\sum_{j=1}^p\sum_{k=1}^p\sum_{l=1}^p E\big\{(1-\lambda_i^{**})\notag \\ &\cdot h(Z^{**(i)}+Z^{*}, 1)Q_{jkl}(Z^{**(i)}+Z^{*})\big\}  d t \nonumber \\
\lesssim& \phi \beta^2\sum_{i=1}^m\int_0^1 \big\{\max_{1\leq j, k, l \leq p} E(|\dot{Z}_{ij}^{**}{Z}_{ik}^{**}{Z}_{il}^{**}|)\big\}E\big\{(1-\lambda_i^{**})\notag \\ &\cdot h(Z^{**(i)}+Z^{*}, 1)\big\}  d t \nonumber \\
\lesssim& \phi \beta^2\sum_{i=1}^m\int_0^1 \big\{\max_{1\leq j, k, l \leq p} E(|\dot{Z}_{ij}^{**}{Z}_{ik}^{**}{Z}_{il}^{**}|)\big\}\big\{E(1-\lambda_i^{**})\big\}  d t \nonumber \\
\lesssim& m^{-1/2}\phi \beta^2 |\delta_{n, m}|^3 \Big[\int_0^1\{t^{1/2}\wedge (1-t)^{1/2}\}^{-1} d t\Big]\notag \\ &\cdot   \big\{M_m^Y(\phi)+M_m^G(\phi)+M_m^W(\phi)\big\} \nonumber \\
\lesssim& m^{-1/2}\phi \beta^2 |\delta_{n, m}|^3  \big\{M_m^Y(\phi)+M_m^G(\phi)\big\} \nonumber \\
\lesssim& m^{-1/2}\phi \beta^2 |\delta_{n, m}|^3 M_m^*(\phi),
\end{align}
where the second inequality is based on Lemma~1, and the fourth inequality follows from (\ref{inequality:term39}), (\ref{inequality:term40}) and Lemma B.1 in \cite{chern:16:1}.
For $\Delta_2^{**}$, it is clear that
\begin{align}\label{inequality:term44}
\Delta_2^{**}&=\sum_{j=1}^p\sum_{k=1}^p\sum_{l=1}^p\sum_{i=1}^m\int_0^1 E\{\lambda_i^{**} h(Z^{**(i)}+Z^{*}, 1)Q_{jkl}(Z^{**(i)}+Z^{*})\} \notag \\ &\cdot E(|\dot{Z}_{ij}^{**}{Z}_{ik}^{**}{Z}_{il}^{**}|) d t \nonumber \\
&\lesssim \sum_{j=1}^p\sum_{k=1}^p\sum_{l=1}^p\sum_{i=1}^m\int_0^1 E\{\lambda_i^{**} h(Z^{**(i)}+Z^{*}, 1)Q_{jkl}(Z^{**}+Z^{*})\} \notag \\ &\cdot E(|\dot{Z}_{ij}^{**}{Z}_{ik}^{**}{Z}_{il}^{**}|) d t \nonumber \\
&\lesssim \sum_{j=1}^p\sum_{k=1}^p\sum_{l=1}^p\sum_{i=1}^m\int_0^1 E\{\lambda_i^{**} h(Z^{**(i)}+Z^{*}, 1)h(Z^{*}+Z^{**}, 2)\notag \\ &\cdot Q_{jkl}(Z^{**}+Z^{*})\} E(|\dot{Z}_{ij}^{**}{Z}_{ik}^{**}{Z}_{il}^{**}|) d t \nonumber \\
&\lesssim \sum_{j=1}^p\sum_{k=1}^p\sum_{l=1}^p\sum_{i=1}^m\int_0^1 E\{h(Z^{*}+Z^{**}, 2)Q_{jkl}(Z^{**}+Z^{*})\} \notag \\ &\cdot E(|\dot{Z}_{ij}^{**}{Z}_{ik}^{**}{Z}_{il}^{**}|) d t \nonumber \\
&\lesssim \phi \beta^2  \int_0^1 E\{h(Z^{*}+Z^{**}, 2)\}\big\{\max_{1\leq j, k, l\leq p}\sum_{i=1}^m E(|\dot{Z}_{ij}^{**}{Z}_{ik}^{**}{Z}_{il}^{**}|)\big\} d t,
\end{align}
where the first inequality follows from Lemma~1, the second inequality is based on the fact that
$$\lambda_i^{**}h(Z^{**(i)}+Z^{*}, 1)=\lambda_i^{**}h(Z^{**(i)}+Z^{*}, 1)h(Z^{*}+Z^{**}, 2), \quad i=1,\dots,m,$$
and the fourth inequality follows from Holder's inequality and Lemma~1.
Moreover, it is obvious that
\begin{align*}
&\max_{1\leq j, k, l\leq p}\sum_{i=1}^m E(|\dot{Z}_{ij}^{**}{Z}_{ik}^{**}{Z}_{il}^{**}|)\notag \\
\lesssim& m^{-3/2}|\delta_{n, m}|^3\{t^{1/2}\wedge (1-t)^{1/2}\}^{-1}\notag \\ &\cdot \max_{1\leq j \leq p}\sum_{i=1}^m E(|Y_{ij}-\mu_j^Y|^3+|G_{ij}-\mu_j^Y|^3+|W_{ij}-\mu_j^Y|^3)\notag \\
\lesssim&  m^{-3/2}|\delta_{n, m}|^3\{t^{1/2}\wedge (1-t)^{1/2}\}^{-1}\notag \\ &\cdot \big\{\max_{1\leq j \leq p}\sum_{i=1}^m E(|Y_{ij}-\mu_j^Y|^3)+\max_{1\leq j \leq p}\sum_{i=1}^m E(|G_{ij}-\mu_j^Y|^3)\big\}.
\end{align*}
Together with the fact that
$E(|G_{ij}-\mu_j^Y|^3)\lesssim \{E(|G_{ij}-\mu_j^Y|^2)\}^{3/2} \lesssim \{E(|Y_{ij}-\mu_j^Y|^2)\}^{3/2}\lesssim E(|Y_{ij}-\mu_j^Y|^3)$, we have
\begin{align}\label{inequality:term45}
\max_{1\leq j, k, l\leq p}\sum_{i=1}^m E(|\dot{Z}_{ij}^{**}{Z}_{ik}^{**}{Z}_{il}^{**}|) \lesssim m^{-1/2}|\delta_{n, m}|^3\{t^{1/2}\wedge (1-t)^{1/2}\}^{-1}L_m^Y.
\end{align}
Thus, based on (\ref{inequality:term28}), (\ref{inequality:term44}) and (\ref{inequality:term45}), it is apparent that
\begin{align}\label{inequality:term46}
\Delta_2^{**}&\lesssim \phi \beta^2  \int_0^1 E\{h(Z^{*}+Z^{**}, 2)\}\big\{\max_{1\leq j, k, l\leq p}\sum_{i=1}^m E(|\dot{Z}_{ij}^{**}{Z}_{ik}^{**}{Z}_{il}^{**}|)\big\} d t \notag \\
&\lesssim m^{-1/2}\phi \beta^2 |\delta_{n, m}|^3 \big\{L_m^Y\rho_{n, m}+\phi^{-1}L_m^Y(\log p)^{1/2}\big\}.
\end{align}
Then, according to (\ref{inequality:term37}), (\ref{inequality:term41}), (\ref{inequality:term42}), (\ref{inequality:term43}) and (\ref{inequality:term46}), one has
\begin{align}\label{inequality:term47}
|III^{**}|\lesssim& m^{-1/2}\phi \beta^2 |\delta_{n, m}|^3 M_m^*(\phi)+\notag \\ & m^{-1/2}\phi \beta^2 |\delta_{n, m}|^3 \big\{L_m^Y\rho_{n, m}+\phi^{-1}L_m^Y(\log p)^{1/2}\big\}\notag \\
\lesssim& m^{-1/2} \beta^2 |\delta_{n, m}|^3 \big\{\phi L_m^Y\rho_{n, m}+L_m^Y(\log p)^{1/2}+\phi   M_m^*(\phi)\big\}\notag \\
\lesssim& m^{-1/2} \phi^2 (\log p)^2 |\delta_{n, m}|^3 \big\{\phi L_m^Y\rho_{n, m}+L_m^Y(\log p)^{1/2}+\phi   M_m^*(\phi)\big\}\notag \\
\lesssim& m^{-1/2} \phi_2^2 (\log p)^2 |\delta_{n, m}|^3 \big\{\phi_2 L_m^Y\rho_{n, m}+L_m^Y(\log p)^{1/2}+\phi_2   M_m^*(\phi_2)\big\},
\end{align}
where the last inequality is based on the facts that $\phi_2\geq \phi$ and $M_m^*(\phi_2)\geq M_m^*(\phi)$.
Thus, by combining (\ref{inequality:term32}), (\ref{inequality:term33}), (\ref{inequality:term36}) with (\ref{inequality:term47}), we have
\begin{align}\label{inequality:term48}
|E(\mathcal{I}_{n, m}^{**})|\lesssim& m^{-1/2} \phi_2^2 (\log p)^2 |\delta_{n, m}|^3 \notag \\ &\cdot \big\{\phi_2 L_m^Y\rho_{n, m}+L_m^Y(\log p)^{1/2}+\phi_2   M_m^*(\phi_2)\big\}.
\end{align}
To this end, by combining (\ref{inequality:term5}), (\ref{inequality:term31}) with (\ref{inequality:term48}), we conclude that
\begin{align}\label{inequality:term49}
|E(\mathcal{I}_{n, m})|\lesssim& n^{-1/2} \phi_1^2 (\log p)^2 \big\{\phi_1 L_n^X\rho_{n, m}+L_n^X(\log p)^{1/2}+\phi_1   M_n(\phi_1)\big\}+\notag \\
&m^{-1/2} \phi_2^2 (\log p)^2 |\delta_{n, m}|^3 \big\{\phi_2 L_m^Y\rho_{n, m}+L_m^Y(\log p)^{1/2}+\phi_2   M_m^*(\phi_2)\big\}.
\end{align}
Moreover, we have
\begin{align}\label{inequality:term50}
&P\{Z(1)\leq y-\phi^{-1}\}\notag \\
\leq& P\{F_{\beta}(Z(1))\leq 0\}\leq E\{\kappa(Z(1))\}=E\{\kappa(Z(0))\}+\notag \\ & E\{\kappa(Z(1))-\kappa(Z(0))\}\notag \\
=&E\{\kappa(Z(0))\}+E(\mathcal{I}_{n, m})\leq P\{F_{\beta}(Z(0))\leq \phi^{-1}\}+\notag \\ &E(\mathcal{I}_{n, m})\notag \\
\leq& P\{F_{\beta}(Z(0))\leq \phi^{-1}\}+|E(\mathcal{I}_{n, m})|\leq P\{Z(0)\leq y+\phi^{-1}\}+\notag \\ &|E(\mathcal{I}_{n, m})|\notag \\
=&P\{Z(0)\leq y-\phi^{-1}\}+\big[P\{Z(0)\leq y+\phi^{-1}\}-\notag \\ &P\{Z(0)\leq y-\phi^{-1}\}\big]+|E(\mathcal{I}_{n, m})|\notag \\
\leq& P\{Z(0)\leq y-\phi^{-1}\}+c_1\phi^{-1}(\log p)^{-1/2}+|E(\mathcal{I}_{n, m})|
\end{align}
where $c_1>0$ is a universal constant that depends only on $b$ defined in condition (a), the first inequality is by (\ref{inequality:term3}), the second  inequality is based on the fact that $1\{F_{\beta}(Z(1))\leq 0\}\leq \kappa(Z(1))$, the third  inequality is due to the fact that $1\{F_{\beta}(Z(0))> \phi^{-1}\}\leq 1-\kappa(Z(0))$, the fifth inequality is owing to (\ref{inequality:term3}), and the last inequality follows from combining condition (a) with Lemma A.1 in \cite{chern:16:1}. Likewise, we have
\begin{align}\label{inequality:term51}
&1-P\{Z(1)\leq y+\phi^{-1}\}\notag \\
\leq& P\{F_{\beta}(Z(1))> \phi^{-1}\}\leq 1-E\{\kappa(Z(1))\}=1-E\{\kappa(Z(0))\}-\notag \\ &E\{\kappa(Z(1))-\kappa(Z(0))\}\notag \\
=&1-E\{\kappa(Z(0))\}-E(\mathcal{I}_{n, m})\leq 1-P\{Z(0)\leq y-\phi^{-1}\}-E(\mathcal{I}_{n, m})\notag \\
=& 1-P\{Z(0)\leq y+\phi^{-1}\}+\big[P\{Z(0)\leq y+\phi^{-1}\}-\notag \\ &P\{Z(0)\leq y-\phi^{-1}\}\big]-E(\mathcal{I}_{n, m})\notag \\
\leq& 1-P\{Z(0)\leq y+\phi^{-1}\}+c_1\phi^{-1}(\log p)^{-1/2}+|E(\mathcal{I}_{n, m})|.
\end{align}
Finally, by combining (\ref{inequality:term49}), (\ref{inequality:term50}) with (\ref{inequality:term51}), we have
\begin{align*}
\rho_{n, m}\leq& c_1\phi^{-1}(\log p)^{-1/2}+\sup_{v\in [0, 1]}\sup_{y\in \mathbb{R}^p}|E(\mathcal{I}_{n, m})|\lesssim  (\min\{\phi_1, \phi_2\})^{-1}(\log p)^{1/2} \\
&+n^{-1/2}\phi_1^2(\log p)^2\big\{\phi_1 L_n^X \rho_{n,m}+L_n^X(\log p)^{1/2}+\phi_1 M_n(\phi_1)\big\}\\
&+m^{-1/2}\phi_2^2(\log p)^2|\delta_{n, m}|^3\big\{\phi_2 L_m^Y \rho_{n,m}+L_m^Y(\log p)^{1/2}+\phi_2 M_m^*(\phi_2)\big\},
\end{align*}
which completes the proof.
\qed

\noindent{\em Proof of Lemma~3}:\ \
First of all, given the data set $X^n=\{X_1,\dots,X_n\}$, we construct another data set $\tilde{X}^n=\{\tilde{X}_1,\dots,\tilde{X}_n\}$ with each $\tilde{X}_i=(\tilde{X}_{i1},\dots,\tilde{X}_{i,2p})'\in\mathbb{R}^{2p}$  such that $\tilde{X}_{ij}={X}_{ij}$ for $j=1,\dots,p$ and $\tilde{X}_{ij}=-{X}_{i,j-p}$ for $j=p+1,\dots,2p$. We further denote $E(\tilde{X}_i)=\mu^{\tilde{X}}$ for all $i=1,\dots,n$. Likewise, we also construct the data sets $\tilde{F}^n$, $\tilde{Y}^m$ and $\tilde{G}^m$. Based on these newly constructed data sets, we define the corresponding quantities $S_{n}^{\tilde{X}}$, $S_{n}^{\tilde{F}}$, $S_{m}^{\tilde{Y}}$, $S_{m}^{\tilde{G}}$, $L_{n}^{\tilde{X}}$, $L_{n}^{\tilde{F}}$, $L_{m}^{\tilde{Y}}$, $L_{m}^{\tilde{G}}$, $M_{n}^{\tilde{X}}(\phi)$, $M_{n}^{\tilde{F}}(\phi)$, $M_{m}^{\tilde{Y}}(\phi)$, $M_{m}^{\tilde{G}}(\phi)$, $\tilde{M}_{n}(\phi)=M_{n}^{\tilde{X}}(\phi)+M_{n}^{\tilde{F}}(\phi)$, $\tilde{M}_{m}^*(\phi)=M_{m}^{\tilde{Y}}(\phi)+M_{m}^{\tilde{G}}(\phi)$ analogously. Also denote
\begin{align*}
\tilde{\rho}_{n, m}=&\sup_{v\in [0, 1]}\sup_{\tilde{y}\in \mathbb{R}^{2p}}\big|P\big\{v^{1/2}(S_n^{\tilde{X}}+\delta_{n, m}S_m^{\tilde{Y}}-n^{1/2}\mu^{\tilde{X}}-\delta_{n, m}m^{1/2}\mu^{\tilde{Y}})+\notag \\ &(1-v)^{1/2}(S_n^{\tilde{F}}+\delta_{n, m}S_m^{\tilde{G}}-n^{1/2}\mu^{\tilde{X}}-\delta_{n, m}m^{1/2}\mu^{\tilde{Y}})\leq \tilde{y}\big\}-\notag \\ &P(S_n^{\tilde{F}}+\delta_{n, m}S_m^{\tilde{G}}-n^{1/2}\mu^{\tilde{X}}-\delta_{n, m}m^{1/2}\mu^{\tilde{Y}}\leq \tilde{y})\big|.
\end{align*}
It is not difficult to verify that
\begin{align}\label{inequality:term52}
&\min_{1\leq j \leq 2p}E\{(S_{nj}^{\tilde{X}}-n^{1/2}\mu_j^{\tilde{X}}+\delta_{n, m}S_{mj}^{\tilde{Y}}-\delta_{n, m}m^{1/2}\mu_j^{\tilde{Y}})^2\}\notag\\
=&\min_{1\leq j \leq p}E\{(S_{nj}^X-n^{1/2}\mu_j^X+\delta_{n, m}S_{mj}^Y-\delta_{n, m}m^{1/2}\mu_j^Y)^2\},
\end{align}
and
\begin{align}\label{inequality:term53}
L_{n}^{\tilde{X}}=L_{n}^X, \quad L_{m}^{\tilde{Y}}=L_{m}^Y, \quad \tilde{M}_{n}(\phi)={M}_{n}(\phi), \quad \tilde{M}_{m}^*(\phi)={M}_{m}^*(\phi),
\end{align}
for any $\phi\geq 1$.  Given any $A\in\mathcal{A}^{Re}$ with the expression $A=\{w\in \mathbb{R}^p: a_j\leq w_j \leq b_j \quad \text{for all} \quad j=1,\dots,p\}$, we let $\tilde{y}\in (\tilde{y}_1,\dots,\tilde{y}_{2p})'\in\mathbb{R}^{2p}$ such that $\tilde{y}_{j}=b_j$ for $j=1,\dots,p$ and $\tilde{y}_{j}=-a_{j-p}$ for $j=p+1,\dots,2p$. Then, it can be deduced that for any $v\in [0, 1]$,
\begin{align*}
&\big|P\big\{v^{1/2}(S_n^X+\delta_{n, m}S_m^Y-n^{1/2}\mu^X-\delta_{n, m}m^{1/2}\mu^Y)+\notag \\ &(1-v)^{1/2}(S_n^F+\delta_{n, m}S_m^G-n^{1/2}\mu^X-\delta_{n, m}m^{1/2}\mu^Y)\in A\big\}-\notag \\ &P(S_n^F+\delta_{n, m}S_m^G-n^{1/2}\mu^X-\delta_{n, m}m^{1/2}\mu^Y\in A)\big|\\
=&\big|P\big\{v^{1/2}(S_n^{\tilde{X}}+\delta_{n, m}S_m^{\tilde{Y}}-n^{1/2}\mu^{\tilde{X}}-\delta_{n, m}m^{1/2}\mu^{\tilde{Y}})+\notag \\ &(1-v)^{1/2}(S_n^{\tilde{F}}+\delta_{n, m}S_m^{\tilde{G}}-n^{1/2}\mu^{\tilde{X}}-\delta_{n, m}m^{1/2}\mu^{\tilde{Y}})\leq \tilde{y}\big\}-\notag \\ &P(S_n^{\tilde{F}}+\delta_{n, m}S_m^{\tilde{G}}-n^{1/2}\mu^{\tilde{X}}-\delta_{n, m}m^{1/2}\mu^{\tilde{Y}}\leq \tilde{y})\big|.
\end{align*}
This entails that
\begin{align}\label{inequality:term54}
&\rho_{n, m}^*\notag\\
=&\sup_{v\in [0, 1]}\sup_{A\in \mathcal{A}^{Re}}\big|P\big\{v^{1/2}(S_n^X+\delta_{n, m}S_m^Y-n^{1/2}\mu^X-\delta_{n, m}m^{1/2}\mu^Y)+\notag \\ &(1-v)^{1/2}(S_n^F+\delta_{n, m}S_m^G-n^{1/2}\mu^X-\delta_{n, m}m^{1/2}\mu^Y)\in A\big\}-\notag \\ &P(S_n^F+\delta_{n, m}S_m^G-n^{1/2}\mu^X-\delta_{n, m}m^{1/2}\mu^Y\in A)\big|\notag \\
=&\sup_{v\in [0, 1]}\sup_{\tilde{y}\in \mathbb{R}^{2p}}\big|P\big\{v^{1/2}(S_n^{\tilde{X}}+\delta_{n, m}S_m^{\tilde{Y}}-n^{1/2}\mu^{\tilde{X}}-\delta_{n, m}m^{1/2}\mu^{\tilde{Y}})+\notag \\ &(1-v)^{1/2}(S_n^{\tilde{F}}+\delta_{n, m}S_m^{\tilde{G}}-n^{1/2}\mu^{\tilde{X}}-\delta_{n, m}m^{1/2}\mu^{\tilde{Y}})\leq \tilde{y}\big\}-\notag \\ &P(S_n^{\tilde{F}}+\delta_{n, m}S_m^{\tilde{G}}-n^{1/2}\mu^{\tilde{X}}-\delta_{n, m}m^{1/2}\mu^{\tilde{Y}}\leq \tilde{y})\big|\notag \\
=&\tilde{\rho}_{n, m}.
\end{align}
In addition, by combining condition (a), (\ref{inequality:term52}) with Lemma~2, we have
\begin{align}\label{inequality:term55}
\tilde{\rho}_{n, m}\lesssim &n^{-1/2}\phi_1^2\{\log(2p)\}^2\big[\phi_1 L_n^{\tilde{X}} \tilde{\rho}_{n, m}+L_n^{\tilde{X}}\{\log (2p)\}^{1/2}+\phi_1 \tilde{M}_n(\phi_1)\big]+ \notag \\
&m^{-1/2}\phi_2^2\{\log(2p)\}^2|\delta_{n, m}|^3\notag \\ &\cdot \big[\phi_2 L_m^{\tilde{Y}} \tilde{\rho}_{n, m}+L_m^{\tilde{Y}}\{\log (2p)\}^{1/2}+\phi_2 \tilde{M}_m^*(\phi_2)\big]\notag\\
&+(\min\{\phi_1, \phi_2\})^{-1}\{\log(2p)\}^{1/2}.
\end{align}	
Hence by combining (\ref{inequality:term53}), (\ref{inequality:term54}), with (\ref{inequality:term55}), we conclude that
\begin{align*}
\rho_{n,m}^*\leq &K^*\Big[n^{-1/2}\phi_1^2(\log p)^2\big\{\phi_1 L_n^X \rho_{n,m}^*+L_n^X(\log p)^{1/2}+\phi_1 M_n(\phi_1)\big\}+\\
&m^{-1/2}\phi_2^2(\log p)^2|\delta_{n, m}|^3\big\{\phi_2 L_m^Y \rho_{n,m}^*+L_m^Y(\log p)^{1/2}+\phi_2 M_m^*(\phi_2)\big\}+\\
&(\min\{\phi_1, \phi_2\})^{-1}(\log p)^{1/2}\Big],
\end{align*}
for a universal constant $K^*>0$ that depends only on $b$, which completes the proof.	
\qed

\noindent{\em Proof of Lemma~4}:\ \
First, we have
\begin{align}\label{inequality:term56}
\rho_{n, m}^{**}\leq \rho_{n, m}^{*},
\end{align}
by definition.
Second, condition (a) together with  Lemma~3 implies that for every $\phi_1, \phi_2\geq 1$,
\begin{align}\label{inequality:term57}
\rho_{n,m}^*\leq &K^*\Big[8n^{-1/2}\phi_1^2(\log p)^2\big\{\phi_1 L_n^X \rho_{n,m}^*+2^{-1}L_n^X(\log p)^{1/2}+\phi_1 M_n(2\phi_1)\big\}+\notag \\
&8m^{-1/2}\phi_2^2(\log p)^2|\delta_{n, m}|^3\big\{\phi_2 L_m^Y \rho_{n,m}^*+2^{-1}L_m^Y(\log p)^{1/2}+\phi_2 M_m^*(2\phi_2)\big\}\notag \\
&+(2\phi_1)^{-1}(\log p)^{1/2}+(2\phi_2)^{-1}(\log p)^{1/2}\Big],
\end{align}	
for the universal constant $K^*>0$ defined in Lemma~3. Upon condition (b), we substitute $\phi_1=2^{-1}\phi_n^*$ and $\phi_2=2^{-1}\phi_m^{**}$ into (\ref{inequality:term57}) to obtain
\begin{align}\label{inequality:term58}
\rho_{n,m}^*\leq &K^*\Big\{n^{-1/2}(\phi_n^*)^3(\log p)^2 L_n^X \rho_{n,m}^*+n^{-1/2}(\phi_n^*)^2(\log p)^{5/2}L_n^X+\notag \\
&n^{-1/2}(\phi_n^*)^3(\log p)^{2} M_n(\phi_n^*)+m^{-1/2}(\phi_m^{**})^3(\log p)^2 L_m^Y |\delta_{n, m}|^3\rho_{n,m}^*+\notag \\
&m^{-1/2}(\phi_m^{**})^2(\log p)^{5/2}L_m^Y|\delta_{n, m}|^3+m^{-1/2}(\phi_m^{**})^3(\log p)^{2}|\delta_{n, m}|^3 M_m^*(\phi_m^{**})\notag \\
&+(\phi_n^*)^{-1}(\log p)^{1/2}+(\phi_m^{**})^{-1}(\log p)^{1/2}\Big\}\notag \\
\leq &K^*\Big\{n^{-1/2}(\phi_n^*)^3(\log p)^2 \bar{L}_n^* \rho_{n,m}^*+n^{-1/2}(\phi_n^*)^2(\log p)^{5/2}\bar{L}_n^*+\notag \\
&n^{-1/2}(\phi_n^*)^3(\log p)^{2} M_n(\phi_n^*)+m^{-1/2}(\phi_m^{**})^3(\log p)^2 \bar{L}_m^{**} |\delta_{n, m}|^3\rho_{n,m}^*+\notag \\
&m^{-1/2}(\phi_m^{**})^2(\log p)^{5/2}\bar{L}_m^{**}|\delta_{n, m}|^3+m^{-1/2}(\phi_m^{**})^3(\log p)^{2}|\delta_{n, m}|^3 M_m^*(\phi_m^{**})\notag \\
&+(\phi_n^*)^{-1}(\log p)^{1/2}+(\phi_m^{**})^{-1}(\log p)^{1/2}\Big\}\notag \\
=& K^*\Big[K_1^3\rho_{n,m}^*+K_1^2\{(\bar{L}_n^*)^2(\log p)^7/n\}^{1/6}+K_1^3\{M_n(\phi_n^*)/\bar{L}_n^*\}+K_1^3\rho_{n,m}^*\notag\\
&+K_1^2\{(\bar{L}_m^{**})^2(\log p)^7|\delta_{n, m}|^6/m\}^{1/6}+K_1^3\{M_m^*(\phi_m^{**})/\bar{L}_m^{**}\}+\notag\\
&K_1^{-1}\{(\bar{L}_n^*)^2(\log p)^7/n\}^{1/6}+K_1^{-1}\{(\bar{L}_m^{**})^2(\log p)^7|\delta_{n, m}|^6/m\}^{1/6}\Big]\notag \\
=&2K^*K_1^3\rho_{n,m}^*+K^*(K_1^2+K_1^{-1})\{(\bar{L}_n^*)^2(\log p)^7/n\}^{1/6}+\notag \\
&K^*K_1^3\{M_n(\phi_n^*)/\bar{L}_n^*\}+K^*(K_1^2+K_1^{-1})\{(\bar{L}_m^{**})^2(\log p)^7|\delta_{n, m}|^6/m\}^{1/6}\notag \\
&+K^*K_1^3\{M_m^*(\phi_m^{**})/\bar{L}_m^{**}\}\notag \\
\leq& 2^{-1}\rho_{n,m}^*+K^*(K_1^2+K_1^{-1})\{(\bar{L}_n^*)^2(\log p)^7/n\}^{1/6}+4^{-1}\{M_n(\phi_n^*)/\bar{L}_n^*\}\notag \\
&+K^*(K_1^2+K_1^{-1})\{(\bar{L}_m^{**})^2(\log p)^7|\delta_{n, m}|^6/m\}^{1/6}+4^{-1}\{M_m^*(\phi_m^{**})/\bar{L}_m^{**}\},
\end{align}	
where the last inequality is by $K_1\in (0, (K^*\vee 2)^{-1}]$. It then follows from (\ref{inequality:term58})  that
\begin{align}\label{inequality:term59}
\rho_{n,m}^{*}\leq& K_2\big[\{(\bar{L}_n^*)^2(\log p)^7/n\}^{1/6}+\{M_n(\phi_n^*)/\bar{L}_n^*\}+\notag \\
&\{(\bar{L}_m^{**})^2(\log p)^7|\delta_{n, m}|^6/m\}^{1/6} +\{M_m^*(\phi_m^{**})/\bar{L}_m^{**}\}\big],
\end{align}	
where $K_2=\max\{2K^*(K_1^2+K_1^{-1}), 2^{-1}\}$ is a universal constant that depends  on $b$ only. Putting (\ref{inequality:term56}) and (\ref{inequality:term59}) together leads to
\begin{align*}
\rho_{n,m}^{**}\leq& K_2\big[\{(\bar{L}_n^*)^2(\log p)^7/n\}^{1/6}+\{M_n(\phi_n^*)/\bar{L}_n^*\}+\notag \\
&\{(\bar{L}_m^{**})^2(\log p)^7|\delta_{n, m}|^6/m\}^{1/6}+\{M_m^*(\phi_m^{**})/\bar{L}_m^{**}\}\big],
\end{align*}	
which completes the proof.
\qed

\noindent{\em Proof of Lemma~5}:\ \
Given any $\phi>0$, we denote $\beta=\phi \log p$. Given any $y\in \mathbb{R}^p$, we define the function $F_\beta(w)$ as in (\ref{inequality:term2}) that implicitly depends on $y$. Then, piecing $F_\beta(w)$ and $\phi$ together, we construct a  function $\kappa(w)=\varphi_0(\phi F_{\beta}(w))=\varphi(F_{\beta}(w))$ according to (\ref{inequality:term4}). For any sequence of constants $\delta_{n, m}$ that depends on both $n$ and $m$, we denote $\rho_{n, m}^{MBB}$ by
\begin{align*}
\rho_{n, m}^{MBB}=&\sup_{y\in \mathbb{R}^p}\big|P_e(S_n^{eX}+\delta_{n, m}S_m^{eY}\leq y)-\notag\\
&P(S_n^F-n^{1/2}\mu^X+\delta_{n, m}S_m^G-\delta_{n, m}m^{1/2}\mu^Y \leq y)\big|,
\end{align*}
where $P_e(\cdot)$ means the probability with respect to $e^{n+m}$ only.
Now we are in a position to start the proof. First, it is clear that
\begin{align}\label{inequality:term90}
&P_e(S_n^{eX}+\delta_{n, m}S_m^{eY}\leq y-\phi^{-1})\notag\\
\leq&P_e\big\{F_\beta(S_n^{eX}+\delta_{n, m}S_m^{eY})\leq 0\big\}\leq
E_e\big\{\kappa(S_n^{eX}+\delta_{n, m}S_m^{eY})\big\}\notag\\
=&E\big\{\kappa(S_n^{F}+\delta_{n, m}S_m^{G}-n^{1/2}\mu^X-\delta_{n, m}m^{1/2}\mu^Y)\big\}+\big[E_e\big\{\kappa(S_n^{eX}+\delta_{n, m}S_m^{eY})\big\}\notag\\
&-E\big\{\kappa(S_n^{F}+\delta_{n, m}S_m^{G}-n^{1/2}\mu^X-\delta_{n, m}m^{1/2}\mu^Y)\big\}\big] \notag\\
\leq& P\big\{F_\beta(S_n^{F}+\delta_{n, m}S_m^{G}-n^{1/2}\mu^X-\delta_{n, m}m^{1/2}\mu^Y)\leq \phi^{-1}\big\}+\big[E_e\big\{\kappa(S_n^{eX}+\notag\\
&\delta_{n, m}S_m^{eY})\big\}-E\big\{\kappa(S_n^{F}+\delta_{n, m}S_m^{G}-n^{1/2}\mu^X-\delta_{n, m}m^{1/2}\mu^Y)\big\}\big] \notag\\
\leq& P(S_n^{F}+\delta_{n, m}S_m^{G}-n^{1/2}\mu^X-\delta_{n, m}m^{1/2}\mu^Y\leq y+\phi^{-1})+\big[E_e\big\{\kappa(S_n^{eX}+\notag\\
&\delta_{n, m}S_m^{eY})\big\}-E\big\{\kappa(S_n^{F}+\delta_{n, m}S_m^{G}-n^{1/2}\mu^X-\delta_{n, m}m^{1/2}\mu^Y)\big\}\big] \notag\\
\leq& P(S_n^{F}+\delta_{n, m}S_m^{G}-n^{1/2}\mu^X-\delta_{n, m}m^{1/2}\mu^Y\leq y-\phi^{-1})+c_1\phi^{-1}(\log p)^{1/2}+\notag\\
&\big|E_e\big\{\kappa(S_n^{eX}+\delta_{n, m}S_m^{eY})\big\}-E\big\{\kappa(S_n^{F}+\delta_{n, m}S_m^{G}-n^{1/2}\mu^X-\delta_{n, m}m^{1/2}\mu^Y)\big\}\big|,
\end{align}
where $c_1>0$ is a universal constant depending only on $b$, and the last inequality follows from combining condition (a) with Lemma A.1 in \cite{chern:16:1}.
To bound the term $\big|E_e\big\{\kappa(S_n^{eX}+\delta_{n, m}S_m^{eY})\big\}-E\big\{\kappa(S_n^{F}+\delta_{n, m}S_m^{G}-n^{1/2}\mu^X-\delta_{n, m}m^{1/2}\mu^Y)\big\}\big|$ in (\ref{inequality:term90}), one has
\begin{align}\label{inequality:term91}
&\big|E_e\big\{\kappa(S_n^{eX}+\delta_{n, m}S_m^{eY})\big\}-E\big\{\kappa(S_n^{F}+\delta_{n, m}S_m^{G}-n^{1/2}\mu^X-\delta_{n, m}m^{1/2}\mu^Y)\big\}\big|\notag \\
=&\big|E_e\big\{\varphi(F_\beta(S_n^{eX}+\delta_{n, m}S_m^{eY}))\big\}-\notag\\
&E\big\{\varphi(F_\beta(S_n^{F}+\delta_{n, m}S_m^{G}-n^{1/2}\mu^X-\delta_{n, m}m^{1/2}\mu^Y))\big\}\big|\notag \\
\leq& (\|\varphi''\|_{\infty}/2+\beta \|\varphi'\|_{\infty})\hat{\Delta}_{n, m}\notag \\
\lesssim& (\phi^2+\beta \phi)\hat{\Delta}_{n, m}\notag \\
\lesssim& (\phi^2\log p)\hat{\Delta}_{n, m},
\end{align}
where the first inequality follows from Theorem~1 in \cite{chern:14:1}. Then it follows from  (\ref{inequality:term90}) and  (\ref{inequality:term91}) that
\begin{align}\label{inequality:term92}
&P_e(S_n^{eX}+\delta_{n, m}S_m^{eY}\leq y-\phi^{-1})-\notag\\
&P(S_n^{F}+\delta_{n, m}S_m^{G}-n^{1/2}\mu^X-\delta_{n, m}m^{1/2}\mu^Y\leq y-\phi^{-1})\notag \\
\lesssim& \phi^{-1}(\log p)^{1/2}+(\phi^2\log p)\hat{\Delta}_{n, m}.
\end{align}
Similar argument leads to
\begin{align}\label{inequality:term93}
&P(S_n^{F}+\delta_{n, m}S_m^{G}-n^{1/2}\mu^X-\delta_{n, m}m^{1/2}\mu^Y\leq y-\phi^{-1})\notag\\
&-P_e(S_n^{eX}+\delta_{n, m}S_m^{eY}\leq y-\phi^{-1})\notag \\
\lesssim& \phi^{-1}(\log p)^{1/2}+(\phi^2\log p)\hat{\Delta}_{n, m}.
\end{align}
As a consequence, piecing (\ref{inequality:term92}) and  (\ref{inequality:term93}) together yields that
\begin{align}\label{inequality:term94}
\rho_{n, m}^{MBB} \lesssim \phi^{-1}(\log p)^{1/2}+(\phi^2\log p)\hat{\Delta}_{n, m}.
\end{align}
By substituting $\phi=2^{-1/3}(\log p)^{-1/6}(\hat{\Delta}_{n, m})^{1/3}$ into (\ref{inequality:term94}), we obtain
\begin{align*}
\rho_{n, m}^{MBB} \lesssim (\log p)^{2/3}\hat{\Delta}_{n, m}.
\end{align*}
Together with the same reasoning as in the proof of Lemma~3, we conclude that
\begin{align*}
\rho_{n, m}^{MB} \lesssim (\log p)^{2/3}\hat{\Delta}_{n, m},
\end{align*}
which completes the proof.
\qed

\noindent{\em Proof of Lemma~6}:\ \
First, condition (a) entails that
\begin{align}\label{inequality:term115}
\lim_{k_1\to\infty}\lim_{k_2\to\infty}\sum_{n=k_1}^{\infty}\sum_{m\in \varrho(k_2)\cap\sigma(n)}P(A_{n, m})=0.
\end{align}
In addition, one has
\begin{align}\label{inequality:term116}
P\big(\bigcap_{k_1=1}^{\infty}\bigcap_{k_2=1}^{\infty}\bigcup_{n=k_1}^{\infty}\bigcup_{m\in \varrho(k_2)\cap\sigma(n)}A_{n, m}\big)\leq& P\big(\bigcup_{n=k_1}^{\infty}\bigcup_{m\in \varrho(k_2)\cap\sigma(n)}A_{n, m}\big)\notag \\ \leq&\sum_{n=k_1}^{\infty}\sum_{m\in \varrho(k_2)\cap\sigma(n)}P(A_{n, m}),
\end{align}
for all $k_1, k_2\geq 1$. Finally, by combining (\ref{inequality:term115}) with (\ref{inequality:term116}), we conclude that
\begin{align*}
P\big(\bigcap_{k_1=1}^{\infty}\bigcap_{k_2=1}^{\infty}\bigcup_{n=k_1}^{\infty}\bigcup_{m\in \varrho(k_2)\cap\sigma(n)}A_{n, m}\big)\leq& \lim_{k_1\to\infty}\lim_{k_2\to\infty}\sum_{n=k_1}^{\infty}\sum_{m\in \varrho(k_2)\cap\sigma(n)}P(A_{n, m})\notag \\ =&0.
\end{align*}
\qed

\bibliographystyle{imsart-number}
\bibliography{bib-xue,5-31-09-xue}